%% file: paper.tex
\documentclass[sigconf]{acmart}

\input{tex/preamble}

\AtBeginDocument{%
  \providecommand\BibTeX{{%
    \normalfont B\kern-0.5em{\scshape i\kern-0.25em b}\kern-0.8em\TeX}}}

\copyrightyear{2023} 
\acmYear{2023} 
\setcopyright{acmlicensed}\acmConference[ISCA '23]{Proceedings of the 50th Annual International Symposium on Computer Architecture}{June 17--21, 2023}{Orlando, FL, USA}
\acmBooktitle{Proceedings of the 50th Annual International Symposium on Computer Architecture (ISCA '23), June 17--21, 2023, Orlando, FL, USA}
\acmPrice{15.00}
\acmDOI{10.1145/3579371.3589038}
\acmISBN{979-8-4007-0095-8/23/06}

\begin{document}

\title{OliVe: Accelerating Large Language Models via\\Hardware-friendly Outlier-Victim Pair Quantization}


\author{Cong Guo}
\authornote{Contribute equally to this paper.}
\email{guocong@sjtu.edu.cn}
\affiliation{%
  \institution{Shanghai Jiao Tong University}
  \institution{Shanghai Qi Zhi Institute}
  \city{Shanghai}
  \country{China}
}

\author{Jiaming Tang}
\email{sakits_tjm@sjtu.edu.cn}
\authornotemark[1]
\affiliation{%
\institution{Shanghai Jiao Tong University}
\institution{Shanghai Qi Zhi Institute}
\city{Shanghai}
\country{China}
}

\author{Weiming Hu}
\authornote{Work done while affiliated with ShanghaiTech University.}
\email{huweim1120@gmail.com}
\affiliation{%
\institution{Shanghai Jiao Tong University}
\institution{Shanghai Qi Zhi Institute}
\city{Shanghai}
\country{China}
}

\author{Jingwen Leng}
\authornote{Jingwen Leng and Minyi Guo are corresponding authors of this paper.}
\email{leng-jw@cs.sjtu.edu.cn}
\affiliation{%
\institution{Shanghai Jiao Tong University}
\institution{Shanghai Qi Zhi Institute}
\city{Shanghai}
\country{China}
}

\author{Chen Zhang}
\email{chzhang1990@gmail.com}
\affiliation{%
\institution{Microsoft Research}
\city{Beijing}
\country{China}
}

\author{Fan Yang}
\email{fanyang@microsoft.com}
\affiliation{%
\institution{Microsoft Research}
\city{Beijing}
\country{China}
}

\author{Yunxin Liu}
\email{liuyunxin@air.tsinghua.edu.cn}
\affiliation{%
\institution{Institute for AI Industry Research (AIR), Tsinghua University}
\city{Beijing}
\country{China}
\\
\institution{Shanghai Artificial Intelligence Laboratory}
\city{Shanghai}
\country{China}
}

\author{Minyi Guo}
\email{guo-my@cs.sjtu.edu.cn}
\authornotemark[3]
\affiliation{%
\institution{Shanghai Jiao Tong University}
\institution{Shanghai Qi Zhi Institute}
\city{Shanghai}
\country{China}
}

\author{Yuhao Zhu}
\email{yzhu@rochester.edu}
\affiliation{%
\institution{University of Rochester}
  \city{Rochester}
  \state{New York}
  \country{USA}
}

\renewcommand{\shortauthors}{Guo and Tang, et al.}


\input{tex/abstract}



\maketitle

\input{tex/introduction}
\input{tex/motivation}
\input{tex/encoding}
\input{tex/architecture}
\input{tex/evaluation}

\input{tex/conclusion}

\begin{acks}
This work was supported by the National Key R\&D Program of China under Grant 2022YFB4501401, the National Natural Science Foundation of China (NSFC) grant (62222210, and 62072297, and 61832006).
The authors would like to thank the anonymous reviewers for their constructive feedback for improving the work. 
We also thank Tailong Wangliu, Shuangjie Ruan for their continuous support.
\end{acks}

\bibliographystyle{ACM-Reference-Format}
\bibliography{paper}

\end{document}

%% file: tex/preamble.tex
\usepackage{tikz}
\usepackage[normalem]{ulem}
\usepackage{multicol}
\usepackage{pifont}
\usepackage{enumitem}
\usepackage{multirow}
\usepackage{color}
\usepackage{listings}
\usepackage{float}
\usepackage{ragged2e}
\usepackage{xspace}
\usepackage{soul}
\usepackage{lipsum}
\usepackage{makecell}
\usepackage{mathtools}
\usepackage{subcaption}
\usepackage[keeplastbox]{flushend}
\usepackage{listings}
\usepackage{multirow}
\usepackage{wrapfig}
\usepackage{etoolbox}
\usepackage{amsmath,amsfonts}
\usepackage{algorithmic}
\usepackage{graphicx}
\usepackage{textcomp}
\usepackage{xcolor}
\usepackage[ruled,vlined,linesnumbered,algo2e]{algorithm2e}

\newcommand{\proj}{\benchmark{OliVe}\xspace}
\newcommand{\Fig}[1]{Fig.~\ref{#1}}

\newcommand{\Tbl}[1]{Tbl.~\ref{#1}}
\newcommand{\Sec}[1]{Sec.~\ref{#1}}

\newcommand{\benchmark}[1]{{\texttt{#1}}}
\renewcommand{\paragraph}[1]{\vspace*{0.15cm}\noindent\textbf{#1}\hspace*{.1cm}}

\newcommand{\ra}[1]{\renewcommand{\arraystretch}{#1}}

\definecolor{myyellow}{RGB}{254, 230, 153}
\definecolor{myblue}{RGB}{193, 228, 248}
\definecolor{mygreen}{RGB}{24, 154, 76}

\newcommand{\revise}[1]{{\color{black}{#1}}}

%% file: tex/abstract.tex
\begin{abstract} 

Transformer-based large language models (LLMs) have achieved great success with the growing model size. LLMs' size grows by $240\times$ every two years, which outpaces the hardware progress and makes model inference increasingly costly.
Model quantization is a promising approach to mitigate the widening gap between LLM size and hardware capacity.
However, the existence of outliers, values with significant magnitudes, in LLMs makes existing quantization methods less effective.
Prior outlier-aware quantization schemes adopt sparsity encoding techniques to separate outliers from normal values where the process requires \emph{global} coordination (e.g., a global sparsity coordination list). This incurs complex encoding/decoding hardware logics and an extra orchestration controller for the computation between outlier and normal values. 
As such, it is not hardware-efficient and hence only achieves sub-optimal quantization benefits.

We propose \proj{}, an algorithm/architecture co-designed solution that adopts an outlier-victim pair (OVP) quantization and handles outlier values \emph{locally} with low hardware overheads and high performance gains.
The key insight of \proj{} is that outliers are important while the normal values \emph{next} to them are not. Thus those normal values (called victims) can be sacrificed to accommodate outliers.
This enables a memory-aligned OVP encoding scheme, which can be efficiently integrated to the existing hardware accelerators like systolic array and tensor core.
As a result, \proj{}-based accelerator surpasses the existing outlier-aware accelerator, GOBO, by 4.5$\times$ speedup and 4.0$\times$ energy reduction, respectively, with a superior model accuracy.

%

\end{abstract}

%% file: tex/introduction.tex
\section{Introduction}\label{sec:introduction}

Transformer-based large language models (LLMs)~\cite{vaswani2017attention} have demonstrated great success in the past years. 
Such success is often achieved with the increasingly larger model size:
the model size grows by $240\times$ every two years, significantly outpacing the hardware progress ($3.1\times$ per two years)~\cite{gholami2020ai}.
As a result, the inference of LLMs becomes challenging and costly.
For instance, OPT-175B~\cite{zhang2022opt}, a recent Transformer-based LLM, has 175 billion parameters, which cannot fit in the latest high-end H100 GPU with 80GB memory.

Quantization~\cite{zhou2016dorefa, wang2019haq, dong2019hawq, shen2020q, dong2020hawq, cai2020zeroq, cai2020rethinking, sharma2018bit} is one of the most hardware-efficient ways to reduce inference costs for large models. It uses low-precision data types to compress models and accelerate the computation with practical hardware implementations, e.g., TPU~\cite{jouppi2017datacenter} and GPU tensor core~\cite{a100}.


However, existing quantization schemes~\cite{zafrir2019q8bert, shen2020q, dettmers2022llm} are less effective in Transformer-based LLMs.
Recent studies show when the model size exceeds a threshold (e.g., 6 billion), the model performance is vulnerable to only a tiny fraction ($<0.1\%$) of outliers, whose values are much more significant than normal values~\cite{dettmers2022llm}. Indiscriminately clipping both outlier and normal values will lead to significant drops in model accuracy~\cite{dettmers2022llm, wei2022outlier}. As a result, the common practice is to adopt a larger bit-width, e.g., 8-bit or 16-bit, to quantize Transform-based models, compared to convolutional networks (CNNs).


\begin{figure}[t] 
    \centering 
    \includegraphics[width=1\linewidth]{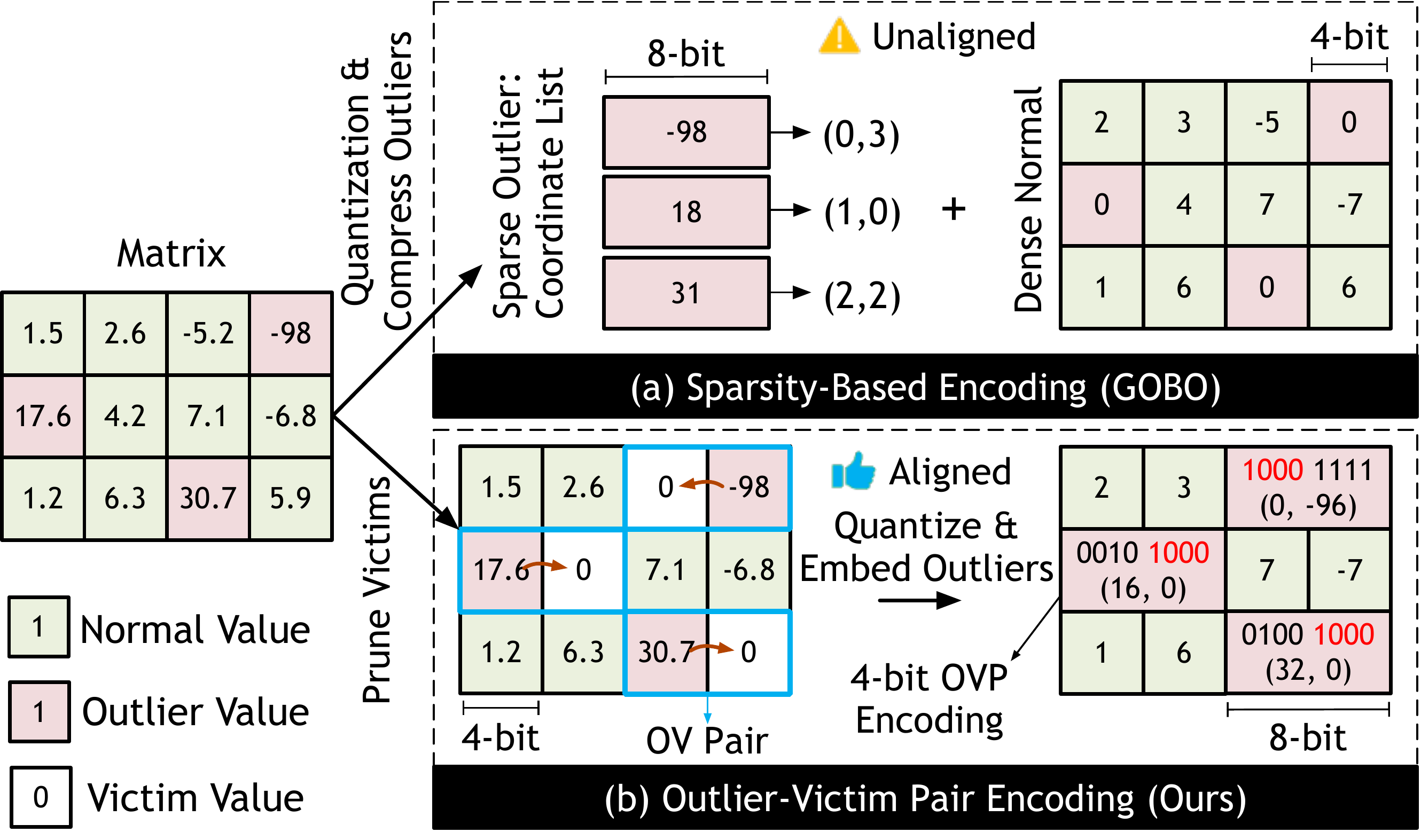}  
    \caption{\revise{Outlier-aware encoding comparison. (a) Prior quantization works adopt sparsity-based encoding that store normal and outlier values separately. (b) Our proposed outlier-victim pair encoding stores normal and outlier values locally.} }
    \label{fig:intro_victor}
    \vspace*{-0.2cm}
\end{figure}

Researchers have proposed various quantization/architecture co-design works~\cite{park2018energy,jain2019biscaled,song2020drq, zadeh2020gobo, wei2022outlier} to deal with the outliers in Transformer models. For example, outlier suppression~\cite{wei2022outlier} proposes to suppress the outliers. But it still has significant accuracy loss in the lower bit-width (4-bit), suggesting the difficulty in accommodating the effects of outliers. 
In addition, architecture researchers have designed sophisticated outlier-aware hardware architectures to store outliers with high precision to maintain model accuracy. 
These outlier-aware quantization frameworks divide the tensor into normal and outlier values, and encode them separately using different ways. 
For normal values, a dense matrix with low precision (e.g., 4-bit) quantization is adopted.
And the sparse and high-precision (e.g., 8-bit and 16-bit) outlier values can be compressed with sparsity-based encoding. Such encoding unfortunately leads to unaligned memory access.
For example, GOBOs~\cite{zadeh2020gobo} and OLAccels~\cite{park2018energy} use the coordinate list to indicate the location of each outlier value in the matrix, as shown in \Fig{fig:intro_victor}a. BiScaled-DNNs~\cite{jain2019biscaled} exploits block sparse indices format to store the outlier indices, and DRQ~\cite{song2020drq} uses the direct bitmap for outliers. 
These outlier-aware solutions require complex architectural designs with significant hardware overheads to accommodate outliers. 
Moreover, due to the random and unaligned memory access, the sparsity-based encoding is incompatible with the memory sub-systems of existing accelerators, such as GPU and TPU. Specifically, GOBO~\cite{zadeh2020gobo} can only de/compress weight tensors on the off-chip DRAM, it still relies on the original on-chip memory and computation architecture of GPU with high precision FP16/32.

The aforementioned outlier-aware architectures separate normal values from outliers in a \textbf{\emph{global}} way. 
For instance, GOBO~\cite{zadeh2020gobo} involves a global sparse coordinate list in the quantization and computation, leading to a large hardware overhead and low performance benefits.
In this work, we aim to design an architecture to handle \revise{outliers} in a \textbf{\emph{localized}} way with high hardware efficiency.
To achieve that, we group two consecutive fixed-size values in a tensor and analyze their impact to model accuracy.
There can be three kinds of \revise{pairs}: i) a normal \revise{pair} with two normal values, ii) one-outlier \revise{pair} with one normal value and one outlier value, iii) two-outlier \revise{pair} with two outlier values.
We observe that the third two-outlier \revise{pair} almost never shows up in well-trained LLMs.
For the second one-outlier \revise{pair}, we find that \emph{only keeping its outlier value while pruning its normal value} (i.e., treating it as zero) is sufficient to maintain the model accuracy.

Based on the above observations, we propose a novel outlier-aware quantization architecture, called \proj{}, based on the outlier-victim pair (OVP) encoding. 
The salient feature of \proj{} is memory-aligned and therefore hardware-friendly.
As illustrated in \Fig{fig:intro_victor}b, \proj{} first prunes normal values that are adjacent to the outliers as zero.
These pruned normal values are called \textbf{victims}, which sacrifice themselves and make space for outliers.
Then, we exploit the extra space provided by victims and embed the 
outliers into the low-precision matrix.


\proj{} is able to maintain a high accuracy for large Transformer models with a low hardware overhead due to the following reasons.
First, \proj{} incorporates victims to tackle outliers in LLMs. The effects of victims resemble model pruning~\cite{han2015deep}. Although clipping \revise{a} few (0.1\%) outliers will lead to \revise{a} disastrous accuracy drop~\cite{dettmers2022llm, wei2022outlier}, pruning the same amount \revise{of} ``normal'' values will only impact model accuracy slightly ($<0.1\%$ drop). Therefore, \proj{} sacrifices (``prunes'') those insignificant values as victims for the outliers, allowing a more aggressive encoding scheme to accommodate extremely significant values.
Second, the OVP encoding follows a specific outlier-victim (or victim-outlier) pattern to achieve memory alignment with little hardware overheads. 
Each victim is adjacent to an outlier, and the outlier-victim pair must align the memory access pattern. For example, in \Fig{fig:intro_victor}b, \revise{right} outlier \revise{$-98$ in the OV pair} needs a left victim, and \revise{left outliers} \revise{$17.6$ and $30.7$} require the right victims. That can align 8-bit (1-byte) memory accesses with high efficiency. 
This design enables a completely localized outlier decoding/encoding process.


To implement \proj{}, different data types are employed for outliers and normal values, which have different dynamic ranges and representation formats, including \revise{\texttt{int4} and \texttt{FP4}}.
\revise{As shown in \Fig{fig:intro_victor}b, we propose a novel encoding method (\Sec{sec:encoding}) for the 4-bit OV pair, which composes a 4-bit outlier and a 4-bit victim into a special 8-bit format and differs from the original \texttt{int8} or \texttt{FP8}.}
Due to its \revise{hardware-friendly and} compatible design, \proj{} can be easily integrated into existing quantization frameworks and accelerator architectures such as systolic array in Google TPUs~\cite{tpuv4isca} and tensor core in NVIDIA GPUs~\cite{v100,a100}.
\proj{} can also inherently support the mixed-precision and mixed-type architecture, showing its flexibility and practicality for larger-scale Transformer models.

To the best of our knowledge, \proj{} is the first work pushing the limit of Transformer post-training quantization (PTQ)~\cite{banner2019post}, which requires no \revise{retraining} after quantization, to the 4-bit level for both the weight and activation tensors with the accuracy loss of $< 1\%$.
Surprisingly, \proj{}'s 4-bit PTQ accuracies for BERT~\cite{devlin2018bert} and BART~\cite{lewis2019bart} models outperform the 6-bit PTQ results of outlier suppression~\cite{wei2022outlier}, a state-of-the-art Transformer quantization method.
\proj{}-based accelerator surpasses the existing outlier-aware accelerators OLAccel~\cite{park2018energy} and GOBO~\cite{zadeh2020gobo} by $3.8\times$ and $4.5\times$ performance improvement, and $2.1\times$ and $4.0\times$ energy reduction, respectively.  
More importantly, the \proj{}-based accelerator has more comprehensive and practical applicability than other outlier-specific architectures.

We make the following contributions in this paper.
\begin{itemize}
    \item We conduct the pair-wise importance analysis and show that outliers are important while their adjacent normal values are not, revealing the algorithmic opportunity of outlier-victim pair (OVP) that sacrifices the colocated normal values (called victims) to accommodate the outliers.  
    \item We propose the OVP-based quantization framework, called \proj{}, which includes an efficient hardware encoding and novel outlier representation data type.
    \item We propose the efficient architectural implementation and integration of \proj{} quantization, and show that its efficiency and benefits outperform the existing outlier-aware quantization algorithms and hardware accelerators.
\end{itemize}

%% file: tex/motivation.tex
\section{Motivation: Aligned Outlier}
\label{sec:motivation}
\begin{figure}[t]
    \centering
    \subfloat[ResNet-18 on ImageNet.]{\label{fig:cnn_sigma}
    \includegraphics[width=0.23\textwidth]{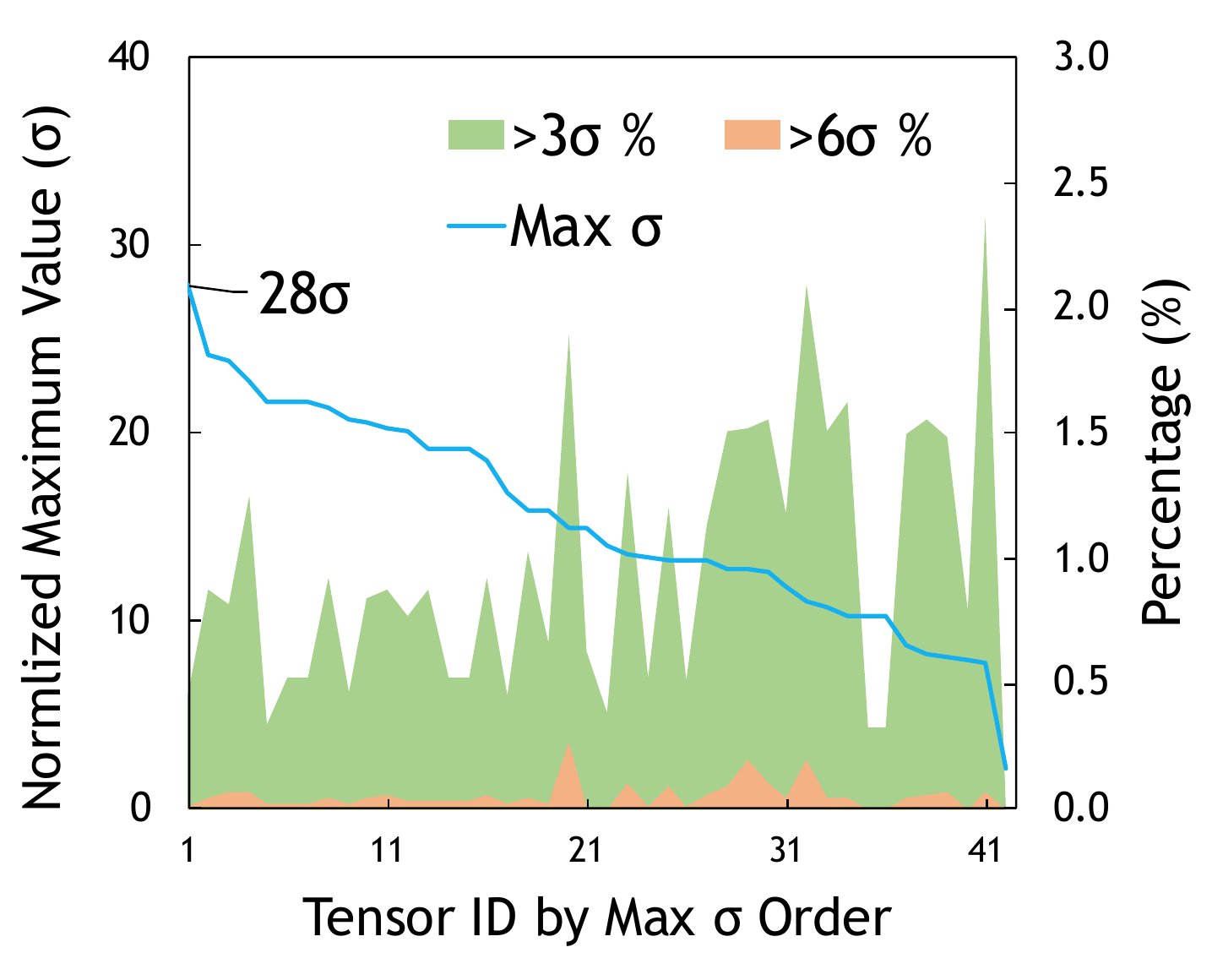}}
    \subfloat[BERT$_{base}$ on MNLI.]{\label{fig:bert_sigma}
    \includegraphics[width=0.23\textwidth]{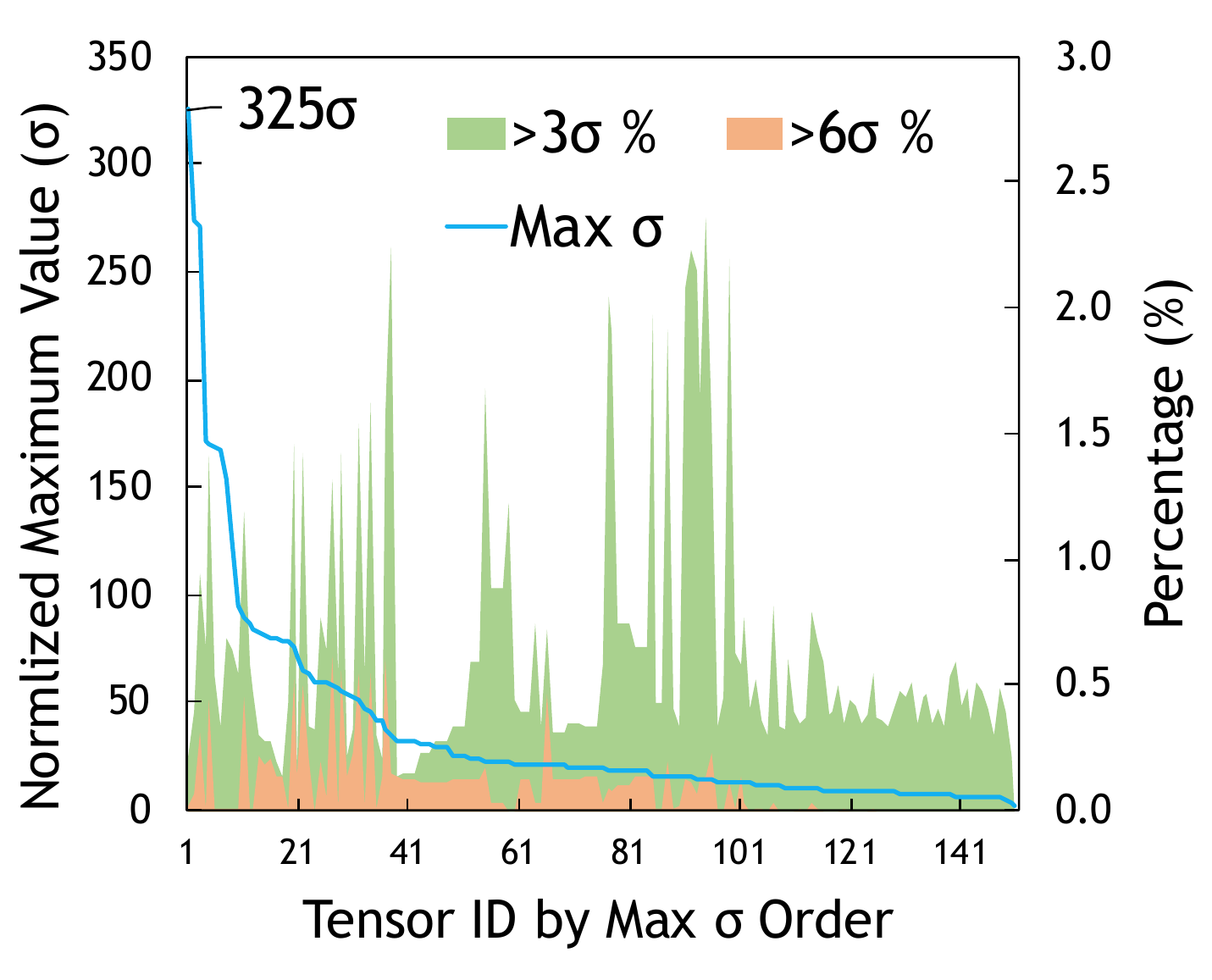} }
    
    \caption{Outlier Comparison of CNN model and Transformer model. The $\sigma$ is the standard deviation of the tensor. We normalize the maximum number by $\sigma$ to plot the Max $\sigma$ curve (left y-axis). The $>3\sigma \%$ and $>6\sigma \%$ (right y-axis) are the percentage of the values of $>3\sigma$ and $>6\sigma$, respectively.}\label{fig:sigma}
    \vspace*{-0.5cm}
\end{figure}

In this section, we first show that the outlier of the Transformer model is much more significant and important compared to convolution neural networks (CNN).
Previous works~\cite{zafrir2019q8bert, shen2020q, zadeh2020gobo,song2020drq} propose the outlier-aware quantization microarchitecture with adaptive bit length to accomplish the low-bit quantization but necessitate substantial hardware resources to deal with the variable-length data, which cause unaligned memory accesses and are incompatible with the memory sub-system of existing accelerators, e.g., GPU~\cite{a100}.
In contrast, we propose a memory-aligned and hardware-friendly method, called outlier-victim pair mechanism, which is inspired by DNN pruning and our outlier group location analysis for Transformers. We can prune some ``victims'' to make space to embed high-precision outliers into the memory-aligned low-bit tensor with ignorable accuracy loss.

\subsection{Outlier Matters}
We visually demonstrate how significant the Transformer's outlier is in \Fig{fig:sigma}. We adopt the empirical \textbf{$3\sigma$ rules}~\cite{enwiki:1116981313} of the normal distribution to divide the values into outlier and normal values.
We employ the ResNet-18~\cite{he2016deep} as the representative for the CNN model and the BERT$_{base}$~\cite{devlin2018bert} for the Transformer model.
We fit the DNN tensors with normal distribution, i.e., Equation~\ref{EQ:pdf}, where $x$ is the value, $\mu$ is the mean, and $\sigma$ is the standard deviation. We convert the tensor into a standard normal distribution.
\begin{equation}\label{EQ:pdf}
    \mathit{f}(x)= \frac{1}{\sigma\sqrt{2\pi} } e^{-\frac{1}{2}\left(\frac{x-\mu}{\sigma}\right)^2}
\end{equation}
We collect all tensors' maximum values and normalize them by the $\sigma$ (Max $\sigma$). We sort and plot the tensors by their Max $\sigma$ in \Fig{fig:sigma}.

Most tensors can fit the normal distribution $3\sigma$ rules, i.e., about 99.7\% of the values lie within three standard deviations of the mean. 
The outlier ($>3\sigma$) ratio of most tensors is lower than $0.5\%$, and the values of $>6\sigma$ are extremely few in tensors.
Therefore, normal values are relatively concentrated, indicating that we can quantize the normal values with a narrow range to enhance the resolution of quantization.

The more obvious observation is that the Max $\sigma$ of the Transformer is larger than that of CNN by one order of magnitude. 
Some research~\cite{choi2018pact,jung2019learning} shows that although the outliers are clipped for CNN models, the accuracy can still be restored to the original value with the \revise{retraining} algorithm under ultra-low-bit precision, e.g., 4-bit. 
However, it is challenging for Transformer models, which have much more significant outliers.
The state-of-the-art quantization works~\cite{dettmers2022llm,wei2022outlier} also demonstrate a similar observation and only can achieve the original accuracy with higher-precision quantization for large-scale Transformer models due to the outliers.
Therefore, keeping the outlier without clipping will significantly benefit quantizing Transformer models.

\subsection{Outlier Is Unaligned}

\begin{table}[t]
    \ra{1.3}
    \centering
    \resizebox{\columnwidth}{!}{
    \begin{tabular}{c|c|c|c}
        \textbf{Accelerator} &
      \textbf{Encoding} &
      \begin{tabular}[c]{@{}c@{}}\textbf{Aligned} \\ \textbf{Memory?}\end{tabular}  &
      \begin{tabular}[c]{@{}c@{}}\textbf{GPU} \\ \textbf{Compatible?}\end{tabular} \\ \Xhline{1.2pt}
      
      OLAccel~\cite{park2018energy}  \begin{tabular}[c]{@{}c@{}}{} \\ {}\end{tabular}  & Coordinate list & {No} & {No} \\ \hline

      \begin{tabular}[c]{@{}c@{}}{BiScaled-} \\ {DNN~\cite{jain2019biscaled}}\end{tabular} & Block sparse index  & \begin{tabular}[c]{@{}c@{}}{Alined data} \\ {Unaligned index}\end{tabular} & {No} \\ \hline
  
      DRQ~\cite{song2020drq}  \begin{tabular}[c]{@{}c@{}}{} \\ {}\end{tabular} & Binary mask map & \begin{tabular}[c]{@{}c@{}}{Unalined data} \\ {Aligned index}\end{tabular} &   {No} \\ \hline

      GOBO~\cite{zadeh2020gobo} \begin{tabular}[c]{@{}c@{}}{} \\ {}\end{tabular}  & Coordinate list & {No} &   {DRAM-only}  \\ \hline
    \textbf{OliVe (Ours)}  \begin{tabular}[c]{@{}c@{}}{} \\ {}\end{tabular}  &\textbf{ Outlier-victim pair} & \textbf{\color{mygreen}Yes} & \textbf{\color{mygreen}Yes} \\
    \end{tabular}
}

\caption{Comparison between existing outlier-aware accelerators and our proposed method \proj{}.}

\label{tab:existing_work}
\vspace*{-.2cm}
\end{table}

\begin{table}[b]
    \vspace*{-0.2cm}
    \ra{1.5}
    \centering
    \resizebox{\columnwidth}{!}{
    \begin{tabular}{c|c|c|c}
        \textbf{Pair Type} &
      \textbf{Normal-Normal} &
      \textbf{Outlier-Normal} &
      \textbf{Outlier-Outlier}\\ \Xhline{1.2pt}
      
      \textbf{BERT$_{base}$}~\cite{devlin2018bert}& 99.12\% & 0.84\% & 0.04\% \\ \hline

      \textbf{BERT$_{large}$}~\cite{devlin2018bert} & 99.24\%  & 0.71\%& 0.05\% \\ \hline
  
      \textbf{GPT2-XL}~\cite{radford2019language} & 98.80\% & 1.14\% &   0.06\% \\ \hline

      \textbf{OPT-6.7B}~\cite{zhang2022opt} & 99.33\% & 0.64\% &   0.03\%  \\ 
    \end{tabular} 
}
\vspace*{0.1cm}
\caption{The percentage of three types of pair.}
\label{tab:outlier_test}
\vspace*{-0.4cm}
\end{table}

The importance of outliers has attracted many research interests, which sparked several outlier-aware architectures, as depicted in \Tbl{tab:existing_work}.
OLAccel~\cite{park2018energy} and GOBO~\cite{zadeh2020gobo} are similar and exploit the coordinate list to indicate the location of outliers, which use high-precision (8-bit or 16-bit) quantization. 
BiScaled-DNN~\cite{jain2019biscaled} and DRQ~\cite{song2020drq} employ block sparse index and bitmap, respectively. BiScaled-DNN quantizes all values with the same bit-width but different scale factors for normal values and outliers, which are aligned. However, the extra index compressed in the block sparsity method is unaligned.
On the contrary, DRQ's bitmap is aligned, but data is stored by mixed and thus unaligned 4- \& 8-bit values.

In summary, prior works design the outlier-aware architecture based on the sparsity of outliers, which leads to unaligned memory storage and accesses.
More seriously, the indices of sparsity-based encoding and the outliers are separate. 
As such, they need the extra outlier controller to parse indices for the outliers and orchestrate the computation between normal values and outlier values.
For example, the extra outlier \revise{controllers} of GOBO and OLAccel count up to $55\%$ and $71\%$ overhead to the total area of the processing element (PE) array~\cite{zadeh2020gobo,park2018energy}.
The sparsity-based encoding for outliers is also \textbf{incompatible} with the memory sub-system of existing accelerators.
For the GOBO design~\cite{zadeh2020gobo}, it can only compress and decompress the memory at the DRAM level for GPU.
This greatly limits the applicability of its proposed outlier-aware architecture.

Therefore, a more hardware-friendly and applicable outlier decoding/encoding method should be proposed to fit the outlier-aware quantization. Our proposed \proj{} architecture is able to align memory accesses and is also compatible with existing accelerators based on the OVP mechanism.

\subsection{Outlier and Victim Analysis}

Generally, the sparsity-based encoding borrowed from DNN pruning is a straightforward and effective solution for sparse outliers.
However, these works ignored that quantization is different from pruning.
For pruning, the pruned zero values do not participate in the computation. As such, the pruning method has to compress the sparse values with sparsity-based encoding.
For quantization, the quantized normal values are the majority and need computation. 
Naturally, the outlier values can exploit the normal values to achieve memory alignment instead of sparsity-based encoding.

\begin{figure}[t]
   \vspace*{-0.1cm}
    \centering
    \includegraphics[width=0.48\textwidth]{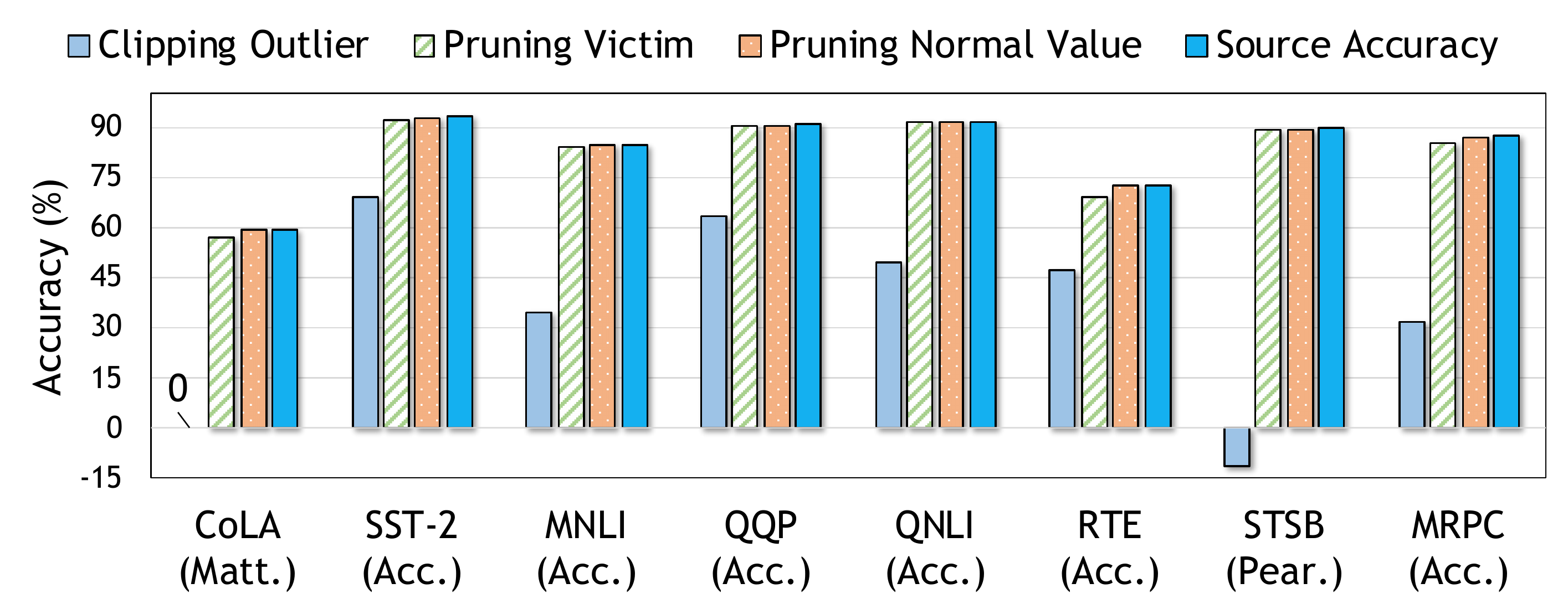} 
    \vspace*{-0.55cm}
    \caption{Accuracy comparison of multiple pruning methods.}\label{fig:outlier_test}
    \vspace*{-0.4cm}
\end{figure}

As depicted in \Fig{fig:intro_victor}b in \Sec{sec:introduction}, we employ the insight of pruning  but with a different perspective from prior works.
The new method employs the \textbf{outlier-victim pair}  (OVP) mechanism.
We first prune some quantized low-precision normal values, which we call \textbf{victims}. These victims are adjacent to the outliers and make extra space for the high-precision outliers. 
Therefore, we can embed the outliers in their original location without explicit sparse indexing. That can avoid the complex indexing hardware and make it compatible with GPU.
To align the memory, we distinguish the ``\revise{right} outlier'' and ``\revise{left} outlier'' according to their \revise{position in the pair}. We assign a right victim for the \revise{left} outlier (e.g., \revise{$17.6$} in \Fig{fig:intro_victor}b) and a left victim for the \revise{right} outlier (e.g., \revise{$-98$} in \Fig{fig:intro_victor}b).

{The OVP mechanism is based on our observation of large Transformer models, including BERT-base~\cite{devlin2018bert}, BERT-large~\cite{devlin2018bert}, GPT2-XL~\cite{radford2019language}, and OPT-6.7B~\cite{zhang2022opt}}.
We collect all tensors, calculate their standard variance $\sigma$, and divide the values into normal values ($<3\sigma$) and outlier values ($>3\sigma$) by the $3\sigma$ rule.
We then pair every two adjacent values (no overlapping), which leads to three types: normal-normal pair, outlier-normal pair, and outlier-outlier pair, as shown in \Tbl{tab:outlier_test}.
These three types have two normal values, one normal value and one outlier value, and two outlier values, respectively.

\Tbl{tab:outlier_test} demonstrates that most (about $99\%$) pairs are normal-normal pairs, with only around 1\% of outlier-normal pairs.
Outlier-outlier pairs need to prune the smaller outlier in the pair.
Fortunately, the outlier-outlier pairs only have an extremely low probability of less than 0.06\% in all studied models.
Therefore, the outlier distribution is extremely dispersed, and we can retain most outliers.

We also conducted the accuracy experiments with the BERT$_{base}$ model~\cite{wei2022outlier} on the GLUE dataset~\cite{wang2018glue}, as depicted in \Fig{fig:outlier_test}.
First, we clip the outliers to the $3\sigma$, where clipping is the common method adopted by quantization.
Then, we prune the victims and normal values to zero. The victims are adjacent to the outliers, and normal values are randomly pruned with the same amount as the outliers. We keep the rest values with full precision (FP32).
Although such few outliers (about 1\%) are clipped, as shown in \Fig{fig:outlier_test} clipping outlier, the accuracy loss is unacceptable for the BERT model. 
The results emphasize the importance of outliers in Transformer-based model.
For comparison, pruning random normal values has almost no accuracy loss than the source accuracy.
The pruning of victim values only shows a negligible accuracy decrease than the pruning of normal \revise{values} because the victims include some outliers due to the outlier-outlier pair and have specific locations corresponding to the adjacent outlier. 

In summary, our analysis indicates that outliers are important while the victims are not, so that we can sacrifice victims to accommodate the outliers.
This motivates us to design the hardware-friendly OVP mechanism that provides aligned outlier-aware quantization to accelerate the large Transformer models.
In the next section, we will introduce the outlier-victim pair encoding design.


%% file: tex/encoding.tex
\section{Outlier-Victim Pair Encoding}
\label{sec:encoding}

\begin{figure}[t] 
    \vspace{2mm}
    \centering 
    \includegraphics[width=1\linewidth]{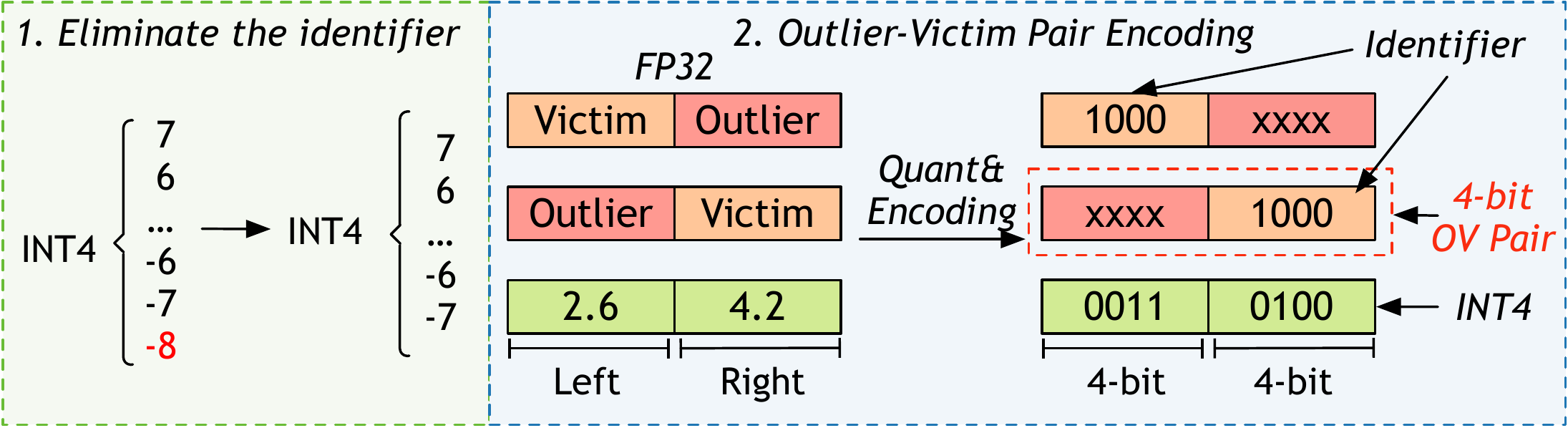} 
    \caption{\revise{The 4-bit outlier-victim pair encoding.} 
    } 
    \label{fig:OV_pair_encoding}
    \vspace*{-0.2cm}
\end{figure}

In this section, we present the details of outlier-victim pair (OVP) encoding that is \textbf{\emph{globally identical but locally distinguishable}} for outlier and normal values.
The OVP encoding can maintain globally aligned memory access and distinguish the outliers locally with ignorable overhead.
For normal values, we can support multiple data types to fit the adaptive data type.
For encoding outliers, we design an outlier-specific data type, adaptive bias float, \texttt{abfloat}, which can avoid range overlapping between normal values and outliers, thus improving the utilization ratio of the numerical representation space of outlier encoding.
Finally, based on the OVP encoding, we propose a framework that can automatically select the outlier threshold for OVP encoding to determine a suitable ratio of the outlier-victim pair.

\subsection{OVP Encoding Algorithm}


\begin{algorithm2e}[b]
    \small
       \DontPrintSemicolon
       \KwIn{
           Values, $val_1, val_2$; Outlier threshold, ${T}$.
       }
       \KwOut{
           OVP encoding, $out_1, out_2$.
       }
   
       \SetKwFunction{FMain}{OVPairEncoding}
       \SetKwProg{Fn}{def}{:}{}
       \Fn{\FMain{$val_1$, $val_2$, $T$}}{
           \If {$val_1 > T$ and $val_1 > val_2$}
           {
               $out_1 =$ OutlierQuantization($val_1$);\\
               $out_2 = 1000_2$; \tcp*[h]{Outlier identifier.}
           }
           \ElseIf {$val_2 > T$}
           {
               $out_1 = 1000_2$ \\
               $out_2 =$ OutlierQuantization($val_2$);
           }
           \Else
           {
               $out_1 =$ NormalQuantization($val_1$); \\
               $out_2 =$ NormalQuantization($val_2$); 
           }
           \Return{$out_1, out_2$}
       }
       \caption{The 4-bit OVP encoding algorithm.}
       \label{alg:OV_pair_encoding}
\end{algorithm2e}

Based on the previous pair-wise tenor value analysis, there are three pair types: normal-normal, outlier-normal, and outlier-outlier. For outlier-normal, the normal value in the pair will be pruned and turned into a victim. For outlier-outlier, we remain the large one and prune the other. Then, we get the normal-normal pairs and outlier-victim pairs in the DNN tensors.

\paragraph{Outlier Identifier.}
To distinguish from the normal-normal pair, we need a special  identifier for the outlier-victim pair.
And this distinct identifier cannot appear in the normal-normal pair, which means we need to eliminate one number in the representation of normal values. 
For example, as shown in \Fig{fig:OV_pair_encoding}, we employ the signed \texttt{int4} (4-bit integer) for the normal value quantization. The original \texttt{int4} can represent the integers in the range of $[-8, 7]$, where $\mathtt{1000_2}$ represents the value of $-8$. 
First, we make $\mathtt{1000_2}$ the outlier identifier and remove the value of $\mathtt{1000_2}$ from \texttt{int4}, whose encoding range becomes $[-7, 7]$.
Second, we quantize the \revise{outlier-victim pairs} with \revise{4-bit} OVP encoding.
We set the victims with the outlier identifier $\mathtt{1000_2}$ and quantize the outlier with the outlier-specific data type \revise{(\Sec{sec:abfloat})}.
Naturally, there are two types of OV pair, i.e., left outlier (O-V) and right outlier (V-O) pair.
Due to the distinct outlier identifier design, we can implicitly distinguish them without using an extra index bit (\Sec{sec:ov_decoder}).

\begin{table}[b]
    \centering
    \ra{1.5}
    \resizebox{\columnwidth}{!}{
      \begin{tabular}{c|c|c}
        \textbf{Data Type} & \textbf{Values} & \textbf{Outlier Identifier} \\ \Xhline{1pt}
        \revise{\textbf{\texttt{int4}}}& $0, \pm 1, \pm 2, \pm 3, \pm 4, \pm 5, \pm 6, \pm 7$  &  $\mathtt{1000_2}$ (-8) \\ \hline
        \revise{\textbf{\texttt{flint4}}}~\cite{guo2022ant}& $0, \pm 1, \pm 2, \pm 3, \pm 4, \pm 6, \pm 8, \pm 16$ &  $\mathtt{1000_2}$ (-0)  \\ \hline
        \revise{\textbf{\texttt{int8}}}&  \revise{$0, \pm 1, \pm 2, \cdots, \pm 126, \pm 127$} &  $\mathtt{1000 0000_2}$ \revise{(-128)} \\
        \end{tabular}
    }
        \caption{Data types for normal values of OVP encoding.}
        \label{tbl:normal_data}
       \vspace*{-0.3cm}
\end{table}

Algo.~\ref{alg:OV_pair_encoding} shows the 4-bit OVP encoding algorithm, which needs to read two values simultaneously, where the requirement is very easy to meet.
For the hardware implementation, we can add a buffer for the encoder. Also, the OVP encoder can be implemented by embedding in the quantization unit with ignorable overheads. 
For the software implementation, we can make a thread handle two values simultaneously.
As a result, the encoding algorithm can be implemented efficiently in both hardware and software, which we describe more details later.

\subsection{Data Type for Normal Values}

For normal values, we build upon prior work~\cite{guo2022ant}, which can support multiple data types, including \texttt{int4}, \texttt{flint4} (4-bit \revise{\texttt{flint}}), and \texttt{int8}, as shown in \Tbl{tbl:normal_data}.
The \texttt{int4} type is one of the most widely used data \revise{types} for 4-bit quantization \revise{with integers in the value range of $[-7, 7]$.}
The \texttt{flint4} type is proposed by prior work ANT~\cite{guo2022ant}, which has shown that selecting the data type according to a tensor's distribution achieves the state-of-the-art performance and accuracy.

Based on the above insights, we also adopt the mixed data types to quantize normal values in our OVP pair encoding.
For \texttt{flint4}, we use the same binary value of  $\mathtt{1000_2}$ as the outlier identifier. Specifically,  $\mathtt{1000_2}$ of \texttt{flint4} corresponds to $-0$, which is not used in the original design. 
In other words, our OVP encoding seamlessly works for \texttt{flint4} without wasting any number representations.
We use the original \texttt{flint4} encoding algorithm~\cite{guo2022ant} to quantize normal values.

Moreover, the OVP encoding can be generally extended to higher-precision quantization, such as the 8-bit. Similarly, the 8-bit normal value also needs to eliminate one number. For instance, \texttt{int8} can represent \revise{$[-128, 127]$} integers, and we can make $1000 0000_2$ the outlier identifier for \texttt{int8} and narrow its range to \revise{$[-127, 127]$}. Similarly, the encoding algorithm can easily extend to read two 8-bit elements simultaneously.



\begin{table}[b]
    \vspace*{-0.2cm}
    \centering
    \ra{1.3}
    \resizebox{0.85\columnwidth}{!}{
      \begin{tabular}{c|c|c|c}
        Binary   & Exponent & Integer & Real Value \\ \Xhline{1pt}
        \uline{00}0& 0 & 0 & 0 \\ \hline
        \uline{00}1& 0   & 3 & $3\times 2^0 = 3$ \\ \hline
        \uline{01}x& 1 & 2, 3 & $2\times 2^1 = 4, 3\times 2^1 = 6$  \\ \hline
        \uline{10}x& 2  & 2, 3 & $2\times 2^2 = 8, 3\times 2^2 = 12$ \\ \hline
        \uline{11}x& 3  & 2, 3 & $2\times 2^3 = 16, 3\times 2^3 = 24$\\ 
        \end{tabular}
    }
        \caption{The 3-bit unsigned E2M1, which means two bits for exponent and one bit for mantissa, with $\mathtt{bias = 0}$.}
        \label{tbl:e2m1}
        \vspace*{-0.3cm}
\end{table}

\begin{figure}[t] 
    \vspace*{-0.1cm}
    \centering 
    \includegraphics[width=1\linewidth]{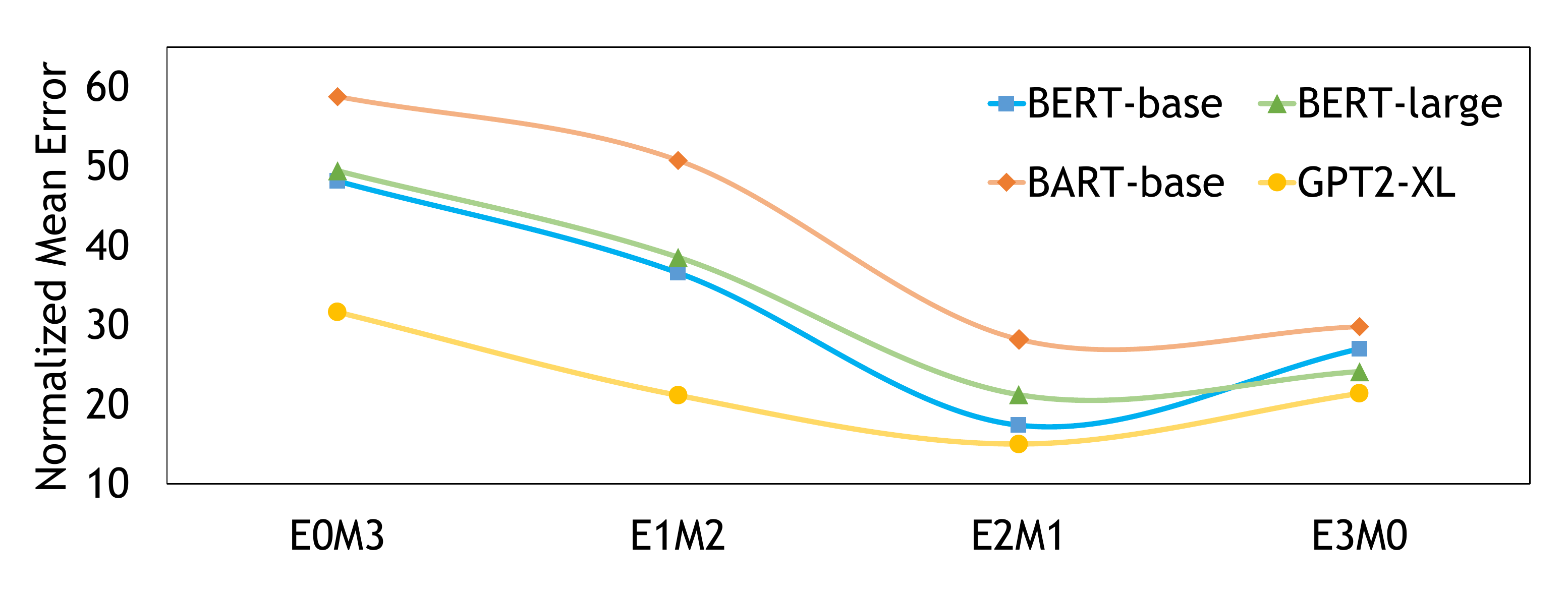}
    \vspace*{-0.7cm}
    \caption{The rounding error of the largest outliers quantized with different data types. 
    Experiments were conducted on BERT-base, BERT-large, BART-base, and GPT2-XL.
    } 
    \label{fig:outliers_mse}
    \vspace*{-0.3cm}
\end{figure}

\subsection{Data Type for Outliers: Abfloat }
\label{sec:abfloat}
Next, we quantize outliers using the outlier-specific data type.
The large outliers usually have a wide range, for which we use float-based data to quantize. 
We propose a data type called \textbf{\underline{a}daptive \underline{b}iased \underline{float}}, \texttt{abfloat} in short. The key idea is that by adding a proper bias to the exponent, all encoded values can skip the interval where normal values lie and provide more range for outliers.



\paragraph{Float-to-Fixed Conversion.}
To accommodate the normal values and avoid fractions, we first convert the floating-point encoding to the fixed point with an exponent. Also, the fixed point is friendly to the hardware implementation and has a lower overhead than the floating point.
We transform the the floating point to fixed point with the following equation,
\begin{equation}    
\mathtt{sign \times (1 \ll mb + mantissa) \ll (exponent + bias) },
\label{eq:e2m1}
\end{equation}
where $\mathtt{mb}$ is the mantissa bit-width. Therefore, this fixed-point encoding scheme is more friendly and efficient for  hardware \revise{implementation}, as it only involves shift operations. \Tbl{tbl:e2m1} shows the example of fixed-point E2M1 data type.

\paragraph{Adaptive Bias.}
Obviously, \Tbl{tbl:normal_data} and \Tbl{tbl:e2m1} show that the range of fixed-point \texttt{abfloat} overlaps with the normal values. For example, \texttt{int4} and E2M1 contain the same numbers, 3, 4, and 6. 
Another example is that \texttt{flint4} and E2M1 have almost the same number range except for 24. 
Therefore, we need the adaptive bias to adjust the range of \texttt{abfloat}.
For example, we set $\mathtt{bias = 2}$ for E2M1, whose real values will be extended to $\{12, \cdots,96\}$, which is \revise{complementary} with the \texttt{int4} normal value. Similarly, we set $\mathtt{bias = 3}$ and extend range to $\{24, \cdots,192\}$ for \texttt{flint4} data type.
We design a new decoder and instruction to implement adaptive bias in accelerators for the \texttt{abfloat} \revise{(\Sec{sec:ov_decoder})}.

\begin{algorithm2e}[t]
    \small
       \DontPrintSemicolon
       \KwIn{
           Element $e$; Bias, $b$;
       }
       \KwOut{
           Quantized Element $q$;
       }
   
       \SetKwFunction{FMain}{AbfloatQuant}
       \SetKwProg{Fn}{def}{:}{}
       \Fn{\FMain{$e$, $b$}}{
           \tcp*[h]{Get exponent and base integer.}\\
           $exp = \lfloor log_2(abs(e)) \rfloor - 1$;\\
           $base\_int = Round[e / 2^{exp}]$; \\
           \If {$base\_int == 4$}
           {
               $exp = exp + 1$; \\
               $base\_int = base\_int - 2$;
           }

           \tcp*[h]{Encoded as abfloat data type.}\\
           $exp = exp - b$; \\
           $base\_int = base\_int\ \&\ 1$; \\
           $unsigned\_q = Concat(exp, base\_int)$; \\

           
           $q = Concat(e < 0, unsigned\_q)$ \\
           \Return{$q$}
       }
       \caption{The \texttt{abfloat} encoding algorithm.}
       \label{alg:abfloat_encoding}
\end{algorithm2e}

\paragraph{E2M1 Abfloat.}
The 4-bit signed float has four possible configurations of exponent and mantissa: E0M3, E1M2, E2M1, and E3M0. They have different ranges and precisions. 
We conduct the following experiments to choose the most appropriate configuration as the final outlier-specific data type. 
To accommodate the broad range of outlier values, we quantize the largest outlier values (i.e., Max $\sigma$ in \Fig{fig:sigma}) in Transformer models using all \texttt{abfloat} types.
Then, we collect the average absolute error, as shown in \Fig{fig:outliers_mse}. We found that E2M1 gives the least error in all tests, which provides both a large enough range and a certain degree of precision, and it also presents the best results in our subsequent evaluations.
Similarly, we adopt signed E4M3 \revise{for 8-bit \texttt{abfloat}}.



Algo.~\ref{alg:abfloat_encoding} shows in detail how an element is encoded as \texttt{abfloat}. 
The outlier encoding is \revise{an} element-wise function, which can be implemented on software and hardware efficiently. 
Outlier encoding should also eliminate the outlier identifier. Otherwise, the decoder cannot distinguish the outlier-victim pair.
\texttt{Abfloat} has two zero numbers: $1000$ (-0) and $0000$ (0). Therefore, we disable the $1000$ and $0000$ for outlier values to avoid conflict with the outlier identifier.



\subsection{Quantization Framework}
We now apply OVP (outlier-victim pair) encoding for quantizing Transformer models.
To decide the scale factor (i.e., outlier-victim threshold), we embed the OVP encoding with the existing mean squared error (MSE) minimization algorithm, which is commonly used by many quantization works~\cite{zhang2018lq, banner2019post, cai2020zeroq}.
The OVP-based quantization algorithm determines the threshold for distinguishing outliers and normal values. 
On one hand, a small threshold would lead to more outlier-victim pairs, which could potentially minimize the quantization error (i.e., MSE).
On the other hand, it also increases the ratio of outlier-outlier pairs, where both values are outliers in the pair.
If there are too many such outlier-outlier pairs, the MSE would increase owing to the pruning of outliers.
Thus, we need to control the ratio of outlier-outlier pairs for better accuracy.

\begin{figure*}[t]
    \centering
    \includegraphics[width=1\textwidth]{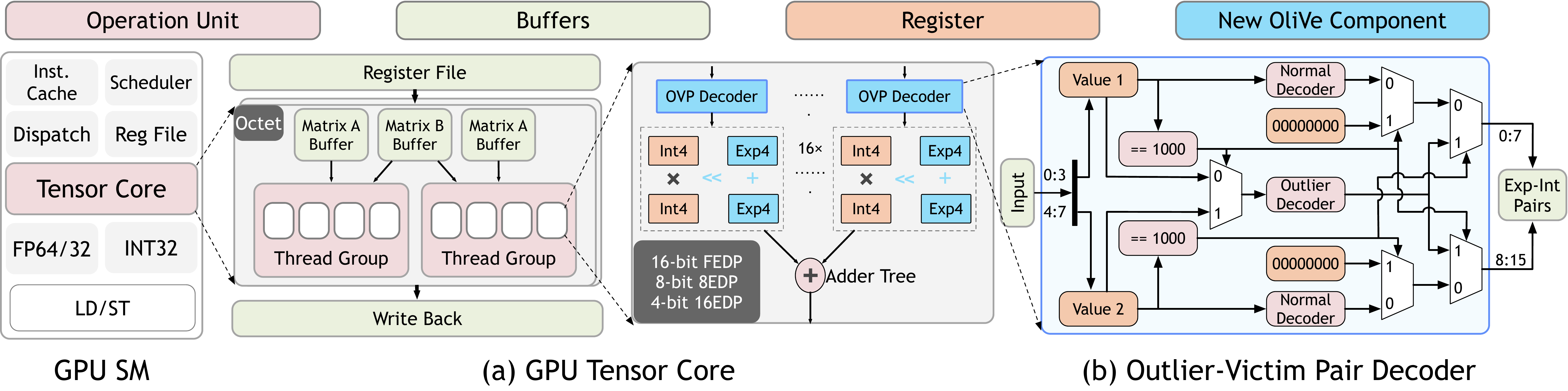} 
    \caption{\revise{\proj{} integration on GPU tensor cores (a), which only requires a set of lightweight OVP decoder (b).}}\label{fig:gpu_arch}
\end{figure*}

In our work, we target the post-training quantization (PTQ)~\cite{nagel2021white}, which does not require \revise{retraining} and hence is best suitable for large models as their trainings are expensive.
However, we still need to use one batch of data from the \textbf{training set} for the scale factor selection.  
Intuitively, inspired by the $3\sigma$ rule, we take $3\sigma$ as the initial scale factor.
Then the algorithm will search for the best scale factor with the smallest MSE within a specific range of this baseline, which shows good results in our evaluations. 
For quantization-aware training (QAT)~\cite{nagel2021white}, we can get a suitable scale factor by \revise{retraining} it with the straight-through estimator (STE)~\cite{bengio2013estimating}.

%% file: tex/architecture.tex
\section{OliVe Architecture}
\label{sec:architecture}

\begin{table}[b]
    \vspace*{-0.3cm}
    \ra{1.5}
    \centering
    \resizebox{0.95\columnwidth}{!}{
    \begin{tabular}{c|c|c|c|c|c}
        \textbf{Architecture}   &
      \textbf{SM} &
      \textbf{TC} &
      \textbf{16-bit Unit} &
      \textbf{8-bit Unit} &
      \textbf{4-bit Unit} 

      \\ \Xhline{1.2pt}
      
      \textbf{Turing}~\cite{turing}& 68 & 544 & 34,816 & 69,632 & 139,264 \\ 
    \end{tabular} 
}
\caption{The Turing GPU architecture.}
\vspace*{-0.5cm}
\label{tab:Turing}
\end{table}

This section presents how to integrate \proj{} in GPU and output-stationary systolic array architecture. We then present the hardware decoder for the aforementioned outlier-victim pair encoding and outlier data type. 
On these architectures, our proposed \proj{} architecture can directly support the mixed precision~\cite{a100,sharma2018bit} and mixed data type~\cite{a100,sharma2018bit}, which are efficient for quantizing DNN tensors that have different importance and distribution.

\subsection{GPU Tensor Core}
We first describe how to integrate the \proj{} design into the tensor core architecture of GPU in the \revise{\Fig{fig:gpu_arch}a}. 
We employ Turing architecture~\cite{turing} as our baseline GPU, which has 68 streaming multiprocessors (SMs), and each SM has eight tensor cores (544 in total), as shown in \Tbl{tab:Turing}. 
According to the modeling of prior work~\cite{raihan2019modeling}, each tensor core has two octets, which have eight FEDPs (four-element dot product). As such, there are $68\times 8 \times 2 \times 8 \times 4 = 34,816$ 16-bit float multipliers.
The Turing architecture can originally support mixed-precision computation. 
For example, the RTX 2080Ti GPU with Turing architecture \cite{turing} provides 107.6,  215.2, and 430.3 TOPS (tera operations per second) for 16-bit float, 8-bit int, and 4-bit int, respectively.
Therefore, we assume that the tensor core can simultaneously support 8-bit 8EDP (eight-element dot product) and 4-bit 16EDP (16-element dot product), as shown in \revise{\Fig{fig:gpu_arch}a}.

We can easily embed our proposed \proj{} architecture in GPU, which adopts the SIMD architecture. We first put the 4-bit outlier-victim pair decoders (\revise{\Fig{fig:gpu_arch}b}) for each 16EDP.
To support the new \proj{} data types, we add an adder and a shifter for each 16EDP.
Similarly, we also design the 8-bit decoder for the 8EDP units. 

\subsection{Decoders}
\label{sec:ov_decoder}

\revise{
\paragraph{Outlier-Victim Pair Decoder.} 
To support outlier-victim pair decoding, we design a new decoder that can be easily embedded in existing accelerators.
As shown in \revise{\Fig{fig:gpu_arch}b}, the decoder reads 1 byte, which is the smallest addressable memory unit in many architectures, and exactly one value pair. 
Then, the decoder transforms the outlier identifier  $\mathtt{1000_2}$ to 0 and decodes the outlier value with the outlier decoder.
To accommodate the computation of the outlier \texttt{abfloat} values, the decoder will generate an exponent-integer pair. Therefore, the decoder needs to append a  $\mathtt{0000_2}$ as the exponent number for the normal \texttt{int4} data type. For \texttt{flint4}, we exploit its original decoder~\cite{guo2022ant} to get the exponent-integer pair.
}

\begin{figure}[b] 
    \centering 
    \includegraphics[width=.9\linewidth]{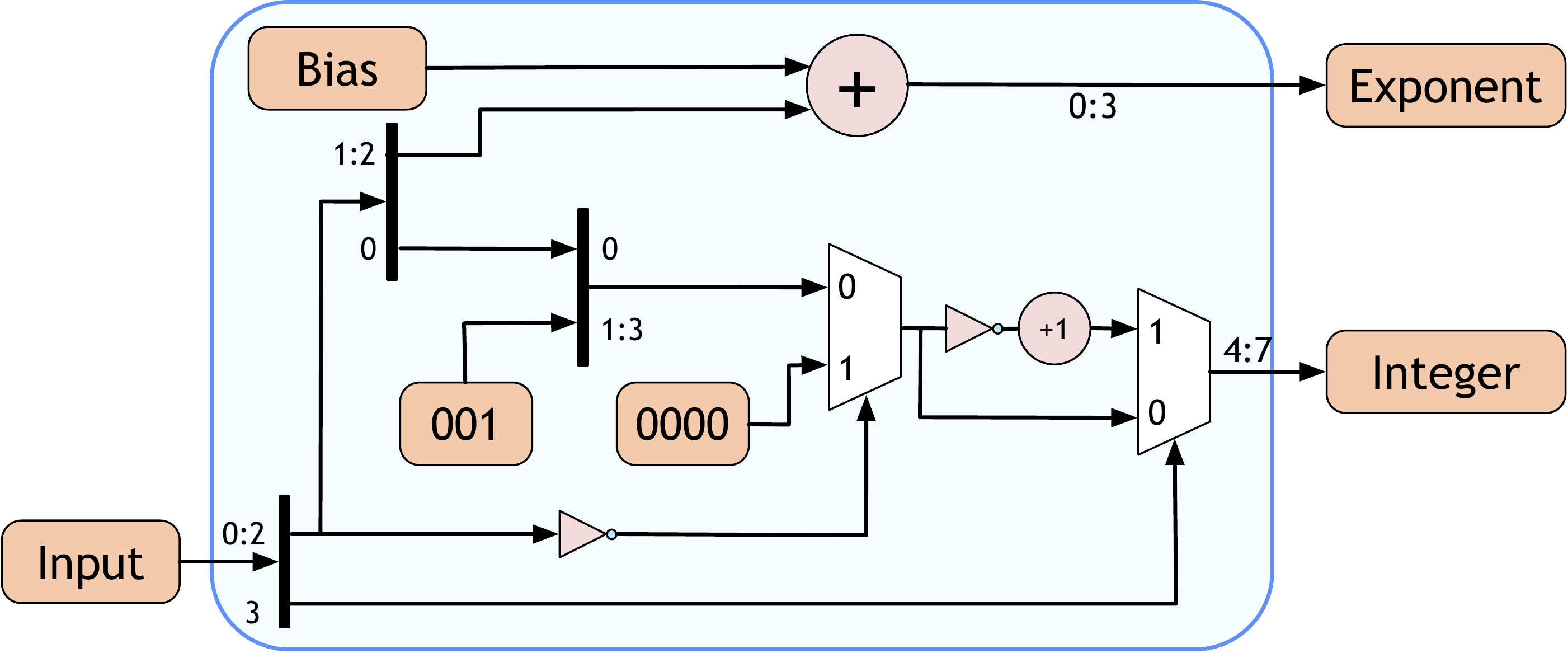}  
    \caption{The 4-bit \texttt{abfloat} decoder for outlier values.} 
    \label{fig:outlier_decoder}
    \vspace*{-0.2cm}
\end{figure}

\paragraph{Outlier Decoder.}
The above OVP decoder contains an outlier decoder for outlier values with the E2M1 \texttt{abfloat} data type. \revise{\Fig{fig:outlier_decoder}} shows the details of the 4-bit \texttt{abfloat} decoder design.
For a 4-bit E2M1 \texttt{abfloat} number $x = (b_2b_1b_0)_2$, following equations decode exponent and integer: 
$$
\mathtt{{exponent} = bias + (b_2b_1)_2}
$$
$$
\mathtt{
{integer} = \begin{cases}
    0 &{if}\  x=000_2 \\
    (1b_0)_2 & {otherwise}
 \end{cases}
}
$$
For example, when the bias is 2, a number $0101_2$ is $48_{10}$, since its exponent is $2_{10} + 10_2 = 4_{10}$ and base integer is $11_2 = 3_{10}$. Therefore, its real value is $3 \ll 4 = 48$.

Similarly, we also design and implement the 8-bit outlier-victim pair decoder and the E4M3 \texttt{abfloat} outlier decoder, which are straightforward extensions of 4-bit instances. As such, we do not present their details due to the limited space.


\begin{figure}[t]
    \centering
    \includegraphics[width=0.43\textwidth]{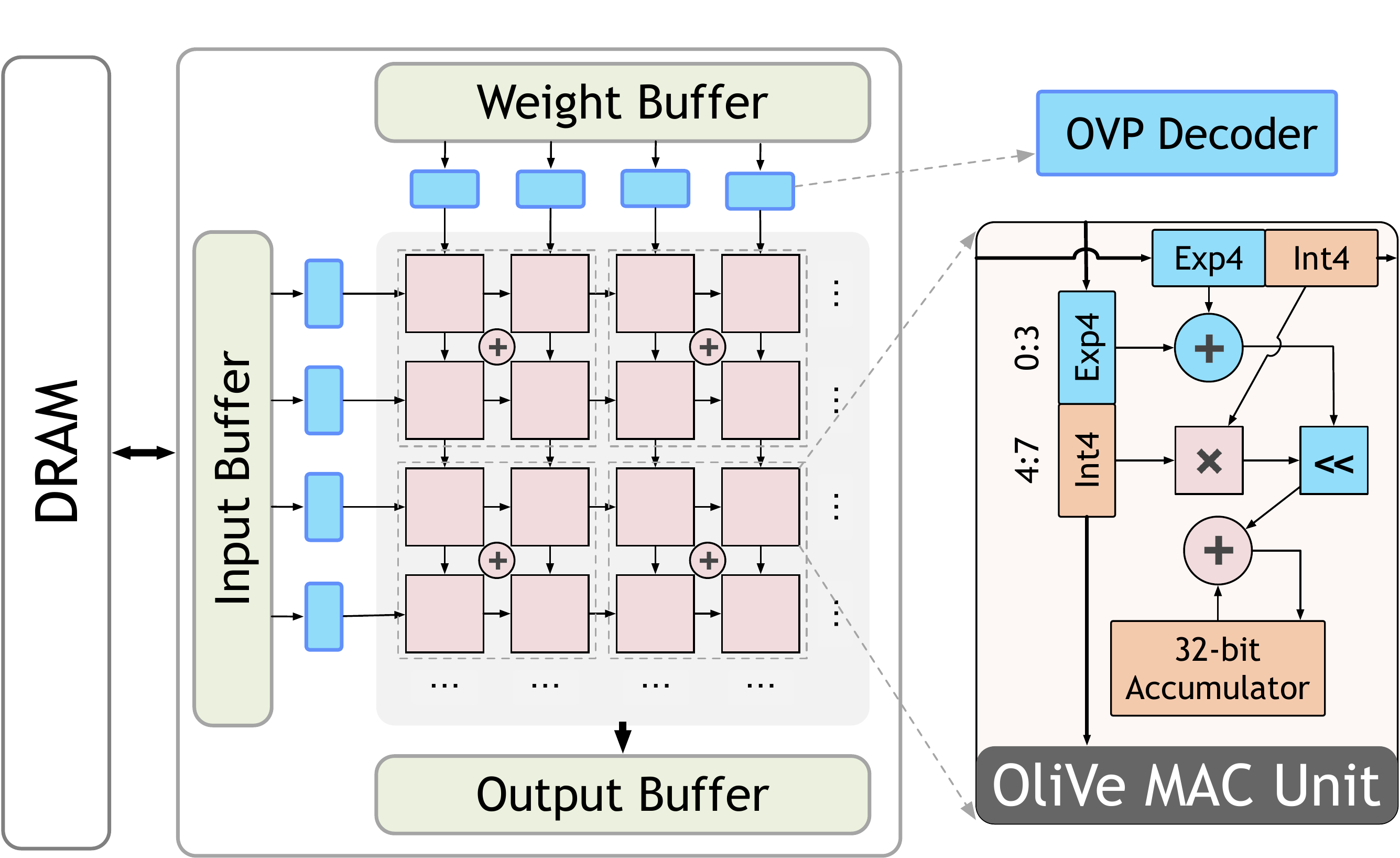} 
    \caption{\revise{\proj{} integration on systolic array.}}\label{fig:sa_arch}
    \vspace*{-0.3cm}
\end{figure}

\revise{
\subsection{Systolic Array}
The systolic array (SA) integration is shown in \Fig{fig:sa_arch}.
SA uses the same outlier-victim pair decoder design (\revise{\Fig{fig:gpu_arch}b}) as GPU, which shows the wide applicability of our design.}
But, unlike GPU, we only place the decoders along the borderlines, which can save most decoders. For example, if the array size is $n\times m$, we  only need $n+m$ instead of $n \times m$ decoders. That is one advantage of SA over the GPU's SIMD architecture.
Our proposed \proj{}-based data type can also support the systolic array processing element (PE) with an extra adder and shifter.
We add an extra adder for every four PEs to support high-precision quantization, e.g., \texttt{int8}.

\subsection{OliVe MAC unit}
After decoding for outlier and normal values, they are all transformed into unified exponent-integer pairs. To support the decoded exponent-integer pair computation, we need to add a shifter and an adder for the fixed-point MAC (multiply and accumulation) unit, as shown in \revise{\Fig{fig:sa_arch}} and the unit of \revise{\Fig{fig:gpu_arch}} 4-bit 16EDP.
For example, we have two exponent-integer pairs $<a, b>$ and $<c, d>$, where $a$ and $c$ are exponents, $b$ and $d$ are integers, and $<a,b>$ represents:
$$<a,b> = b\ll a$$
Then, we can get the result: 
\begin{align*}
    &<a,b> \times <c,d> \\
    = & \ (b \times d) \ll (a+c) \\
 = &<a+c, b\times d>
\end{align*}
Note that the final result can store with a 32-bit int.


\subsection{Mixed Precision} 
As mentioned in \Sec{sec:encoding}, \proj{} quantization can support the \texttt{int8} for normal values and E4M3 \texttt{abfloat} for outlier values. Therefore, we propose the mixed-precision processing element (PE) for the higher precision data types.

\paragraph{8-bit Int.}
For the GPU tensor core architecture, it is originally designed with mixed-precision computation.
For the systolic array, our architecture naturally supports 8-bit computation with four 4-bit PEs~\cite{sharma2018bit}. For an \texttt{int8} number $x$, the higher 4 bits and the lower 4 bits can be split into two 4-bit numbers $h$ and $l$, and the $x$ can be represented by:
 $$x = (h_x \ll 4) + l_x = <4, h_x> + <0, l_x>.$$
We then can multiply two \texttt{int8} numbers of $x$ and $y$: 
 \begin{align*}    
 x\times y =& \underbrace{<4, h_x>\times<4, h_y>}_{PE0} + \underbrace{<4, h_x>\times<0, l_y>}_{PE1} \\
 + &\underbrace{<0, l_x>\times<4, h_y>}_{PE2} + \underbrace{<0, l_x>\times<0, l_y>}_{PE3}
 \end{align*}
Therefore, we can use four 4-bit PEs to calculate the above four multiplications and accumulate the products to get the final product value of $x\times y$. 

\paragraph{8-bit \texttt{Abfloat}}
Similarly, multiplication of 8-bit \texttt{abfloat} can be supported using the same approach. For an 8-bit \texttt{abfloat} number $z$, it is first decoded into an exponent $e_z$ and an integer $i_z$. For $i_z$, we similarly split it into $i_z = (h_z << 4) + l_z$, then $z = <4 + e_z, h_z> + <e_z, l_z>$. Hence the same method can be used to perform 8-bit \texttt{abfloat} multiplication with four 4-bit PEs, where the \texttt{abfloat} has an extra $e_z$ than \texttt{int8}.

In the most extreme case, two outliers with \texttt{abfloat} may be multiplied together.
Because we adopt the 32-bit int as the accumulator, the maximum multiplicand should not be over $\sqrt{2^{31} - 1}$. 
Therefore, for the outlier value with the \texttt{abfloat} type, we will clip the absolute value of the outlier within $2^{15} < \sqrt{2^{31} - 1}$ to avoid the overﬂow for the \revise{\texttt{int32}} accumulators.
Our experiments show that the outlier values of the Transformer models are much smaller than $2^{15}$.
Specifically, $2^{15}$ is about $768\sigma$ after normalization and quantization. As shown in \Fig{fig:sigma}, the maximum value of outliers does not exceed $325\sigma$. Thus, we observe that no outlier is truncated in practice.

\subsection{Instruction Set}
For 4-bit tensor cores, the Turing GPU architecture adopts the instruction $\mathtt{mma.s32.s4.s4.s32}$. These four operands are matrices $D$ (\texttt{int32}), $A$ (\texttt{int4}), $B$ (\texttt{int4}), and $C$ (\texttt{int32}), and $D = A\times B + C$. To support the OVP-based computation on GPU, we design a new instruction called $\mathtt{mmaovp}$:
$$
\mathtt{\underbrace{\textcolor{mygreen}{mmaovp}}_{OVP-MMA}.s32.\underbrace{\textcolor{mygreen}{ovpi4}}_{int4}.\underbrace{\textcolor{mygreen}{ovpf4}}_{flint4}.s32.\underbrace{\textcolor{mygreen}{s4}}_{bias} }.
$$

Moreover, because of the memory-aligned design of the data type, \proj{} maintains the original programming interface for GPUs. We can replace the original \texttt{int}-based instruction with OVP-based instruction (e.g., $\mathtt{mmaovp}$) to easily construct the OVP-supported DNN quantization framework. Therefore, our \proj{} framework has comprehensive and practical applicability, which is the most significant advantage of \proj{}.

%% file: tex/evaluation.tex
\section{Evaluation}
\label{sec:evaluation}

In this section, we evaluate the LLM's accuracy with \proj{} quantization. We also demonstrate \proj{}'s area overhead, speedup, and energy efficiency on GPU and systolic array, respectively.

\subsection{Methodology}
\label{subsec:eval:methodology}

\paragraph{Framework and Evaluation Models.} 
To evaluate our \proj{} quantization framework, we implement it in Pytorch~\cite{paszke2019pytorch}. 
We evaluate BERT-base~\cite{devlin2018bert}, BERT-large~\cite{devlin2018bert}, and BART-base~\cite{lewis2019bart}, the three most commonly used language models, on eight datasets of the GLUE benchmark~\cite{wang2018glue}.
In addition, we evaluate BERT-base~\cite{devlin2018bert} and BART-base~\cite{lewis2019bart} on the summarization tasks SQuAD v1.1 and SQuAD v2.0~\cite{rajpurkar-etal-2016-squad}. 
To valid our \revise{quantization} framework on large language models, we also evaluate GPT2-XL~\cite{radford2019language}, BLOOM-7B1~\cite{scao2022bloom}, and OPT-6.7B~\cite{zhang2022opt} on Wikitext103~\cite{enwiki:1116981313} and C4~\cite{jesse2021c4} datasets. 
For all models mentioned above, we use state-of-the-art checkpoints from the huggingface repositories~\cite{huggingfacerepo}.

\paragraph{\revise{Quantization} Baselines.}
We compare \proj{} with existing quantization works, including GOBO~\cite{zadeh2020gobo}, Outlier Suppression~\cite{wei2022outlier}, Q8BERT~\cite{zafrir2019q8bert}, and ANT~\cite{guo2022ant}.
Outlier suppression~\cite{wei2022outlier} is the state-of-the-art Transformer quantization work.
GOBO~\cite{zadeh2020gobo} is also an outlier-aware quantization work. 
Q8BERT~\cite{zafrir2019q8bert} is a method for quantizing GEMM operations to 8-bit.
ANT~\cite{guo2022ant} is a hardware-friendly quantization framework that achieves state-of-the-art results in both performance and accuracy.

\revise{
\paragraph{Accelerator Baselines.}
We compare the performance and energy of \proj{} against five DNN quantization accelerators, including OLAccel~\cite{park2018energy}, AdaptivFloat~\cite{tambe2020algorithm} (shorted as AdaFloat), GOBO~\cite{park2018energy}, ANT~\cite{guo2022ant}, and original \texttt{int8} tensor cores in GPU~\cite{turing}.
OLAccel~\cite{park2018energy} first proposed the outlier-aware quantization architecture for CNNs. 
We extend OLAccel to the Transformer-based models with element-wise mixed-precision weight and activation quantization.
AdaFloat~\cite{tambe2020algorithm} extends the \texttt{float} type with a tensor-wise exponent bias.
GOBO~\cite{zadeh2020gobo} is similar to OLAccel, but only supports the weight quantization for Transformer-based networks.
}

\revise{
\paragraph{OliVe Implementation.}} We implement our decoder in Verilog RTL and synthesize it with Synopsys design compiler~\cite{kurup2012logic} with a 22~nm TSMC technology library to estimate its area, latency, and power. 
\revise{We use CACTI~\cite{muralimanohar2009cacti} to estimate the area and power of on-chip memories. We integrate \proj{} into GPU and hardware accelerator for the end-to-end performance and energy evaluation.}

\revise{For the GPU integration and evaluation, we modify and extend GPGPU-Sim 4.0~\cite{gpgpu-sim2009} and AccelSim~\cite{gpgpu-sim4.0} with the configuration of NVIDIA 2080 Ti architecture.} 
We use AccelWattch~\cite{khairy2020accel}, GPUWattch~\cite{leng2013gpuwattch}, and CACTI~\cite{muralimanohar2009cacti} for the energy estimation. 
The majority of Transformer layers are matrix multiplication operations. 
For GEMM implementation on the tensor core, we use CUTLASS~\cite{Kerr_CUTLASS_2022}, which is NVIDIA's open-source implementation.

\revise{For the accelerator evaluation, we compare AdaFloat, OLAccel and ANT with \proj{}.
We develop a cycle-level simulator to estimate the overall performance of \proj{} based on DnnWeaver~\cite{sharma2016high}.
Although DnnWeaver~\cite{sharma2016high} is a FPGA tool set, prior DNN quantization accelerators, which include the BitFusion~\cite{sharma2018bit}, and ANT~\cite{guo2022ant}, have extended its frontend to add the ASIC performance and energy simulation.
As \proj{} does not redesign the baseline accelerator architecture, we can directly embed new \proj{}-related instructions and data format in the simulator without breaking the original simulation flow.
In other words, we have used and modified the open-sourced implementaions of BitFusion~\cite{sharma2018bit, sharma2018bitrepo}, and ANT~\cite{guo2022ant,guo2022antrepo}.
}

\begin{table}[t]
    \centering
    \small
    \renewcommand{\arraystretch}{1.2}
    \begin{tabular}{lcccccccccc}
        \hline
          {\bf Method} & {\bf Algorithm} & \textbf{CoLA}  & \textbf{SST-2} & \textbf{MNLI}  & \textbf{QQP } & \textbf{MRPC}   \\
        \hline
        BERT$_{base}$            &  32-bit  & 59.60 & 93.35 & 84.94 & 90.91 & 87.75  \\
        \hline
        \hspace{0.5em} \textbf{Ours} & \bf 4-bit PTQ             & \textbf{59.30} & \textbf{92.43} & \textbf{84.10} & \textbf{90.36} & \textbf{87.99}  \\
        \hspace{0.5em} ANT & 4-bit QAT         & 53.91  & \revise{\textbf{92.43}}  & 83.45  & -  & -  \\
         \hspace{0.5em} ANT & 4-bit PTQ          & 42.90  & 90.48  & 73.36  & 78.04  & 68.87 \\
         \hspace{0.5em} OS & 4-bit QAT         & 50.56  & 91.86  & 83.05  & 90.33  & 84.31  \\
         \hspace{0.5em} OS & 6-bit PTQ          & 54.40 & 91.86 & 82.02 & 88.94 & 83.33 \\
         \hspace{0.5em} Q8 & 8-bit QAT         & 58.48  & 92.24  & -  & -  & -\\
         \hline
         BERT$_{large}$  & 32-bit & 63.35 & 93.46 & 86.65 & 91.07 & 87.99    \\
        \hline
         \hspace{0.5em} \textbf{Ours}  & \bf 4-bit PTQ           & \textbf{63.99} & \textbf{92.89} & \textbf{84.89}& \textbf{90.14} & \textbf{86.52}  \\
         \hline
        BART$_{base}$    & 32-bit  & 56.32 & 93.35 & 86.45 & 91.34 & 87.50    \\ 
        \hline
        \hspace{0.5em} \bf Ours&\bf  4-bit PTQ             & \textbf{54.30} & \textbf{92.89} & \textbf{85.33}& \textbf{91.23} & \revise{86.76} \\
        \hspace{0.5em} OS  & 4-bit QAT  & 50.83 & 92.43 & 84.57 & 90.93  & \revise{\textbf{87.01}}  \\
        \hspace{0.5em} OS & 6-bit PTQ    & 44.51& 90.94& 82.98& 88.45 & 80.88 \\
         \hline
    \end{tabular}
    \vspace*{0.1cm}
    \caption[]{Results on GLUE datasets. Q8 and OS are Q8BERT~\cite{zafrir2019q8bert} and outlier suppression~\cite{wei2022outlier} for short, respectively. 
    Prior works do not report results in BERT$_{large}$ so we only compare against the original full-precision model.
    }
    \label{tbl:GLUE_results}
    \vspace*{-0.6cm}
\end{table}

\subsection{Accuracy Results}

We first evaluate the accuracy of \proj quantization framework on different tasks and datasets, which is the prerequisite for applying it to reduce the inference cost of large language models (LLMs).

\paragraph{GLUE Dataset.} 
We evaluate BERT-base~\cite{devlin2018bert}, BERT-large~\cite{devlin2018bert} and BART-base~\cite{lewis2019bart} on eight datasets of GLUE benchmark, but due to space limitation, we only show the results on CoLA, SST-2, MNLI, QQP and MRPC datasets in \Fig{tbl:GLUE_results}.
For the BERT-base model, our 4-bit PTQ method accuracy drop less than 1\% compared to the original full precision model on all eight datasets and outperforms all studied methods including 4-bit, 6-bit, and 8-bit PTQ and QAT methods. Since GOBO~\cite{zadeh2020gobo} only quantizes weights, we use the same method to compare with it and the result is shown in \Tbl{tbl:GOBO_comparison}.
Our method also outperforms the GOBO under the weight-only quantization setting.
In addition, we evaluate the BERT-large model, which is evaluated by few prior quantization works due to the larger number of parameters and hence much more challenging compared to BERT-base.
The results in \Tbl{tbl:GLUE_results} show the accuracy loss for BERT-large is around 1\% on the five presented datasets and similar results are found on other datasets.
For the BART-base model, our 4-bit PTQ results in \Tbl{tbl:GLUE_results} show around 2\% accuracy loss compared to the accuracy of original full-precision in all datasets.
In the above evaluation, our 4-bit PTQ results are better than all the PTQ and most of the QAT results reported by prior works.

\begin{table}[b]
    \centering
    \renewcommand{\arraystretch}{1.2}
    \begin{tabular}{lcccccccccc}
        \hline
          {\bf Method} & {\bf Bits} & \textbf{MNLI}  & \textbf{STSB}(Pear.)  \\
        \hline
        BERT$_{base}$ &  32 & 84.94 & 89.70 \\
        \hline
        \hspace{0.5em} \textbf{Ours (weights only)}            & \bf 4 & \textbf{84.75}& \textbf{89.62}\\
         \hspace{0.5em} GOBO$^*$(weights only) & 4 & 84.45  & 88.33  \\
        \hline
    \end{tabular}
    \vspace*{0.1cm}
    \caption[]{Comparison with GOBO on the MNLI and STSB dataset. $^*$The accuracy of our GOBO implementation slightly differs from the number reported in the original paper~\cite{zadeh2020gobo}.}
    \label{tbl:GOBO_comparison}
    \vspace*{-0.2cm}
\end{table}

\paragraph{SQuAD Dataset.} 
We also evaluate the accuracy of \proj{} quantization on summarization task SQuAD~\cite{rajpurkar-etal-2016-squad}, which is more challenging than the previous GLUE dataset. \Tbl{tbl:SQUAD_results} shows the results on SQuAD v1.1 and SQuAD v2.0 datasets.
On both datasets, our 4-bit PTQ method obtains a less than 2\% accuracy loss on the BERT-base model and around 3\% accuracy loss on the BART-base model, which is better than the 6-bit PTQ method of the state-of-the-art quantization work outlier suppression.
\vspace*{-0.2cm}

\begin{table}[t]
    \centering
    \renewcommand{\arraystretch}{1.2}
    \begin{tabular}{lcccccccccc}
        \hline
          {\bf Method} & {\bf Bits} & \textbf{SQuAD v1.1}  & \textbf{SQuAD v2.0}  \\
        \hline
        BERT$_{base}$ &  32 & 88.28/80.82 & 77.34/73.60 \\
        \hline
        \hspace{0.5em} \textbf{Ours}            & \bf 4 & \textbf{86.38/78.16}& \textbf{75.90/72.08}\\
        
         \hspace{0.5em} Outlier Suppression & 6 & 84.48/75.53  & 74.69/70.55  \\
         \hline
        BART$_{base}$ &  32 & 91.63/84.79 & 80.82/77.41  \\
        \hline
        \hspace{0.5em} \textbf{Ours}            & \bf 4  & \textbf{88.15/79.87} & \textbf{77.37/73.69} \\
         \hspace{0.5em} Outlier Suppression & 6 & 83.68/75.34 & 74.44/70.36  \\
        \hline
    \end{tabular}
    \vspace*{0.1cm}
    \caption[]{PTQ results on SQuAD datasets.}
    \vspace*{-1cm}
    \label{tbl:SQUAD_results}
\end{table}

\begin{center}
    \begin{table}[b]
        \centering
        \renewcommand{\arraystretch}{1.2}
        \begin{tabular}{cccccccccc}
            \hline
            \multirow{2}{*}{\textbf{Method}} & \multicolumn{2}{c}{\textbf{GPT2-XL}} & \multicolumn{2}{c}{\textbf{BLOOM-7B1}} & \multicolumn{2}{c}{\textbf{OPT-6.7B}}  \\
            & Wiki  & C4  & Wiki  & C4  & Wiki  & C4  \\
            \hline
            FP32 & 17.48 & 16.30  & 13.05 & 14.94 & 22.14 & 10.63 \\
            \hline
             \hspace{0.5em} \texttt{int8} & 18.29 & 17.35  & 14.04 & 16.18 & 37.45 & 74.30 \\
             \hspace{0.5em} \textbf{8-bit OliVe}  & 17.49  & 16.37    & 13.13  & 15.04 & 22.34 & 10.73 \\
            \hline
            \hspace{0.5em} \texttt{int4} & 1E+4 & 9E+3  & 3E+6 & 9E+6 & 5E+2 & 1E+2 \\
            \hspace{0.5em} 4-bit ANT & 27.79 & 27.35  & 23.22 & 27.36 & 4E+4 & 4E+4 \\
            \hspace{0.5em} \textbf{4-bit OliVe}  & 19.11 & 18.08  & 15.16 & 17.18 & 55.44 & 32.41 \\
            \hline
        \end{tabular}
        \vspace*{0.1cm}
        \caption[]{PTQ results on large language models. The accuracy metric is perplexity, and lower is better.}
        \label{tbl:LLM_results}
        \vspace*{-0.6cm}
    \end{table}
\end{center}

\paragraph{Large Language Models.} 
We evaluate the accuracy of \proj{} for LLMs under the PTQ setting.
LLMs' inference is challenging as it requires significant memory, which makes their \revise{retraining} even more resource-consuming.
Thus, the PTQ method without \revise{retraining} is more desirable than the QAT method for LLMs. 


The recent work~\cite{dettmers2022llm} has shown that the \texttt{int8} quantization has a significant accuracy drop when the number of parameters of the OPT model grows to 6.7B. As shown in \Tbl{tbl:LLM_results}, our 8-bit PTQ method has only a negligible perplexity increase on OPT-6.7B (lower is better), while the accuracy of the \texttt{int8}-based quantization method has a significant degradation and is worse than our 4-bit PTQ method on the C4 dataset. On GPT2-XL and BLOOM-7B1 models, our 8-bit PTQ method essentially achieves the original perplexity, and the 4-bit PTQ method achieves the performance close to \texttt{int8}. For comparison, the accuracy results of \texttt{int4} and 4-bit ANT are unacceptable (10-1000$\times$ worse than FP32 model).

To summarize, our \proj{} quantization framework pushes the limit of 4-bit quantization to a new state-of-the-art, as it is able to achieve nearly original accuracy for the commonly used language models including BERT-base, BERT-large, and BART-base on most datasets.
Moreover, \proj{} also gives the state-of-the-art results of 4-bit and 8-bit quantization on large language models like GPT2-XL, BLOOM-7B1, and OPT-6.7B.

\vspace*{-0.5cm}
\subsection{GPU Performance and Energy}

\begin{figure}[t]
    \hspace{-5mm}
    \centering
    \subfloat[Speedup on GPU.]{\label{fig:gpu_s}
    \includegraphics[width=\linewidth]{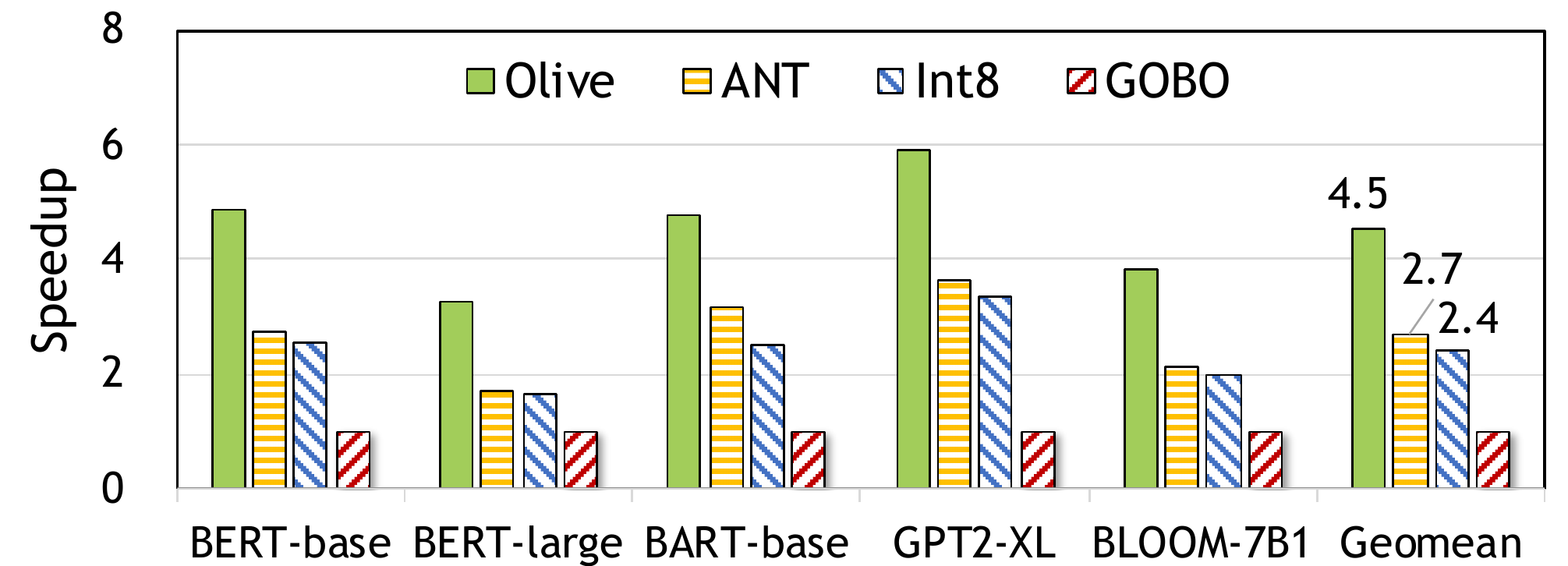}}
    \\
    \subfloat[Normalized energy on GPU.]{\label{fig:gpu_e}
    \hspace{-5mm}
    \includegraphics[width=\linewidth]{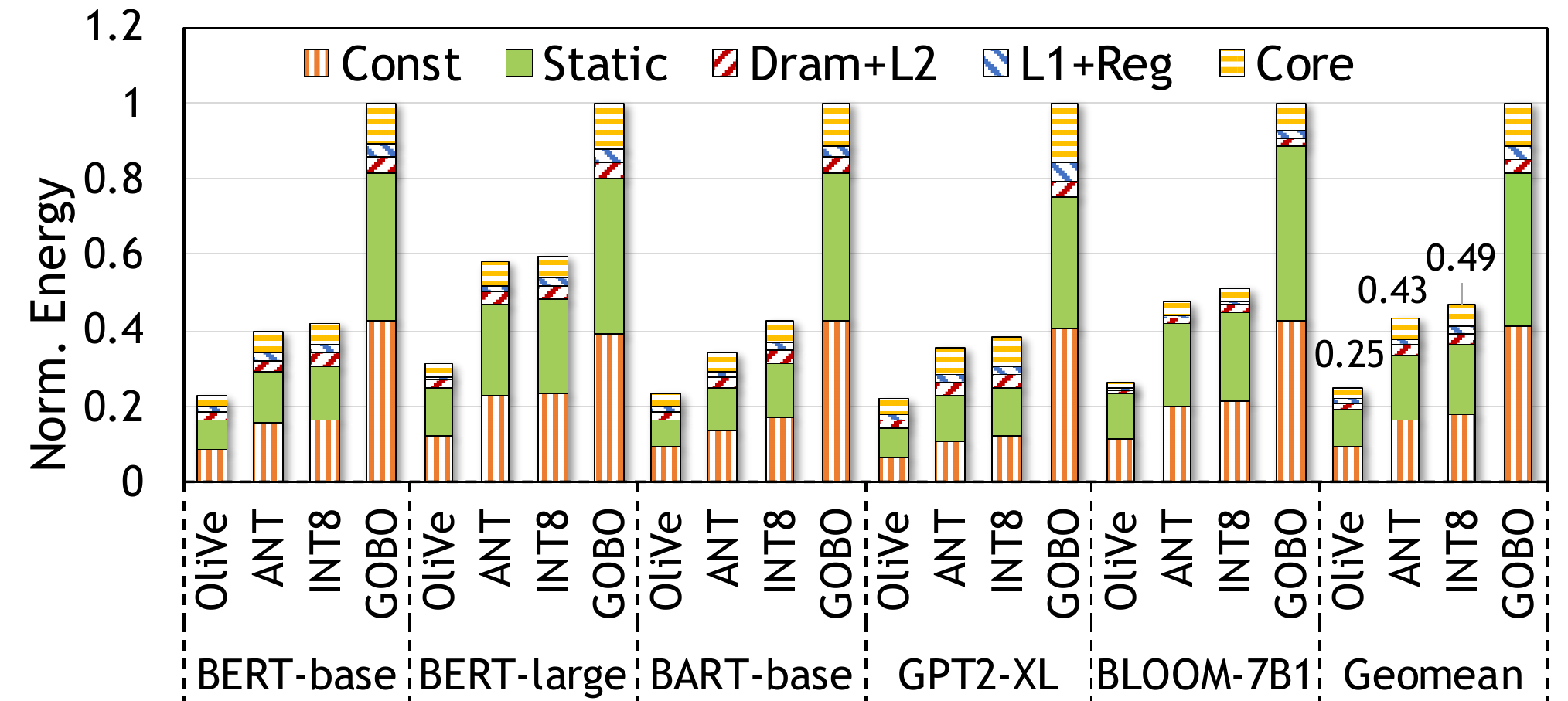} }
    \caption{\revise{Comparison of four different designs on GPU.}}\label{fig:gpu_res}
\end{figure}

We evaluate LLMs on the GPU simulator, where the batch size is set to \revise{2} for GPT-like models and \revise{16} for BERT-like models. 
For \proj{}, 4-bit quantization can limit the loss to a relatively small error range. 
GOBO~\cite{zadeh2020gobo} can achieve the original accuracy of all models but has a significant overhead on compressing weight in DRAM. Note that GOBO only quantizes the weight tensors and computes with FP16.
We implemented GOBO's memory organization in the GPU.
For ANT~\cite{guo2022ant}, we make all models close to the original accuracy or perplexity by mixed precision (BERT-like models~\cite{devlin2018bert, lewis2019bart} with $<$ 1\% loss and GPT-like models~\cite{zhang2022opt, scao2022bloom, radford2019language} with $<$ 3 perplexity) with the PTQ setting. 
In addition, we also compare the original \texttt{int8} of GPU, which has unacceptable accuracy loss, just for performance and energy comparison to GPU baseline.
We compare the GPU architecture integrated with our \proj{} design against various baselines.
The performance and energy results are shown in \Fig{fig:gpu_res}.

\paragraph{Performance.}
\Fig{fig:gpu_s} compares the speedup values of different quantization methods on GPUs. \proj{} achieves the best performance and has higher speedups on the larger language models than GOBO.
Due to the FP16 computation and weight-only quantization, GOBO~\cite{zadeh2020gobo} achieves the lowest performance among all studied designs.
In contrast, \proj{} quantizes both activation and weight to low bits and does not increase the memory access overhead. This avoids performance degradation when the number of parameters increases. 
The PTQ seriously degrades the accuracy of ANT~\cite{guo2022ant} as it cannot handle outliers.
In ANT, 80\% of layers ends up using \texttt{int8} quantization so the performance results between ANT and \texttt{int8} are close.
On average, \proj{} achieves 4.5$\times$, 2.7$\times$, and 2.4$\times$ speedup values over GOBO, \texttt{int8}, and ANT, respectively.

\paragraph{Energy.}
\Fig{fig:gpu_e} shows the normalized energy comparison of different designs, including constant, static, and dynamic power. And the dynamic power includes DRAM, L2 cache, L1 data cache, shared memory, register file, and processing elements (CUDA core and tensor core). The L1 contains the sum of the L1 cache and shared memory energy.
\proj{} has the lowest energy due to the aligned 4-bit design and GPU compatibility.  
Due to the worse accuracy result of the mixed precision, ANT is also close to \texttt{int8} on the energy. Overall, 4-bit \proj{} is very hardware-friendly so that it can take full advantage of the energy savings with lower bits. \proj{} achieves average 4.0$\times$, 2.3$\times$, and 2.0$\times$ energy reduction over GOBO, \texttt{int8}, and ANT, respectively.

\paragraph{Area.}
To measure the overhead of \proj{} decoder on the GPU, we scale the \proj{} decoder to 12~$nm$, which is the same manufacturing process as RTX 2080 Ti~\cite{turing} and calculate the tile area.
According to \Tbl{tab:Turing}, there are 139,264 4-bit decoders and 69,632 8-bit decoders on the GPU die and their area \revise{is} shown in \Tbl{tab:GPU_area}. Since the GPU die size of RTX 2080 Ti is 754 $mm^2$, the 4-bit decoder and 8-bit decoder only account for 0.250\% and 0.166\% of the entire GPU area respectively, which we believe is a tiny and worthy overhead.

\begin{table}[b]
    \vspace*{-0.2cm}
    \ra{1.5}
    \resizebox{\columnwidth}{!}{%
    \begin{tabular}{c|c|c|c}
     Component & Number & Area ($mm^2$) & Area Ratio \\ \Xhline{1.2pt}
    
    4-bit Decoder (13.53$\mu m^2$) & 139,264 & 1.88 & 0.250\%\\ \hline
    8-bit Decoder (18.00$\mu m^2$) & 69,632 & 1.25 & 0.166\% \\
    
    \end{tabular}%
    }
    \vspace*{1mm}
    \caption{The area of \proj{} decoder on RTX 2080 Ti.}
    \label{tab:GPU_area}
    \vspace*{-.2cm}
\end{table}

 \subsection{Accelerator Performance and Energy}
 
As explained in \Sec{subsec:eval:methodology}, we also integrate \proj{} to the systolic-array-based hardware accelerator and compare its performance and energy against existing designs of ANT~\cite{guo2022ant}, OLAccel~\cite{park2018energy}, and AdaFloat~\cite{tambe2020algorithm}. 
Similar to its GPU implementation, ANT is a mixed-precision design. Since AdaFloat does not support mixed precision, we only provide the 8-bit quantization results. 
All accelerators can achieve close to original accuracy for all Transformer models.

\begin{figure}[t]
    \hspace{-5mm}
    \centering
    \subfloat[Speedup on hardware accelerator.]{\label{fig:sa_s}
    \includegraphics[width=\linewidth]{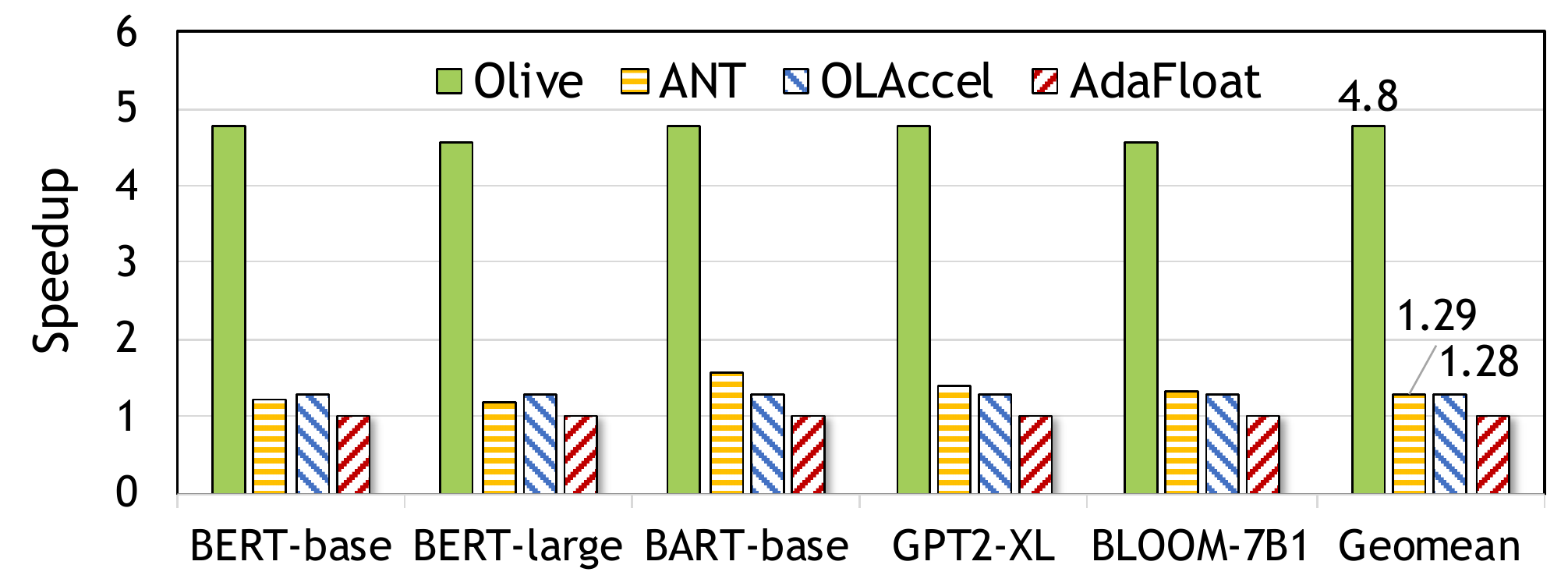}}
    \\
    \subfloat[Normalized energy on hardware accelerator.]{\label{fig:sa_e}
    \hspace{-5mm}
    \includegraphics[width=\linewidth]{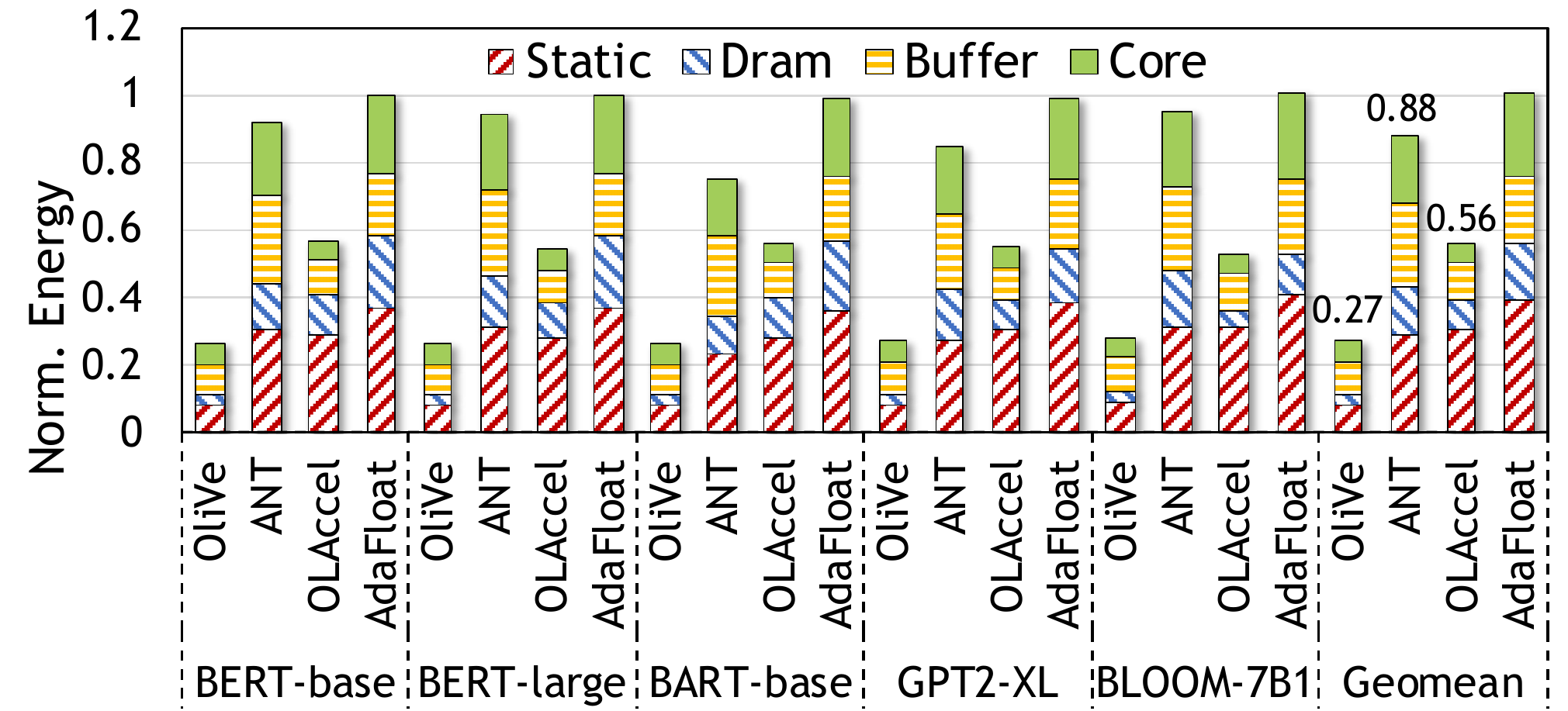} }
    \caption{\revise{Comparison of different designs on accelerators.}}\label{fig:sa_res}
\end{figure}

\paragraph{Performance.}
As shown in \Fig{fig:sa_s}, \proj{} has the most significant advantage in latency speedup.
Owing to its inability to deal with outliers, the performance of ANT is similar to OLAccel on most models.
The speedup values of \proj{} are very similar on all models, and they do not change with the increasing number of model parameters. 
On average, \proj{} achieves 4.8$\times$, 3.8$\times$, and 3.7$\times$ speedup value over AdaFloat, OLAccel, and ANT, respectively.

 \paragraph{Energy.}
 \Fig{fig:sa_e} shows the normalized energy consumption of different designs composed of static and dynamic energy (DRAM, on-chip buffer, and core). 
 \proj{} has the lowest energy consumption. Compared to OLAccel, \proj{} has a significant advantage in terms of static and DRAM. 
 Worse mixed-precision results increase ANT energy consumption, which is even close to AdaFloat in BLOOM-7B1 model. 
 On average, \proj{} achieves 3.7$\times$, 2.1$\times$, and 3.3$\times$ energy reduction over AdaFloat, OLAccel, and ANT, respectively.

 \paragraph{Area.}
\Tbl{tab:ASIC_area} shows the area breakdown of \proj{}-based systolic array architecture under 22~$nm$ process.
In this scenario, the 4-bit and 8-bit decoders \revise{introduce} about 2.2\% and 1.5\% overhead of the core area, respectively, which is inconsiderable compared to the area of PEs in the array. 
Considering on-chip memory structures, the overall area overhead would be even smaller.
In addition, we also scale other accelerators to 22~$nm$ using DeepScaleTool~\cite{sarangi2021deepscaletool} and get similar results to those numbers. 
Note that we implement all accelerators with a similar area size.
\revise{
The small area overhead of our \proj{} directly benefits from the carefully-designed outlier-victim pair (OVP) encoding.
}

\begin{table}[b]
    \vspace*{-0.2cm}
    \ra{1.5}
    \resizebox{\columnwidth}{!}{%
    \begin{tabular}{c|c|c|c}
     Component & Number & Area ($mm^2$) & Area Ratio \\ \Xhline{1.2pt}
    
    4-bit Decoder (37.22$\mu m^2$) & 128 & 0.00476 & 2.2\%\\ \hline
    8-bit Decoder (49.50$\mu m^2$) & 64 & 0.00317 & 1.5\% \\ \hline
    4-bit PE (50.01$\mu m^2$) & 4096 & 0.205 & 96.3\% \\ 
    \end{tabular}%
    }
    \vspace*{1mm}
    \caption{Area breakdown of \proj{} under 22~$nm$ process.}
    \label{tab:ASIC_area}
    \vspace*{-.2cm}
\end{table}

%% file: tex/conclusion.tex
\section{Related Work and Discussion}

This section presents and discusses research on DNN acceleration and compression.
With the growing computation requirements of DNN models, it is crucial to design the algorithms and architecture to accelerate DNN models.
Various compression methods, such as pruning and quantization, have been proposed to exploit the redundancy property of DNNs. 

\paragraph{DNN Acceleration.}
In the past few years, various architectures~\cite{chen2014diannao, chen2014dadiannao, liu2015pudiannao, du2015shidiannao, chen2016eyeriss, peemen2013memory, zhang2015optimizing, gokhale2014240, gupta2015deep, guo2020balancing, qin2020sigma, zhou2021characterizing} have been proposed to match the computation characteristics of DNN models. To accelerate the DNN system, most optimizations focus on compilation~\cite{TVM, FlexTensor, Ansor, Roller, TASO, zhou2023ugrapher} and scheduling~\cite{Baymax, Prophet, Ebird, LazyBatch, DVABatch, BubbleUp, BubbleFlux, Heracles, VELTAIR, guo2022nesting}.

The DNN acceleration highly relies on the performance of matrix multiplication.
Therefore, several works focus on improving data reuse and simplifying control logic through a tailored dataflow architecture for matrix multiplication\cite{chen2014diannao, chen2014dadiannao, tpuv4isca, peemen2013memory, zhang2015optimizing, gokhale2014240, gupta2015deep, guo2020balancing, qin2020sigma, zhou2021characterizing}.
TPU~\cite{tpuv4isca} introduces a highly optimized dataflow architecture that efficiently reuses data across multiple computation stages.
Modern GPUs~\cite{a100} now incorporate matrix multiplication accelerators, such as tensor core, optimized for SIMD operations to enhance DNN workload acceleration further.

\paragraph{Pruning.}
Pruning means removing a portion of weight, input, or output of DNN layers, resulting in a sparse model with reduced model size.
However, a significant reduction leads to irregular memory accesses, which are negative for the acceleration of inference and training.
To address this issue, researchers propose several sparse optimizations in algorithms and hardware architectures to reduce inefficient computation~\cite{han2015deep, albericio2016cnvlutin, zhang2016cambricon, zhou2018cambricon,  zhu2019sparse,Qiu_2019_CVPR, guan2020far, qin2020sigma, guo2020accelerating,  wang2021dual, guan2022block, guan2022transkimmer}.
In addition, a sparse tensor core is introduced in NVIDIA Ampere GPU architecture~\cite{2020ampere} to support the 2:4 structured sparsity.

\paragraph{Quantization.}
Quantization is another effective and efficient way to reduce the DNN model size and computation burden.
There are two popular quantization methods, i.e., quantization-aware training (QAT)~\cite{jacob2018quantization, wang2019learning, zhuang2021effective, li2022efficient} and post-training quantization (PTQ)~\cite{gupta2015deep, jacob2018quantization, wang2019learning, guo2022squant}.
QAT allows the model to adapt to quantization noise by retraining. PTQ is very effective to implement since it converts the original FP32 model directly into a lower-bit model without the training data and pipeline. Thus, PTQ is more feasible for language models at billion scales.

By quantizing data to low bit-width, quantization accelerators can significantly reduce memory bandwidth requirements and increase the computation speed.
BitFusion~\cite{sharma2018bit} combines the low-bit PEs to support different bit-width quantization.
OLAccel~\cite{park2018energy} utilizes 16-bit MAC to the first layer and 4-bit MAC to other layers.
DRQ~\cite{song2020drq} quantizes data in sensitive and insensitive areas with different precision, which is similar to outlier-aware quantization. 
GOBO~\cite{zadeh2020gobo} is an accelerator that takes advantage of outlier-aware quantization, which quantizes the outliers of weights with higher precision. 
However, the outlier-aware quantization accelerators mentioned above have unaligned memory accesses, resulting in additional overhead and a limited computing speed. 
ANT~\cite{guo2022ant} proposes a fixed-length adaptive quantization framework but only takes the distribution of tensors into account and ignores the importance of outliers. 
In contrast, our proposed novel \proj{} quantization framework can handle outlier values in a memory-aligned and hardware-friendly way.

AdaptivFloat~\cite{tambe2020algorithm} is similar to \texttt{abfloat} in adding a bias to the exponent, but the motivations and how the bias is determined are different. AdaptivFloat is to adapt to the dynamic ranges of different layers and calculates the optimal bias at a layer granularity using its algorithm. 
Our \texttt{abfloat} is to make full use of the encoding range, so it simply adds a uniform bias to all encoding values to skip the range of normal values, which is simpler to implement.

\paragraph{GPU Architecture.} 
NVIDIA has been updating its new generations of GPUs, e.g., Ampere architecture~\cite{2020ampere}, which adds the sparse tensor core for structured sparsity in DNNs and compute data compression to increase the memory access bandwidth.
The structured sparsity for tensor cores is orthogonal to our proposed quantization as our element-wise quantization does not affect (sparse) tensor core dataflow. 
Ampere GPU's compute data compression can compress zero values and similar bytes in DRAM and L2 cache.
As such, it is lossless and therefore general-purpose.
It is also transparent and orthogonal to \proj{}, which does not modify the memory system. 
In contrast, prior quantization work~\cite{zadeh2020gobo} perform compression at the DRAM-level, which could be impacted by the data compression in Ampere GPUs.

On the other hand, DNN quantization is a lossy compression. 
We believe the strictly lossless compression would have limited benefits for DNN quantization. 
Thus, our work could complement Ampere's current compute data compression as a special-purpose solution.
Since existing GPU simulators~\cite{gpgpu-sim2009, gpgpu-sim4.0} cannot support data compression, we will continue to follow up and study this problem in the future work.

\section{Conclusion}

In this work, we propose a novel outlier-victim pair (OVP) quantization, which can handle outlier values with low hardware overhead and achieve high performance gains.
The key insight is to sacrifice the normal values next to those essential outliers (called victims) to accommodate them. The OVP encoding designed based on this idea is able to make outliers and normal values globally identical but locally distinguishable.
To the best of our knowledge, \proj{} pushes the limit of 4-bit quantization to a new state-of-the-art, as it is able to achieve nearly original accuracy for commonly used language models.
Moreover, our architecture design can be efficiently integrated into existing hardware accelerators such as tensor core and systolic array.
Finally, \proj{}-based accelerator surpasses the existing outlier-aware accelerator, GOBO, by 4.5$\times$ speedup and 4.0$\times$ energy reduction, respectively.

%

%% file: paper.bbl

\begin{thebibliography}{99}


\ifx \showCODEN    \undefined \def \showCODEN     #1{\unskip}     \fi
\ifx \showDOI      \undefined \def \showDOI       #1{#1}\fi
\ifx \showISBNx    \undefined \def \showISBNx     #1{\unskip}     \fi
\ifx \showISBNxiii \undefined \def \showISBNxiii  #1{\unskip}     \fi
\ifx \showISSN     \undefined \def \showISSN      #1{\unskip}     \fi
\ifx \showLCCN     \undefined \def \showLCCN      #1{\unskip}     \fi
\ifx \shownote     \undefined \def \shownote      #1{#1}          \fi
\ifx \showarticletitle \undefined \def \showarticletitle #1{#1}   \fi
\ifx \showURL      \undefined \def \showURL       {\relax}        \fi
\providecommand\bibfield[2]{#2}
\providecommand\bibinfo[2]{#2}
\providecommand\natexlab[1]{#1}
\providecommand\showeprint[2][]{arXiv:#2}

\bibitem[\protect\citeauthoryear{NVIDIA}{202}{2020}]%
        {2020ampere}
 \bibinfo{year}{2020}\natexlab{}.
\newblock \bibinfo{title}{Nvidia ampere architecture whitepaper}.
\newblock
  \bibinfo{howpublished}{\url{https://images.nvidia.com/aem-dam/en-zz/Solutions/data-center/nvidia-ampere-architecture-whitepaper.pdf}}.
\newblock


\bibitem[\protect\citeauthoryear{Albericio, Judd, Hetherington, Aamodt, Jerger,
  and Moshovos}{Albericio et~al\mbox{.}}{2016}]%
        {albericio2016cnvlutin}
\bibfield{author}{\bibinfo{person}{Jorge Albericio}, \bibinfo{person}{Patrick
  Judd}, \bibinfo{person}{Tayler Hetherington}, \bibinfo{person}{Tor Aamodt},
  \bibinfo{person}{Natalie~Enright Jerger}, {and} \bibinfo{person}{Andreas
  Moshovos}.} \bibinfo{year}{2016}\natexlab{}.
\newblock \showarticletitle{Cnvlutin: Ineffectual-neuron-free deep neural
  network computing}.
\newblock \bibinfo{journal}{\emph{ACM SIGARCH Computer Architecture News}}
  \bibinfo{volume}{44}, \bibinfo{number}{3} (\bibinfo{year}{2016}),
  \bibinfo{pages}{1--13}.
\newblock


\bibitem[\protect\citeauthoryear{Bakhoda, Yuan, Fung, Wong, and Aamodt}{Bakhoda
  et~al\mbox{.}}{2009}]%
        {gpgpu-sim2009}
\bibfield{author}{\bibinfo{person}{Ali Bakhoda}, \bibinfo{person}{George Yuan},
  \bibinfo{person}{Wilson Fung}, \bibinfo{person}{Henry Wong}, {and}
  \bibinfo{person}{Tor Aamodt}.} \bibinfo{year}{2009}\natexlab{}.
\newblock \showarticletitle{Analyzing CUDA workloads using a detailed GPU
  simulator}.
\newblock \bibinfo{journal}{\emph{ISPASS 2009 - International Symposium on
  Performance Analysis of Systems and Software}}, \bibinfo{pages}{163 -- 174}.
\newblock
\urldef\tempurl%
\url{https://doi.org/10.1109/ISPASS.2009.4919648}
\showDOI{\tempurl}


\bibitem[\protect\citeauthoryear{Banner, Nahshan, and Soudry}{Banner
  et~al\mbox{.}}{2019}]%
        {banner2019post}
\bibfield{author}{\bibinfo{person}{Ron Banner}, \bibinfo{person}{Yury Nahshan},
  {and} \bibinfo{person}{Daniel Soudry}.} \bibinfo{year}{2019}\natexlab{}.
\newblock \showarticletitle{Post training 4-bit quantization of convolutional
  networks for rapid-deployment}.
\newblock \bibinfo{journal}{\emph{Advances in Neural Information Processing
  Systems}}  \bibinfo{volume}{32} (\bibinfo{year}{2019}).
\newblock


\bibitem[\protect\citeauthoryear{Bengio, L{\'e}onard, and Courville}{Bengio
  et~al\mbox{.}}{2013}]%
        {bengio2013estimating}
\bibfield{author}{\bibinfo{person}{Yoshua Bengio}, \bibinfo{person}{Nicholas
  L{\'e}onard}, {and} \bibinfo{person}{Aaron Courville}.}
  \bibinfo{year}{2013}\natexlab{}.
\newblock \showarticletitle{Estimating or propagating gradients through
  stochastic neurons for conditional computation}.
\newblock \bibinfo{journal}{\emph{arXiv preprint arXiv:1308.3432}}
  (\bibinfo{year}{2013}).
\newblock


\bibitem[\protect\citeauthoryear{Cai, Yao, Dong, Gholami, Mahoney, and
  Keutzer}{Cai et~al\mbox{.}}{2020}]%
        {cai2020zeroq}
\bibfield{author}{\bibinfo{person}{Yaohui Cai}, \bibinfo{person}{Zhewei Yao},
  \bibinfo{person}{Zhen Dong}, \bibinfo{person}{Amir Gholami},
  \bibinfo{person}{Michael~W Mahoney}, {and} \bibinfo{person}{Kurt Keutzer}.}
  \bibinfo{year}{2020}\natexlab{}.
\newblock \showarticletitle{Zeroq: A novel zero shot quantization framework}.
  In \bibinfo{booktitle}{\emph{Proceedings of the IEEE/CVF Conference on
  Computer Vision and Pattern Recognition}}. \bibinfo{pages}{13169--13178}.
\newblock


\bibitem[\protect\citeauthoryear{Cai and Vasconcelos}{Cai and
  Vasconcelos}{2020}]%
        {cai2020rethinking}
\bibfield{author}{\bibinfo{person}{Zhaowei Cai} {and} \bibinfo{person}{Nuno
  Vasconcelos}.} \bibinfo{year}{2020}\natexlab{}.
\newblock \showarticletitle{Rethinking differentiable search for
  mixed-precision neural networks}. In \bibinfo{booktitle}{\emph{Proceedings of
  the IEEE/CVF Conference on Computer Vision and Pattern Recognition}}.
  \bibinfo{pages}{2349--2358}.
\newblock


\bibitem[\protect\citeauthoryear{Chen, Yang, Guo, Kannan, Mars, and Tang}{Chen
  et~al\mbox{.}}{2017}]%
        {Prophet}
\bibfield{author}{\bibinfo{person}{Quan Chen}, \bibinfo{person}{Hailong Yang},
  \bibinfo{person}{Minyi Guo}, \bibinfo{person}{Ram~Srivatsa Kannan},
  \bibinfo{person}{Jason Mars}, {and} \bibinfo{person}{Lingjia Tang}.}
  \bibinfo{year}{2017}\natexlab{}.
\newblock \showarticletitle{Prophet: Precise QoS Prediction on Non-Preemptive
  Accelerators to Improve Utilization in Warehouse-Scale Computers}. In
  \bibinfo{booktitle}{\emph{Proceedings of the Twenty-Second International
  Conference on Architectural Support for Programming Languages and Operating
  Systems, {ASPLOS} 2017, Xi'an, China, April 8-12, 2017}}.
  \bibinfo{publisher}{{ACM}}, \bibinfo{pages}{17--32}.
\newblock
\urldef\tempurl%
\url{https://doi.org/10.1145/3037697.3037700}
\showDOI{\tempurl}


\bibitem[\protect\citeauthoryear{Chen, Yang, Mars, and Tang}{Chen
  et~al\mbox{.}}{2016b}]%
        {Baymax}
\bibfield{author}{\bibinfo{person}{Quan Chen}, \bibinfo{person}{Hailong Yang},
  \bibinfo{person}{Jason Mars}, {and} \bibinfo{person}{Lingjia Tang}.}
  \bibinfo{year}{2016}\natexlab{b}.
\newblock \showarticletitle{Baymax: QoS Awareness and Increased Utilization for
  Non-Preemptive Accelerators in Warehouse Scale Computers}. In
  \bibinfo{booktitle}{\emph{Proceedings of the Twenty-First International
  Conference on Architectural Support for Programming Languages and Operating
  Systems, {ASPLOS} 2016, Atlanta, GA, USA, April 2-6, 2016}}.
  \bibinfo{publisher}{{ACM}}, \bibinfo{pages}{681--696}.
\newblock
\urldef\tempurl%
\url{https://doi.org/10.1145/2872362.2872368}
\showDOI{\tempurl}


\bibitem[\protect\citeauthoryear{Chen, Du, Sun, Wang, Wu, Chen, and Temam}{Chen
  et~al\mbox{.}}{2014a}]%
        {chen2014diannao}
\bibfield{author}{\bibinfo{person}{Tianshi Chen}, \bibinfo{person}{Zidong Du},
  \bibinfo{person}{Ninghui Sun}, \bibinfo{person}{Jia Wang},
  \bibinfo{person}{Chengyong Wu}, \bibinfo{person}{Yunji Chen}, {and}
  \bibinfo{person}{Olivier Temam}.} \bibinfo{year}{2014}\natexlab{a}.
\newblock \showarticletitle{Diannao: A small-footprint high-throughput
  accelerator for ubiquitous machine-learning}.
\newblock \bibinfo{journal}{\emph{ACM SIGARCH Computer Architecture News}}
  \bibinfo{volume}{42}, \bibinfo{number}{1} (\bibinfo{year}{2014}),
  \bibinfo{pages}{269--284}.
\newblock


\bibitem[\protect\citeauthoryear{Chen, Moreau, Jiang, Zheng, Yan, Shen, Cowan,
  Wang, Hu, Ceze, Guestrin, and Krishnamurthy}{Chen et~al\mbox{.}}{2018}]%
        {TVM}
\bibfield{author}{\bibinfo{person}{Tianqi Chen}, \bibinfo{person}{Thierry
  Moreau}, \bibinfo{person}{Ziheng Jiang}, \bibinfo{person}{Lianmin Zheng},
  \bibinfo{person}{Eddie~Q. Yan}, \bibinfo{person}{Haichen Shen},
  \bibinfo{person}{Meghan Cowan}, \bibinfo{person}{Leyuan Wang},
  \bibinfo{person}{Yuwei Hu}, \bibinfo{person}{Luis Ceze},
  \bibinfo{person}{Carlos Guestrin}, {and} \bibinfo{person}{Arvind
  Krishnamurthy}.} \bibinfo{year}{2018}\natexlab{}.
\newblock \showarticletitle{{TVM:} An Automated End-to-End Optimizing Compiler
  for Deep Learning}. In \bibinfo{booktitle}{\emph{13th {USENIX} Symposium on
  Operating Systems Design and Implementation, {OSDI} 2018, Carlsbad, CA, USA,
  October 8-10, 2018}}. \bibinfo{publisher}{{USENIX} Association},
  \bibinfo{pages}{578--594}.
\newblock
\urldef\tempurl%
\url{https://doi.org/10.5555/3291168.3291211}
\showDOI{\tempurl}


\bibitem[\protect\citeauthoryear{Chen, Luo, Liu, Zhang, He, Wang, Li, Chen, Xu,
  Sun, et~al\mbox{.}}{Chen et~al\mbox{.}}{2014b}]%
        {chen2014dadiannao}
\bibfield{author}{\bibinfo{person}{Yunji Chen}, \bibinfo{person}{Tao Luo},
  \bibinfo{person}{Shaoli Liu}, \bibinfo{person}{Shijin Zhang},
  \bibinfo{person}{Liqiang He}, \bibinfo{person}{Jia Wang},
  \bibinfo{person}{Ling Li}, \bibinfo{person}{Tianshi Chen},
  \bibinfo{person}{Zhiwei Xu}, \bibinfo{person}{Ninghui Sun}, {et~al\mbox{.}}}
  \bibinfo{year}{2014}\natexlab{b}.
\newblock \showarticletitle{Dadiannao: A machine-learning supercomputer}. In
  \bibinfo{booktitle}{\emph{2014 47th Annual IEEE/ACM International Symposium
  on Microarchitecture}}. IEEE, \bibinfo{pages}{609--622}.
\newblock


\bibitem[\protect\citeauthoryear{Chen, Krishna, Emer, and Sze}{Chen
  et~al\mbox{.}}{2016a}]%
        {chen2016eyeriss}
\bibfield{author}{\bibinfo{person}{Yu-Hsin Chen}, \bibinfo{person}{Tushar
  Krishna}, \bibinfo{person}{Joel~S Emer}, {and} \bibinfo{person}{Vivienne
  Sze}.} \bibinfo{year}{2016}\natexlab{a}.
\newblock \showarticletitle{Eyeriss: An energy-efficient reconfigurable
  accelerator for deep convolutional neural networks}.
\newblock \bibinfo{journal}{\emph{IEEE journal of solid-state circuits}}
  \bibinfo{volume}{52}, \bibinfo{number}{1} (\bibinfo{year}{2016}),
  \bibinfo{pages}{127--138}.
\newblock


\bibitem[\protect\citeauthoryear{Choi, Wang, Venkataramani, Chuang, Srinivasan,
  and Gopalakrishnan}{Choi et~al\mbox{.}}{2018}]%
        {choi2018pact}
\bibfield{author}{\bibinfo{person}{Jungwook Choi}, \bibinfo{person}{Zhuo Wang},
  \bibinfo{person}{Swagath Venkataramani}, \bibinfo{person}{Pierce I-Jen
  Chuang}, \bibinfo{person}{Vijayalakshmi Srinivasan}, {and}
  \bibinfo{person}{Kailash Gopalakrishnan}.} \bibinfo{year}{2018}\natexlab{}.
\newblock \showarticletitle{Pact: Parameterized clipping activation for
  quantized neural networks}.
\newblock \bibinfo{journal}{\emph{arXiv preprint arXiv:1805.06085}}
  (\bibinfo{year}{2018}).
\newblock


\bibitem[\protect\citeauthoryear{Choi, Kim, and Rhu}{Choi
  et~al\mbox{.}}{2021}]%
        {LazyBatch}
\bibfield{author}{\bibinfo{person}{Yujeong Choi}, \bibinfo{person}{Yunseong
  Kim}, {and} \bibinfo{person}{Minsoo Rhu}.} \bibinfo{year}{2021}\natexlab{}.
\newblock \showarticletitle{Lazy Batching: An SLA-aware Batching System for
  Cloud Machine Learning Inference}. In \bibinfo{booktitle}{\emph{{IEEE}
  International Symposium on High-Performance Computer Architecture, {HPCA}
  2021, Seoul, South Korea, February 27 - March 3, 2021}}.
  \bibinfo{publisher}{{IEEE}}, \bibinfo{pages}{493--506}.
\newblock
\urldef\tempurl%
\url{https://doi.org/10.1109/HPCA51647.2021.00049}
\showDOI{\tempurl}


\bibitem[\protect\citeauthoryear{Cui, Wei, Chen, Tang, Leng, Li, and Guo}{Cui
  et~al\mbox{.}}{2019}]%
        {Ebird}
\bibfield{author}{\bibinfo{person}{Weihao Cui}, \bibinfo{person}{Mengze Wei},
  \bibinfo{person}{Quan Chen}, \bibinfo{person}{Xiaoxin Tang},
  \bibinfo{person}{Jingwen Leng}, \bibinfo{person}{Li Li}, {and}
  \bibinfo{person}{Mingyi Guo}.} \bibinfo{year}{2019}\natexlab{}.
\newblock \showarticletitle{Ebird: Elastic Batch for Improving Responsiveness
  and Throughput of Deep Learning Services}. In \bibinfo{booktitle}{\emph{37th
  {IEEE} International Conference on Computer Design, {ICCD} 2019, Abu Dhabi,
  United Arab Emirates, November 17-20, 2019}}. \bibinfo{publisher}{{IEEE}},
  \bibinfo{pages}{497--505}.
\newblock
\urldef\tempurl%
\url{https://doi.org/10.1109/ICCD46524.2019.00075}
\showDOI{\tempurl}


\bibitem[\protect\citeauthoryear{Cui, Zhao, Chen, Wei, Li, Zeng, Li, and
  Guo}{Cui et~al\mbox{.}}{2022}]%
        {DVABatch}
\bibfield{author}{\bibinfo{person}{Weihao Cui}, \bibinfo{person}{Han Zhao},
  \bibinfo{person}{Quan Chen}, \bibinfo{person}{Hao Wei},
  \bibinfo{person}{Zirui Li}, \bibinfo{person}{Deze Zeng},
  \bibinfo{person}{Chao Li}, {and} \bibinfo{person}{Minyi Guo}.}
  \bibinfo{year}{2022}\natexlab{}.
\newblock \showarticletitle{{DVABatch}: Diversity-aware {Multi-Entry}
  {Multi-Exit} Batching for Efficient Processing of {DNN} Services on {GPUs}}.
  In \bibinfo{booktitle}{\emph{2022 USENIX Annual Technical Conference (USENIX
  ATC 22)}}. \bibinfo{pages}{183--198}.
\newblock


\bibitem[\protect\citeauthoryear{Dettmers, Lewis, Belkada, and
  Zettlemoyer}{Dettmers et~al\mbox{.}}{2022}]%
        {dettmers2022llm}
\bibfield{author}{\bibinfo{person}{Tim Dettmers}, \bibinfo{person}{Mike Lewis},
  \bibinfo{person}{Younes Belkada}, {and} \bibinfo{person}{Luke Zettlemoyer}.}
  \bibinfo{year}{2022}\natexlab{}.
\newblock \showarticletitle{Llm. int8 (): 8-bit matrix multiplication for
  transformers at scale}.
\newblock \bibinfo{journal}{\emph{arXiv preprint arXiv:2208.07339}}
  (\bibinfo{year}{2022}).
\newblock


\bibitem[\protect\citeauthoryear{Devlin, Chang, Lee, and Toutanova}{Devlin
  et~al\mbox{.}}{2018}]%
        {devlin2018bert}
\bibfield{author}{\bibinfo{person}{Jacob Devlin}, \bibinfo{person}{Ming-Wei
  Chang}, \bibinfo{person}{Kenton Lee}, {and} \bibinfo{person}{Kristina
  Toutanova}.} \bibinfo{year}{2018}\natexlab{}.
\newblock \showarticletitle{Bert: Pre-training of deep bidirectional
  transformers for language understanding}.
\newblock \bibinfo{journal}{\emph{arXiv preprint arXiv:1810.04805}}
  (\bibinfo{year}{2018}).
\newblock


\bibitem[\protect\citeauthoryear{Dodge, Sap, Marasović, Agnew, Ilharco,
  Groeneveld, Mitchell, and Gardner}{Dodge et~al\mbox{.}}{2021}]%
        {jesse2021c4}
\bibfield{author}{\bibinfo{person}{Jesse Dodge}, \bibinfo{person}{Maarten Sap},
  \bibinfo{person}{Ana Marasović}, \bibinfo{person}{William Agnew},
  \bibinfo{person}{Gabriel Ilharco}, \bibinfo{person}{Dirk Groeneveld},
  \bibinfo{person}{Margaret Mitchell}, {and} \bibinfo{person}{Matt Gardner}.}
  \bibinfo{year}{2021}\natexlab{}.
\newblock \bibinfo{title}{Documenting Large Webtext Corpora: A Case Study on
  the Colossal Clean Crawled Corpus}.
\newblock
\newblock
\showeprint{arXiv:2104.08758}


\bibitem[\protect\citeauthoryear{Dong, Yao, Arfeen, Gholami, Mahoney, and
  Keutzer}{Dong et~al\mbox{.}}{2020}]%
        {dong2020hawq}
\bibfield{author}{\bibinfo{person}{Zhen Dong}, \bibinfo{person}{Zhewei Yao},
  \bibinfo{person}{Daiyaan Arfeen}, \bibinfo{person}{Amir Gholami},
  \bibinfo{person}{Michael~W Mahoney}, {and} \bibinfo{person}{Kurt Keutzer}.}
  \bibinfo{year}{2020}\natexlab{}.
\newblock \showarticletitle{Hawq-v2: Hessian aware trace-weighted quantization
  of neural networks}.
\newblock \bibinfo{journal}{\emph{Advances in neural information processing
  systems}}  \bibinfo{volume}{33} (\bibinfo{year}{2020}),
  \bibinfo{pages}{18518--18529}.
\newblock


\bibitem[\protect\citeauthoryear{Dong, Yao, Gholami, Mahoney, and Keutzer}{Dong
  et~al\mbox{.}}{2019}]%
        {dong2019hawq}
\bibfield{author}{\bibinfo{person}{Zhen Dong}, \bibinfo{person}{Zhewei Yao},
  \bibinfo{person}{Amir Gholami}, \bibinfo{person}{Michael~W Mahoney}, {and}
  \bibinfo{person}{Kurt Keutzer}.} \bibinfo{year}{2019}\natexlab{}.
\newblock \showarticletitle{Hawq: Hessian aware quantization of neural networks
  with mixed-precision}. In \bibinfo{booktitle}{\emph{Proceedings of the
  IEEE/CVF International Conference on Computer Vision}}.
  \bibinfo{pages}{293--302}.
\newblock


\bibitem[\protect\citeauthoryear{Du, Fasthuber, Chen, Ienne, Li, Luo, Feng,
  Chen, and Temam}{Du et~al\mbox{.}}{2015}]%
        {du2015shidiannao}
\bibfield{author}{\bibinfo{person}{Zidong Du}, \bibinfo{person}{Robert
  Fasthuber}, \bibinfo{person}{Tianshi Chen}, \bibinfo{person}{Paolo Ienne},
  \bibinfo{person}{Ling Li}, \bibinfo{person}{Tao Luo},
  \bibinfo{person}{Xiaobing Feng}, \bibinfo{person}{Yunji Chen}, {and}
  \bibinfo{person}{Olivier Temam}.} \bibinfo{year}{2015}\natexlab{}.
\newblock \showarticletitle{ShiDianNao: Shifting vision processing closer to
  the sensor}. In \bibinfo{booktitle}{\emph{Proceedings of the 42nd Annual
  International Symposium on Computer Architecture}}. \bibinfo{pages}{92--104}.
\newblock


\bibitem[\protect\citeauthoryear{Gholami, Yao, Kim, Mahoney, and
  Keutzer}{Gholami et~al\mbox{.}}{2021}]%
        {gholami2020ai}
\bibfield{author}{\bibinfo{person}{Amir Gholami}, \bibinfo{person}{Zhewei Yao},
  \bibinfo{person}{Sehoon Kim}, \bibinfo{person}{Michael~W Mahoney}, {and}
  \bibinfo{person}{Kurt Keutzer}.} \bibinfo{year}{2021}\natexlab{}.
\newblock \showarticletitle{AI and Memory Wall}.
\newblock \bibinfo{journal}{\emph{RiseLab Medium Post}} (\bibinfo{year}{2021}).
\newblock


\bibitem[\protect\citeauthoryear{Gokhale, Jin, Dundar, Martini, and
  Culurciello}{Gokhale et~al\mbox{.}}{2014}]%
        {gokhale2014240}
\bibfield{author}{\bibinfo{person}{Vinayak Gokhale}, \bibinfo{person}{Jonghoon
  Jin}, \bibinfo{person}{Aysegul Dundar}, \bibinfo{person}{Berin Martini},
  {and} \bibinfo{person}{Eugenio Culurciello}.}
  \bibinfo{year}{2014}\natexlab{}.
\newblock \showarticletitle{A 240 g-ops/s mobile coprocessor for deep neural
  networks}. In \bibinfo{booktitle}{\emph{Proceedings of the IEEE conference on
  computer vision and pattern recognition workshops}}.
  \bibinfo{pages}{682--687}.
\newblock


\bibitem[\protect\citeauthoryear{Guan, Leng, Li, Chen, and Guo}{Guan
  et~al\mbox{.}}{2020}]%
        {guan2020far}
\bibfield{author}{\bibinfo{person}{Yue Guan}, \bibinfo{person}{Jingwen Leng},
  \bibinfo{person}{Chao Li}, \bibinfo{person}{Quan Chen}, {and}
  \bibinfo{person}{Minyi Guo}.} \bibinfo{year}{2020}\natexlab{}.
\newblock \showarticletitle{How Far Does BERT Look At: Distance-based
  Clustering and Analysis of BERT $'$ s Attention}.
\newblock \bibinfo{journal}{\emph{arXiv preprint arXiv:2011.00943}}
  (\bibinfo{year}{2020}).
\newblock


\bibitem[\protect\citeauthoryear{Guan, Li, Leng, Lin, and Guo}{Guan
  et~al\mbox{.}}{2022a}]%
        {guan2022transkimmer}
\bibfield{author}{\bibinfo{person}{Yue Guan}, \bibinfo{person}{Zhengyi Li},
  \bibinfo{person}{Jingwen Leng}, \bibinfo{person}{Zhouhan Lin}, {and}
  \bibinfo{person}{Minyi Guo}.} \bibinfo{year}{2022}\natexlab{a}.
\newblock \showarticletitle{Transkimmer: Transformer Learns to Layer-wise
  Skim}.
\newblock \bibinfo{journal}{\emph{arXiv preprint arXiv:2205.07324}}
  (\bibinfo{year}{2022}).
\newblock


\bibitem[\protect\citeauthoryear{Guan, Li, Lin, Zhu, Leng, and Guo}{Guan
  et~al\mbox{.}}{2022b}]%
        {guan2022block}
\bibfield{author}{\bibinfo{person}{Yue Guan}, \bibinfo{person}{Zhengyi Li},
  \bibinfo{person}{Zhouhan Lin}, \bibinfo{person}{Yuhao Zhu},
  \bibinfo{person}{Jingwen Leng}, {and} \bibinfo{person}{Minyi Guo}.}
  \bibinfo{year}{2022}\natexlab{b}.
\newblock \showarticletitle{Block-skim: Efficient question answering for
  transformer}. In \bibinfo{booktitle}{\emph{Proceedings of the AAAI Conference
  on Artificial Intelligence}}, Vol.~\bibinfo{volume}{36}.
  \bibinfo{pages}{10710--10719}.
\newblock


\bibitem[\protect\citeauthoryear{Guo, Hsueh, Leng, Qiu, Guan, Wang, Jia, Li,
  Guo, and Zhu}{Guo et~al\mbox{.}}{2020a}]%
        {guo2020accelerating}
\bibfield{author}{\bibinfo{person}{Cong Guo}, \bibinfo{person}{Bo~Yang Hsueh},
  \bibinfo{person}{Jingwen Leng}, \bibinfo{person}{Yuxian Qiu},
  \bibinfo{person}{Yue Guan}, \bibinfo{person}{Zehuan Wang},
  \bibinfo{person}{Xiaoying Jia}, \bibinfo{person}{Xipeng Li},
  \bibinfo{person}{Minyi Guo}, {and} \bibinfo{person}{Yuhao Zhu}.}
  \bibinfo{year}{2020}\natexlab{a}.
\newblock \showarticletitle{Accelerating sparse dnn models without
  hardware-support via tile-wise sparsity}. In \bibinfo{booktitle}{\emph{SC20:
  International Conference for High Performance Computing, Networking, Storage
  and Analysis}}. IEEE, \bibinfo{pages}{1--15}.
\newblock


\bibitem[\protect\citeauthoryear{Guo, Qiu, Leng, Gao, Zhang, Liu, Yang, Zhu,
  and Guo}{Guo et~al\mbox{.}}{2022a}]%
        {guo2022squant}
\bibfield{author}{\bibinfo{person}{Cong Guo}, \bibinfo{person}{Yuxian Qiu},
  \bibinfo{person}{Jingwen Leng}, \bibinfo{person}{Xiaotian Gao},
  \bibinfo{person}{Chen Zhang}, \bibinfo{person}{Yunxin Liu},
  \bibinfo{person}{Fan Yang}, \bibinfo{person}{Yuhao Zhu}, {and}
  \bibinfo{person}{Minyi Guo}.} \bibinfo{year}{2022}\natexlab{a}.
\newblock \showarticletitle{{SQ}uant: On-the-Fly Data-Free Quantization via
  Diagonal Hessian Approximation}. In \bibinfo{booktitle}{\emph{International
  Conference on Learning Representations}}.
\newblock
\urldef\tempurl%
\url{https://openreview.net/forum?id=JXhROKNZzOc}
\showURL{%
\tempurl}


\bibitem[\protect\citeauthoryear{Guo, Qiu, Leng, Zhang, Cao, Zhang, Liu, Yang,
  and Guo}{Guo et~al\mbox{.}}{2022b}]%
        {guo2022nesting}
\bibfield{author}{\bibinfo{person}{Cong Guo}, \bibinfo{person}{Yuxian Qiu},
  \bibinfo{person}{Jingwen Leng}, \bibinfo{person}{Chen Zhang},
  \bibinfo{person}{Ying Cao}, \bibinfo{person}{Quanlu Zhang},
  \bibinfo{person}{Yunxin Liu}, \bibinfo{person}{Fan Yang}, {and}
  \bibinfo{person}{Minyi Guo}.} \bibinfo{year}{2022}\natexlab{b}.
\newblock \showarticletitle{Nesting Forward Automatic Differentiation for
  Memory-Efficient Deep Neural Network Training}. In
  \bibinfo{booktitle}{\emph{2022 IEEE 40th International Conference on Computer
  Design (ICCD)}}. IEEE, \bibinfo{pages}{738--745}.
\newblock


\bibitem[\protect\citeauthoryear{Guo, Zhang, Leng, Liu, Yang, Liu, Guo, and
  Zhu}{Guo et~al\mbox{.}}{2022c}]%
        {guo2022ant}
\bibfield{author}{\bibinfo{person}{Cong Guo}, \bibinfo{person}{Chen Zhang},
  \bibinfo{person}{Jingwen Leng}, \bibinfo{person}{Zihan Liu},
  \bibinfo{person}{Fan Yang}, \bibinfo{person}{Yunxin Liu},
  \bibinfo{person}{Minyi Guo}, {and} \bibinfo{person}{Yuhao Zhu}.}
  \bibinfo{year}{2022}\natexlab{c}.
\newblock \showarticletitle{ANT: Exploiting Adaptive Numerical Data Type for
  Low-bit Deep Neural Network Quantization}. In \bibinfo{booktitle}{\emph{2022
  55th IEEE/ACM International Symposium on Microarchitecture (MICRO)}}. IEEE,
  \bibinfo{pages}{1414--1433}.
\newblock


\bibitem[\protect\citeauthoryear{Guo, Zhang, Leng, Liu, Yang, Liu, Guo, and
  Zhu}{Guo et~al\mbox{.}}{2022d}]%
        {guo2022antrepo}
\bibfield{author}{\bibinfo{person}{Cong Guo}, \bibinfo{person}{Chen Zhang},
  \bibinfo{person}{Jingwen Leng}, \bibinfo{person}{Zihan Liu},
  \bibinfo{person}{Fan Yang}, \bibinfo{person}{Yunxin Liu},
  \bibinfo{person}{Minyi Guo}, {and} \bibinfo{person}{Yuhao Zhu}.}
  \bibinfo{year}{2022}\natexlab{d}.
\newblock \bibinfo{title}{ANT github repository}.
\newblock
  \bibinfo{howpublished}{\url{https://github.com/clevercool/ANT_Micro22}}.
\newblock


\bibitem[\protect\citeauthoryear{Guo, Zhou, Leng, Zhu, Du, Chen, Li, Yao, and
  Guo}{Guo et~al\mbox{.}}{2020b}]%
        {guo2020balancing}
\bibfield{author}{\bibinfo{person}{Cong Guo}, \bibinfo{person}{Yangjie Zhou},
  \bibinfo{person}{Jingwen Leng}, \bibinfo{person}{Yuhao Zhu},
  \bibinfo{person}{Zidong Du}, \bibinfo{person}{Quan Chen},
  \bibinfo{person}{Chao Li}, \bibinfo{person}{Bin Yao}, {and}
  \bibinfo{person}{Minyi Guo}.} \bibinfo{year}{2020}\natexlab{b}.
\newblock \showarticletitle{{Balancing Efficiency and Flexibility for DNN
  Acceleration via Temporal GPU-Systolic Array Integration}}. In
  \bibinfo{booktitle}{\emph{2020 57th ACM/IEEE Design Automation Conference
  (DAC)}}. \bibinfo{pages}{1--6}.
\newblock


\bibitem[\protect\citeauthoryear{Gupta, Agrawal, Gopalakrishnan, and
  Narayanan}{Gupta et~al\mbox{.}}{2015}]%
        {gupta2015deep}
\bibfield{author}{\bibinfo{person}{Suyog Gupta}, \bibinfo{person}{Ankur
  Agrawal}, \bibinfo{person}{Kailash Gopalakrishnan}, {and}
  \bibinfo{person}{Pritish Narayanan}.} \bibinfo{year}{2015}\natexlab{}.
\newblock \showarticletitle{Deep learning with limited numerical precision}. In
  \bibinfo{booktitle}{\emph{International conference on machine learning}}.
  PMLR, \bibinfo{pages}{1737--1746}.
\newblock


\bibitem[\protect\citeauthoryear{Han, Mao, and Dally}{Han
  et~al\mbox{.}}{2015}]%
        {han2015deep}
\bibfield{author}{\bibinfo{person}{Song Han}, \bibinfo{person}{Huizi Mao},
  {and} \bibinfo{person}{William~J Dally}.} \bibinfo{year}{2015}\natexlab{}.
\newblock \showarticletitle{Deep compression: Compressing deep neural networks
  with pruning, trained quantization and huffman coding}.
\newblock \bibinfo{journal}{\emph{arXiv preprint arXiv:1510.00149}}
  (\bibinfo{year}{2015}).
\newblock


\bibitem[\protect\citeauthoryear{He, Zhang, Ren, and Sun}{He
  et~al\mbox{.}}{2016}]%
        {he2016deep}
\bibfield{author}{\bibinfo{person}{Kaiming He}, \bibinfo{person}{Xiangyu
  Zhang}, \bibinfo{person}{Shaoqing Ren}, {and} \bibinfo{person}{Jian Sun}.}
  \bibinfo{year}{2016}\natexlab{}.
\newblock \showarticletitle{Deep residual learning for image recognition}. In
  \bibinfo{booktitle}{\emph{Proceedings of the IEEE conference on computer
  vision and pattern recognition}}. \bibinfo{pages}{770--778}.
\newblock


\bibitem[\protect\citeauthoryear{Jacob, Kligys, Chen, Zhu, Tang, Howard, Adam,
  and Kalenichenko}{Jacob et~al\mbox{.}}{2018}]%
        {jacob2018quantization}
\bibfield{author}{\bibinfo{person}{Benoit Jacob}, \bibinfo{person}{Skirmantas
  Kligys}, \bibinfo{person}{Bo Chen}, \bibinfo{person}{Menglong Zhu},
  \bibinfo{person}{Matthew Tang}, \bibinfo{person}{Andrew Howard},
  \bibinfo{person}{Hartwig Adam}, {and} \bibinfo{person}{Dmitry Kalenichenko}.}
  \bibinfo{year}{2018}\natexlab{}.
\newblock \showarticletitle{Quantization and training of neural networks for
  efficient integer-arithmetic-only inference}. In
  \bibinfo{booktitle}{\emph{Proceedings of the IEEE conference on computer
  vision and pattern recognition}}. \bibinfo{pages}{2704--2713}.
\newblock


\bibitem[\protect\citeauthoryear{Jain, Venkataramani, Srinivasan, Choi,
  Gopalakrishnan, and Chang}{Jain et~al\mbox{.}}{2019}]%
        {jain2019biscaled}
\bibfield{author}{\bibinfo{person}{Shubham Jain}, \bibinfo{person}{Swagath
  Venkataramani}, \bibinfo{person}{Vijayalakshmi Srinivasan},
  \bibinfo{person}{Jungwook Choi}, \bibinfo{person}{Kailash Gopalakrishnan},
  {and} \bibinfo{person}{Leland Chang}.} \bibinfo{year}{2019}\natexlab{}.
\newblock \showarticletitle{BiScaled-DNN: Quantizing long-tailed datastructures
  with two scale factors for deep neural networks}. In
  \bibinfo{booktitle}{\emph{2019 56th ACM/IEEE Design Automation Conference
  (DAC)}}. IEEE, \bibinfo{pages}{1--6}.
\newblock


\bibitem[\protect\citeauthoryear{Jia, Padon, Thomas, Warszawski, Zaharia, and
  Aiken}{Jia et~al\mbox{.}}{2019}]%
        {TASO}
\bibfield{author}{\bibinfo{person}{Zhihao Jia}, \bibinfo{person}{Oded Padon},
  \bibinfo{person}{James~J. Thomas}, \bibinfo{person}{Todd Warszawski},
  \bibinfo{person}{Matei Zaharia}, {and} \bibinfo{person}{Alex Aiken}.}
  \bibinfo{year}{2019}\natexlab{}.
\newblock \showarticletitle{{TASO:} optimizing deep learning computation with
  automatic generation of graph substitutions}. In
  \bibinfo{booktitle}{\emph{Proceedings of the 27th {ACM} Symposium on
  Operating Systems Principles ({SOSP})}}. \bibinfo{publisher}{{ACM}},
  \bibinfo{pages}{47--62}.
\newblock
\urldef\tempurl%
\url{https://doi.org/10.1145/3341301.3359630}
\showDOI{\tempurl}


\bibitem[\protect\citeauthoryear{Jouppi, Hyun~Yoon, Ashcraft, Gottscho, Jablin,
  Kurian, Laudon, Li, Ma, Ma, Norrie, Patil, Prasad, Young, Zhou, and
  Patterson}{Jouppi et~al\mbox{.}}{2021}]%
        {tpuv4isca}
\bibfield{author}{\bibinfo{person}{Norman~P. Jouppi}, \bibinfo{person}{Doe
  Hyun~Yoon}, \bibinfo{person}{Matthew Ashcraft}, \bibinfo{person}{Mark
  Gottscho}, \bibinfo{person}{Thomas~B. Jablin}, \bibinfo{person}{George
  Kurian}, \bibinfo{person}{James Laudon}, \bibinfo{person}{Sheng Li},
  \bibinfo{person}{Peter Ma}, \bibinfo{person}{Xiaoyu Ma},
  \bibinfo{person}{Thomas Norrie}, \bibinfo{person}{Nishant Patil},
  \bibinfo{person}{Sushma Prasad}, \bibinfo{person}{Cliff Young},
  \bibinfo{person}{Zongwei Zhou}, {and} \bibinfo{person}{David Patterson}.}
  \bibinfo{year}{2021}\natexlab{}.
\newblock \showarticletitle{Ten Lessons From Three Generations Shaped Google's
  TPUv4i: Industrial Product}. In \bibinfo{booktitle}{\emph{2021 ACM/IEEE 48th
  Annual International Symposium on Computer Architecture (ISCA)}}.
\newblock


\bibitem[\protect\citeauthoryear{Jouppi, Young, Patil, Patterson, Agrawal,
  Bajwa, Bates, Bhatia, Boden, Borchers, et~al\mbox{.}}{Jouppi
  et~al\mbox{.}}{2017}]%
        {jouppi2017datacenter}
\bibfield{author}{\bibinfo{person}{Norman~P Jouppi}, \bibinfo{person}{Cliff
  Young}, \bibinfo{person}{Nishant Patil}, \bibinfo{person}{David Patterson},
  \bibinfo{person}{Gaurav Agrawal}, \bibinfo{person}{Raminder Bajwa},
  \bibinfo{person}{Sarah Bates}, \bibinfo{person}{Suresh Bhatia},
  \bibinfo{person}{Nan Boden}, \bibinfo{person}{Al Borchers}, {et~al\mbox{.}}}
  \bibinfo{year}{2017}\natexlab{}.
\newblock \showarticletitle{In-datacenter performance analysis of a tensor
  processing unit}. In \bibinfo{booktitle}{\emph{Proceedings of the 44th annual
  international symposium on computer architecture}}. \bibinfo{pages}{1--12}.
\newblock


\bibitem[\protect\citeauthoryear{Jung, Son, Lee, Son, Han, Kwak, Hwang, and
  Choi}{Jung et~al\mbox{.}}{2019}]%
        {jung2019learning}
\bibfield{author}{\bibinfo{person}{Sangil Jung}, \bibinfo{person}{Changyong
  Son}, \bibinfo{person}{Seohyung Lee}, \bibinfo{person}{Jinwoo Son},
  \bibinfo{person}{Jae-Joon Han}, \bibinfo{person}{Youngjun Kwak},
  \bibinfo{person}{Sung~Ju Hwang}, {and} \bibinfo{person}{Changkyu Choi}.}
  \bibinfo{year}{2019}\natexlab{}.
\newblock \showarticletitle{Learning to quantize deep networks by optimizing
  quantization intervals with task loss}. In
  \bibinfo{booktitle}{\emph{Proceedings of the IEEE/CVF Conference on Computer
  Vision and Pattern Recognition}}. \bibinfo{pages}{4350--4359}.
\newblock


\bibitem[\protect\citeauthoryear{Kerr, Wu, Gupta, Blasig, Ramini, Merrill,
  Shivam, Majcher, Springer, Hohnerbach, Wang, and Nicely}{Kerr
  et~al\mbox{.}}{2022}]%
        {Kerr_CUTLASS_2022}
\bibfield{author}{\bibinfo{person}{Andrew Kerr}, \bibinfo{person}{Haicheng Wu},
  \bibinfo{person}{Manish Gupta}, \bibinfo{person}{Dustyn Blasig},
  \bibinfo{person}{Pradeep Ramini}, \bibinfo{person}{Duane Merrill},
  \bibinfo{person}{Aniket Shivam}, \bibinfo{person}{Piotr Majcher},
  \bibinfo{person}{Paul Springer}, \bibinfo{person}{Markus Hohnerbach},
  \bibinfo{person}{Jin Wang}, {and} \bibinfo{person}{Matt Nicely}.}
  \bibinfo{year}{2022}\natexlab{}.
\newblock \bibinfo{booktitle}{\emph{{CUTLASS}}}.
\newblock
\urldef\tempurl%
\url{https://github.com/NVIDIA/cutlass}
\showURL{%
\tempurl}


\bibitem[\protect\citeauthoryear{Khairy, Shen, Aamodt, and Rogers}{Khairy
  et~al\mbox{.}}{2020a}]%
        {gpgpu-sim4.0}
\bibfield{author}{\bibinfo{person}{Mahmoud Khairy}, \bibinfo{person}{Zhesheng
  Shen}, \bibinfo{person}{Tor Aamodt}, {and} \bibinfo{person}{Timothy Rogers}.}
  \bibinfo{year}{2020}\natexlab{a}.
\newblock \showarticletitle{Accel-Sim: An Extensible Simulation Framework for
  Validated GPU Modeling}. \bibinfo{pages}{473--486}.
\newblock
\urldef\tempurl%
\url{https://doi.org/10.1109/ISCA45697.2020.00047}
\showDOI{\tempurl}


\bibitem[\protect\citeauthoryear{Khairy, Shen, Aamodt, and Rogers}{Khairy
  et~al\mbox{.}}{2020b}]%
        {khairy2020accel}
\bibfield{author}{\bibinfo{person}{Mahmoud Khairy}, \bibinfo{person}{Zhesheng
  Shen}, \bibinfo{person}{Tor~M Aamodt}, {and} \bibinfo{person}{Timothy~G
  Rogers}.} \bibinfo{year}{2020}\natexlab{b}.
\newblock \showarticletitle{Accel-Sim: An extensible simulation framework for
  validated GPU modeling}. In \bibinfo{booktitle}{\emph{2020 ACM/IEEE 47th
  Annual International Symposium on Computer Architecture (ISCA)}}. IEEE,
  \bibinfo{pages}{473--486}.
\newblock


\bibitem[\protect\citeauthoryear{Kurup and Abbasi}{Kurup and Abbasi}{2012}]%
        {kurup2012logic}
\bibfield{author}{\bibinfo{person}{Pran Kurup} {and} \bibinfo{person}{Taher
  Abbasi}.} \bibinfo{year}{2012}\natexlab{}.
\newblock \bibinfo{booktitle}{\emph{Logic synthesis using
  Synopsys{\textregistered}}}.
\newblock \bibinfo{publisher}{Springer Science \& Business Media}.
\newblock


\bibitem[\protect\citeauthoryear{Leng, Hetherington, ElTantawy, Gilani, Kim,
  Aamodt, and Janapa~Reddi}{Leng et~al\mbox{.}}{2013}]%
        {leng2013gpuwattch}
\bibfield{author}{\bibinfo{person}{Jingwen Leng}, \bibinfo{person}{Tayler
  Hetherington}, \bibinfo{person}{Ahmed ElTantawy}, \bibinfo{person}{Syed
  Gilani}, \bibinfo{person}{Nam Kim}, \bibinfo{person}{Tor Aamodt}, {and}
  \bibinfo{person}{Vijay Janapa~Reddi}.} \bibinfo{year}{2013}\natexlab{}.
\newblock \showarticletitle{GPUWattch: enabling energy optimizations in
  GPGPUs}.
\newblock \bibinfo{journal}{\emph{ACM SIGARCH Computer Architecture News}}
  \bibinfo{volume}{41} (\bibinfo{date}{07} \bibinfo{year}{2013}).
\newblock
\showISBNx{978-1-4503-2079-5}
\urldef\tempurl%
\url{https://doi.org/10.1145/2508148.2485964}
\showDOI{\tempurl}


\bibitem[\protect\citeauthoryear{Lewis, Liu, Goyal, Ghazvininejad, Mohamed,
  Levy, Stoyanov, and Zettlemoyer}{Lewis et~al\mbox{.}}{2019}]%
        {lewis2019bart}
\bibfield{author}{\bibinfo{person}{Mike Lewis}, \bibinfo{person}{Yinhan Liu},
  \bibinfo{person}{Naman Goyal}, \bibinfo{person}{Marjan Ghazvininejad},
  \bibinfo{person}{Abdelrahman Mohamed}, \bibinfo{person}{Omer Levy},
  \bibinfo{person}{Ves Stoyanov}, {and} \bibinfo{person}{Luke Zettlemoyer}.}
  \bibinfo{year}{2019}\natexlab{}.
\newblock \showarticletitle{Bart: Denoising sequence-to-sequence pre-training
  for natural language generation, translation, and comprehension}.
\newblock \bibinfo{journal}{\emph{arXiv preprint arXiv:1910.13461}}
  (\bibinfo{year}{2019}).
\newblock


\bibitem[\protect\citeauthoryear{Li, Guo, Zhu, Zhou, Qiu, Gao, Leng, and
  Guo}{Li et~al\mbox{.}}{2022}]%
        {li2022efficient}
\bibfield{author}{\bibinfo{person}{Zhengyi Li}, \bibinfo{person}{Cong Guo},
  \bibinfo{person}{Zhanda Zhu}, \bibinfo{person}{Yangjie Zhou},
  \bibinfo{person}{Yuxian Qiu}, \bibinfo{person}{Xiaotian Gao},
  \bibinfo{person}{Jingwen Leng}, {and} \bibinfo{person}{Minyi Guo}.}
  \bibinfo{year}{2022}\natexlab{}.
\newblock \showarticletitle{Efficient Activation Quantization via Adaptive
  Rounding Border for Post-Training Quantization}.
\newblock \bibinfo{journal}{\emph{arXiv preprint arXiv:2208.11945}}
  (\bibinfo{year}{2022}).
\newblock


\bibitem[\protect\citeauthoryear{Liu, Chen, Liu, Zhou, Zhou, Teman, Feng, Zhou,
  and Chen}{Liu et~al\mbox{.}}{2015}]%
        {liu2015pudiannao}
\bibfield{author}{\bibinfo{person}{Daofu Liu}, \bibinfo{person}{Tianshi Chen},
  \bibinfo{person}{Shaoli Liu}, \bibinfo{person}{Jinhong Zhou},
  \bibinfo{person}{Shengyuan Zhou}, \bibinfo{person}{Olivier Teman},
  \bibinfo{person}{Xiaobing Feng}, \bibinfo{person}{Xuehai Zhou}, {and}
  \bibinfo{person}{Yunji Chen}.} \bibinfo{year}{2015}\natexlab{}.
\newblock \showarticletitle{PuDianNao: A Polyvalent Machine Learning
  Accelerator}. In \bibinfo{booktitle}{\emph{Proceedings of the Twentieth
  International Conference on Architectural Support for Programming Languages
  and Operating Systems}}. \bibinfo{pages}{369–381}.
\newblock


\bibitem[\protect\citeauthoryear{Liu, Leng, Zhang, Chen, Li, and Guo}{Liu
  et~al\mbox{.}}{2022}]%
        {VELTAIR}
\bibfield{author}{\bibinfo{person}{Zihan Liu}, \bibinfo{person}{Jingwen Leng},
  \bibinfo{person}{Zhihui Zhang}, \bibinfo{person}{Quan Chen},
  \bibinfo{person}{Chao Li}, {and} \bibinfo{person}{Minyi Guo}.}
  \bibinfo{year}{2022}\natexlab{}.
\newblock \showarticletitle{{VELTAIR:} towards high-performance multi-tenant
  deep learning services via adaptive compilation and scheduling}. In
  \bibinfo{booktitle}{\emph{{ASPLOS} '22: 27th {ACM} International Conference
  on Architectural Support for Programming Languages and Operating Systems,
  Lausanne, Switzerland, 28 February 2022 - 4 March 2022}},
  \bibfield{editor}{\bibinfo{person}{Babak Falsafi}, \bibinfo{person}{Michael
  Ferdman}, \bibinfo{person}{Shan Lu}, {and} \bibinfo{person}{Thomas~F.
  Wenisch}} (Eds.). \bibinfo{publisher}{{ACM}}, \bibinfo{pages}{388--401}.
\newblock
\urldef\tempurl%
\url{https://doi.org/10.1145/3503222.3507752}
\showDOI{\tempurl}


\bibitem[\protect\citeauthoryear{Lo, Cheng, Govindaraju, Ranganathan, and
  Kozyrakis}{Lo et~al\mbox{.}}{2015}]%
        {Heracles}
\bibfield{author}{\bibinfo{person}{David Lo}, \bibinfo{person}{Liqun Cheng},
  \bibinfo{person}{Rama Govindaraju}, \bibinfo{person}{Parthasarathy
  Ranganathan}, {and} \bibinfo{person}{Christos Kozyrakis}.}
  \bibinfo{year}{2015}\natexlab{}.
\newblock \showarticletitle{Heracles: improving resource efficiency at scale}.
  In \bibinfo{booktitle}{\emph{Proceedings of the 42nd Annual International
  Symposium on Computer Architecture ({ISCA})}}.
\newblock
\urldef\tempurl%
\url{https://doi.org/10.1145/2749469.2749475}
\showDOI{\tempurl}


\bibitem[\protect\citeauthoryear{Mars, Tang, Hundt, Skadron, and Soffa}{Mars
  et~al\mbox{.}}{2011}]%
        {BubbleUp}
\bibfield{author}{\bibinfo{person}{Jason Mars}, \bibinfo{person}{Lingjia Tang},
  \bibinfo{person}{Robert Hundt}, \bibinfo{person}{Kevin Skadron}, {and}
  \bibinfo{person}{Mary~Lou Soffa}.} \bibinfo{year}{2011}\natexlab{}.
\newblock \showarticletitle{Bubble-Up: increasing utilization in modern
  warehouse scale computers via sensible co-locations}. In
  \bibinfo{booktitle}{\emph{{IEEE/ACM} International Symposium on
  Microarchitecture ({MICRO})}}.
\newblock
\urldef\tempurl%
\url{https://doi.org/10.1145/2155620.2155650}
\showDOI{\tempurl}


\bibitem[\protect\citeauthoryear{ModelTC}{ModelTC}{2022}]%
        {huggingfacerepo}
\bibfield{author}{\bibinfo{person}{ModelTC}.} \bibinfo{year}{2022}\natexlab{}.
\newblock \bibinfo{title}{repositories}.
\newblock \bibinfo{howpublished}{\url{https://huggingface.co/ModelTC}}.
\newblock


\bibitem[\protect\citeauthoryear{Muralimanohar, Balasubramonian, and
  Jouppi}{Muralimanohar et~al\mbox{.}}{2009}]%
        {muralimanohar2009cacti}
\bibfield{author}{\bibinfo{person}{Naveen Muralimanohar},
  \bibinfo{person}{Rajeev Balasubramonian}, {and} \bibinfo{person}{Norman~P
  Jouppi}.} \bibinfo{year}{2009}\natexlab{}.
\newblock \showarticletitle{CACTI 6.0: A tool to model large caches}.
\newblock \bibinfo{journal}{\emph{HP laboratories}}  \bibinfo{volume}{27}
  (\bibinfo{year}{2009}), \bibinfo{pages}{28}.
\newblock


\bibitem[\protect\citeauthoryear{Nagel, Fournarakis, Amjad, Bondarenko, van
  Baalen, and Blankevoort}{Nagel et~al\mbox{.}}{2021}]%
        {nagel2021white}
\bibfield{author}{\bibinfo{person}{Markus Nagel}, \bibinfo{person}{Marios
  Fournarakis}, \bibinfo{person}{Rana~Ali Amjad}, \bibinfo{person}{Yelysei
  Bondarenko}, \bibinfo{person}{Mart van Baalen}, {and} \bibinfo{person}{Tijmen
  Blankevoort}.} \bibinfo{year}{2021}\natexlab{}.
\newblock \showarticletitle{A white paper on neural network quantization}.
\newblock \bibinfo{journal}{\emph{arXiv preprint arXiv:2106.08295}}
  (\bibinfo{year}{2021}).
\newblock


\bibitem[\protect\citeauthoryear{Nvidia}{Nvidia}{2017}]%
        {v100}
\bibfield{author}{\bibinfo{person}{Nvidia}.} \bibinfo{year}{2017}\natexlab{}.
\newblock \showarticletitle{NVIDIA Tesla V100 GPU Architecture}. In
  \bibinfo{booktitle}{\emph{Technical report}}. NVIDIA.
\newblock


\bibitem[\protect\citeauthoryear{Nvidia}{Nvidia}{2018}]%
        {turing}
\bibfield{author}{\bibinfo{person}{Nvidia}.} \bibinfo{year}{2018}\natexlab{}.
\newblock \showarticletitle{NVIDIA Turing GPU Architecture}. In
  \bibinfo{booktitle}{\emph{Technical report}}. NVIDIA.
\newblock


\bibitem[\protect\citeauthoryear{Nvidia}{Nvidia}{2020}]%
        {a100}
\bibfield{author}{\bibinfo{person}{Nvidia}.} \bibinfo{year}{2020}\natexlab{}.
\newblock \showarticletitle{NVIDIA A100 tensor core architecture}. In
  \bibinfo{booktitle}{\emph{Technical report}}. NVIDIA.
\newblock


\bibitem[\protect\citeauthoryear{Park, Kim, and Yoo}{Park
  et~al\mbox{.}}{2018}]%
        {park2018energy}
\bibfield{author}{\bibinfo{person}{Eunhyeok Park}, \bibinfo{person}{Dongyoung
  Kim}, {and} \bibinfo{person}{Sungjoo Yoo}.} \bibinfo{year}{2018}\natexlab{}.
\newblock \showarticletitle{Energy-efficient neural network `accel'erator based
  on outlier-aware low-precision computation}. In
  \bibinfo{booktitle}{\emph{2018 ACM/IEEE 45th Annual International Symposium
  on Computer Architecture (ISCA)}}. IEEE, \bibinfo{pages}{688--698}.
\newblock


\bibitem[\protect\citeauthoryear{Paszke, Gross, Massa, Lerer, Bradbury, Chanan,
  Killeen, Lin, Gimelshein, Antiga, et~al\mbox{.}}{Paszke
  et~al\mbox{.}}{2019}]%
        {paszke2019pytorch}
\bibfield{author}{\bibinfo{person}{Adam Paszke}, \bibinfo{person}{Sam Gross},
  \bibinfo{person}{Francisco Massa}, \bibinfo{person}{Adam Lerer},
  \bibinfo{person}{James Bradbury}, \bibinfo{person}{Gregory Chanan},
  \bibinfo{person}{Trevor Killeen}, \bibinfo{person}{Zeming Lin},
  \bibinfo{person}{Natalia Gimelshein}, \bibinfo{person}{Luca Antiga},
  {et~al\mbox{.}}} \bibinfo{year}{2019}\natexlab{}.
\newblock \showarticletitle{Pytorch: An imperative style, high-performance deep
  learning library}.
\newblock \bibinfo{journal}{\emph{Advances in neural information processing
  systems}}  \bibinfo{volume}{32} (\bibinfo{year}{2019}),
  \bibinfo{pages}{8026--8037}.
\newblock


\bibitem[\protect\citeauthoryear{Peemen, Setio, Mesman, and Corporaal}{Peemen
  et~al\mbox{.}}{2013}]%
        {peemen2013memory}
\bibfield{author}{\bibinfo{person}{Maurice Peemen}, \bibinfo{person}{Arnaud~AA
  Setio}, \bibinfo{person}{Bart Mesman}, {and} \bibinfo{person}{Henk
  Corporaal}.} \bibinfo{year}{2013}\natexlab{}.
\newblock \showarticletitle{Memory-centric accelerator design for convolutional
  neural networks}. In \bibinfo{booktitle}{\emph{2013 IEEE 31st International
  Conference on Computer Design (ICCD)}}. IEEE, \bibinfo{pages}{13--19}.
\newblock


\bibitem[\protect\citeauthoryear{Qin, Samajdar, Kwon, Nadella, Srinivasan, Das,
  Kaul, and Krishna}{Qin et~al\mbox{.}}{2020}]%
        {qin2020sigma}
\bibfield{author}{\bibinfo{person}{Eric Qin}, \bibinfo{person}{Ananda
  Samajdar}, \bibinfo{person}{Hyoukjun Kwon}, \bibinfo{person}{Vineet Nadella},
  \bibinfo{person}{Sudarshan Srinivasan}, \bibinfo{person}{Dipankar Das},
  \bibinfo{person}{Bharat Kaul}, {and} \bibinfo{person}{Tushar Krishna}.}
  \bibinfo{year}{2020}\natexlab{}.
\newblock \showarticletitle{Sigma: A sparse and irregular gemm accelerator with
  flexible interconnects for dnn training}. In \bibinfo{booktitle}{\emph{2020
  IEEE International Symposium on High Performance Computer Architecture
  (HPCA)}}. IEEE, \bibinfo{pages}{58--70}.
\newblock


\bibitem[\protect\citeauthoryear{Qiu, Leng, Guo, Chen, Li, Guo, and Zhu}{Qiu
  et~al\mbox{.}}{2019}]%
        {Qiu_2019_CVPR}
\bibfield{author}{\bibinfo{person}{Yuxian Qiu}, \bibinfo{person}{Jingwen Leng},
  \bibinfo{person}{Cong Guo}, \bibinfo{person}{Quan Chen},
  \bibinfo{person}{Chao Li}, \bibinfo{person}{Minyi Guo}, {and}
  \bibinfo{person}{Yuhao Zhu}.} \bibinfo{year}{2019}\natexlab{}.
\newblock \showarticletitle{Adversarial Defense Through Network Profiling Based
  Path Extraction}. In \bibinfo{booktitle}{\emph{Proceedings of the IEEE/CVF
  Conference on Computer Vision and Pattern Recognition (CVPR)}}.
\newblock


\bibitem[\protect\citeauthoryear{Radford, Wu, Child, Luan, Amodei, and
  Sutskever}{Radford et~al\mbox{.}}{2019}]%
        {radford2019language}
\bibfield{author}{\bibinfo{person}{Alec Radford}, \bibinfo{person}{Jeff Wu},
  \bibinfo{person}{Rewon Child}, \bibinfo{person}{David Luan},
  \bibinfo{person}{Dario Amodei}, {and} \bibinfo{person}{Ilya Sutskever}.}
  \bibinfo{year}{2019}\natexlab{}.
\newblock \showarticletitle{Language Models are Unsupervised Multitask
  Learners}.
\newblock  (\bibinfo{year}{2019}).
\newblock


\bibitem[\protect\citeauthoryear{Raihan, Goli, and Aamodt}{Raihan
  et~al\mbox{.}}{2019}]%
        {raihan2019modeling}
\bibfield{author}{\bibinfo{person}{Md~Aamir Raihan}, \bibinfo{person}{Negar
  Goli}, {and} \bibinfo{person}{Tor~M Aamodt}.}
  \bibinfo{year}{2019}\natexlab{}.
\newblock \showarticletitle{Modeling deep learning accelerator enabled gpus}.
  In \bibinfo{booktitle}{\emph{2019 IEEE International Symposium on Performance
  Analysis of Systems and Software (ISPASS)}}. IEEE, \bibinfo{pages}{79--92}.
\newblock


\bibitem[\protect\citeauthoryear{Rajpurkar, Zhang, Lopyrev, and
  Liang}{Rajpurkar et~al\mbox{.}}{2016}]%
        {rajpurkar-etal-2016-squad}
\bibfield{author}{\bibinfo{person}{Pranav Rajpurkar}, \bibinfo{person}{Jian
  Zhang}, \bibinfo{person}{Konstantin Lopyrev}, {and} \bibinfo{person}{Percy
  Liang}.} \bibinfo{year}{2016}\natexlab{}.
\newblock \showarticletitle{{SQ}u{AD}: 100,000+ Questions for Machine
  Comprehension of Text}. In \bibinfo{booktitle}{\emph{Proceedings of the 2016
  Conference on Empirical Methods in Natural Language Processing}}.
  \bibinfo{publisher}{Association for Computational Linguistics}.
\newblock


\bibitem[\protect\citeauthoryear{Sarangi and Baas}{Sarangi and Baas}{2021}]%
        {sarangi2021deepscaletool}
\bibfield{author}{\bibinfo{person}{Satyabrata Sarangi} {and}
  \bibinfo{person}{Bevan Baas}.} \bibinfo{year}{2021}\natexlab{}.
\newblock \showarticletitle{DeepScaleTool: A Tool for the Accurate Estimation
  of Technology Scaling in the Deep-Submicron Era}. In
  \bibinfo{booktitle}{\emph{2021 IEEE International Symposium on Circuits and
  Systems (ISCAS)}}. IEEE, \bibinfo{pages}{1--5}.
\newblock


\bibitem[\protect\citeauthoryear{Scao, Fan, Akiki, Pavlick, Ili{\'c}, Hesslow,
  Castagn{\'e}, Luccioni, Yvon, Gall{\'e}, et~al\mbox{.}}{Scao
  et~al\mbox{.}}{2022}]%
        {scao2022bloom}
\bibfield{author}{\bibinfo{person}{Teven~Le Scao}, \bibinfo{person}{Angela
  Fan}, \bibinfo{person}{Christopher Akiki}, \bibinfo{person}{Ellie Pavlick},
  \bibinfo{person}{Suzana Ili{\'c}}, \bibinfo{person}{Daniel Hesslow},
  \bibinfo{person}{Roman Castagn{\'e}}, \bibinfo{person}{Alexandra~Sasha
  Luccioni}, \bibinfo{person}{Fran{\c{c}}ois Yvon}, \bibinfo{person}{Matthias
  Gall{\'e}}, {et~al\mbox{.}}} \bibinfo{year}{2022}\natexlab{}.
\newblock \showarticletitle{BLOOM: A 176B-Parameter Open-Access Multilingual
  Language Model}.
\newblock \bibinfo{journal}{\emph{arXiv preprint arXiv:2211.05100}}
  (\bibinfo{year}{2022}).
\newblock


\bibitem[\protect\citeauthoryear{Sharma, Park, Mahajan, Amaro, Kim, Shao,
  Mishra, and Esmaeilzadeh}{Sharma et~al\mbox{.}}{2016}]%
        {sharma2016high}
\bibfield{author}{\bibinfo{person}{Hardik Sharma}, \bibinfo{person}{Jongse
  Park}, \bibinfo{person}{Divya Mahajan}, \bibinfo{person}{Emmanuel Amaro},
  \bibinfo{person}{Joon~Kyung Kim}, \bibinfo{person}{Chenkai Shao},
  \bibinfo{person}{Asit Mishra}, {and} \bibinfo{person}{Hadi Esmaeilzadeh}.}
  \bibinfo{year}{2016}\natexlab{}.
\newblock \showarticletitle{From high-level deep neural models to FPGAs}. In
  \bibinfo{booktitle}{\emph{2016 49th Annual IEEE/ACM International Symposium
  on Microarchitecture (MICRO)}}. IEEE, \bibinfo{pages}{1--12}.
\newblock


\bibitem[\protect\citeauthoryear{Sharma, Park, Suda, Lai, Chau, Chandra, and
  Esmaeilzadeh}{Sharma et~al\mbox{.}}{2018a}]%
        {sharma2018bit}
\bibfield{author}{\bibinfo{person}{Hardik Sharma}, \bibinfo{person}{Jongse
  Park}, \bibinfo{person}{Naveen Suda}, \bibinfo{person}{Liangzhen Lai},
  \bibinfo{person}{Benson Chau}, \bibinfo{person}{Vikas Chandra}, {and}
  \bibinfo{person}{Hadi Esmaeilzadeh}.} \bibinfo{year}{2018}\natexlab{a}.
\newblock \showarticletitle{Bit fusion: Bit-level dynamically composable
  architecture for accelerating deep neural network}. In
  \bibinfo{booktitle}{\emph{2018 ACM/IEEE 45th Annual International Symposium
  on Computer Architecture (ISCA)}}. IEEE, \bibinfo{pages}{764--775}.
\newblock


\bibitem[\protect\citeauthoryear{Sharma, Park, Suda, Lai, Chau, Chandra, and
  Esmaeilzadeh}{Sharma et~al\mbox{.}}{2018b}]%
        {sharma2018bitrepo}
\bibfield{author}{\bibinfo{person}{Hardik Sharma}, \bibinfo{person}{Jongse
  Park}, \bibinfo{person}{Naveen Suda}, \bibinfo{person}{Liangzhen Lai},
  \bibinfo{person}{Benson Chau}, \bibinfo{person}{Vikas Chandra}, {and}
  \bibinfo{person}{Hadi Esmaeilzadeh}.} \bibinfo{year}{2018}\natexlab{b}.
\newblock \bibinfo{title}{Bitfusion github repository}.
\newblock \bibinfo{howpublished}{\url{https://github.com/hsharma35/bitfusion}}.
\newblock


\bibitem[\protect\citeauthoryear{Shen, Dong, Ye, Ma, Yao, Gholami, Mahoney, and
  Keutzer}{Shen et~al\mbox{.}}{2020}]%
        {shen2020q}
\bibfield{author}{\bibinfo{person}{Sheng Shen}, \bibinfo{person}{Zhen Dong},
  \bibinfo{person}{Jiayu Ye}, \bibinfo{person}{Linjian Ma},
  \bibinfo{person}{Zhewei Yao}, \bibinfo{person}{Amir Gholami},
  \bibinfo{person}{Michael~W Mahoney}, {and} \bibinfo{person}{Kurt Keutzer}.}
  \bibinfo{year}{2020}\natexlab{}.
\newblock \showarticletitle{Q-bert: Hessian based ultra low precision
  quantization of bert}. In \bibinfo{booktitle}{\emph{Proceedings of the AAAI
  Conference on Artificial Intelligence}}, Vol.~\bibinfo{volume}{34}.
  \bibinfo{pages}{8815--8821}.
\newblock


\bibitem[\protect\citeauthoryear{Song, Fu, Wu, Jiang, Jiang, Jing, and
  Liang}{Song et~al\mbox{.}}{2020}]%
        {song2020drq}
\bibfield{author}{\bibinfo{person}{Zhuoran Song}, \bibinfo{person}{Bangqi Fu},
  \bibinfo{person}{Feiyang Wu}, \bibinfo{person}{Zhaoming Jiang},
  \bibinfo{person}{Li Jiang}, \bibinfo{person}{Naifeng Jing}, {and}
  \bibinfo{person}{Xiaoyao Liang}.} \bibinfo{year}{2020}\natexlab{}.
\newblock \showarticletitle{Drq: dynamic region-based quantization for deep
  neural network acceleration}. In \bibinfo{booktitle}{\emph{2020 ACM/IEEE 47th
  Annual International Symposium on Computer Architecture (ISCA)}}. IEEE,
  \bibinfo{pages}{1010--1021}.
\newblock


\bibitem[\protect\citeauthoryear{Tambe, Yang, Wan, Deng, Reddi, Rush, Brooks,
  and Wei}{Tambe et~al\mbox{.}}{2020}]%
        {tambe2020algorithm}
\bibfield{author}{\bibinfo{person}{Thierry Tambe}, \bibinfo{person}{En-Yu
  Yang}, \bibinfo{person}{Zishen Wan}, \bibinfo{person}{Yuntian Deng},
  \bibinfo{person}{Vijay~Janapa Reddi}, \bibinfo{person}{Alexander Rush},
  \bibinfo{person}{David Brooks}, {and} \bibinfo{person}{Gu-Yeon Wei}.}
  \bibinfo{year}{2020}\natexlab{}.
\newblock \showarticletitle{Algorithm-hardware co-design of adaptive
  floating-point encodings for resilient deep learning inference}. In
  \bibinfo{booktitle}{\emph{2020 57th ACM/IEEE Design Automation Conference
  (DAC)}}. IEEE, \bibinfo{pages}{1--6}.
\newblock


\bibitem[\protect\citeauthoryear{Vaswani, Shazeer, Parmar, Uszkoreit, Jones,
  Gomez, Kaiser, and Polosukhin}{Vaswani et~al\mbox{.}}{2017}]%
        {vaswani2017attention}
\bibfield{author}{\bibinfo{person}{Ashish Vaswani}, \bibinfo{person}{Noam
  Shazeer}, \bibinfo{person}{Niki Parmar}, \bibinfo{person}{Jakob Uszkoreit},
  \bibinfo{person}{Llion Jones}, \bibinfo{person}{Aidan~N Gomez},
  \bibinfo{person}{{\L}ukasz Kaiser}, {and} \bibinfo{person}{Illia
  Polosukhin}.} \bibinfo{year}{2017}\natexlab{}.
\newblock \showarticletitle{Attention is all you need}.
\newblock \bibinfo{journal}{\emph{Advances in neural information processing
  systems}}  \bibinfo{volume}{30} (\bibinfo{year}{2017}).
\newblock


\bibitem[\protect\citeauthoryear{Wang, Singh, Michael, Hill, Levy, and
  Bowman}{Wang et~al\mbox{.}}{2018}]%
        {wang2018glue}
\bibfield{author}{\bibinfo{person}{Alex Wang}, \bibinfo{person}{Amanpreet
  Singh}, \bibinfo{person}{Julian Michael}, \bibinfo{person}{Felix Hill},
  \bibinfo{person}{Omer Levy}, {and} \bibinfo{person}{Samuel~R Bowman}.}
  \bibinfo{year}{2018}\natexlab{}.
\newblock \showarticletitle{GLUE: A multi-task benchmark and analysis platform
  for natural language understanding}.
\newblock \bibinfo{journal}{\emph{arXiv preprint arXiv:1804.07461}}
  (\bibinfo{year}{2018}).
\newblock


\bibitem[\protect\citeauthoryear{Wang, Liu, Lin, Lin, and Han}{Wang
  et~al\mbox{.}}{2019a}]%
        {wang2019haq}
\bibfield{author}{\bibinfo{person}{Kuan Wang}, \bibinfo{person}{Zhijian Liu},
  \bibinfo{person}{Yujun Lin}, \bibinfo{person}{Ji Lin}, {and}
  \bibinfo{person}{Song Han}.} \bibinfo{year}{2019}\natexlab{a}.
\newblock \showarticletitle{Haq: Hardware-aware automated quantization with
  mixed precision}. In \bibinfo{booktitle}{\emph{Proceedings of the IEEE/CVF
  Conference on Computer Vision and Pattern Recognition}}.
  \bibinfo{pages}{8612--8620}.
\newblock


\bibitem[\protect\citeauthoryear{Wang, Zhang, Xie, Guo, Liu, and Leng}{Wang
  et~al\mbox{.}}{2021}]%
        {wang2021dual}
\bibfield{author}{\bibinfo{person}{Yang Wang}, \bibinfo{person}{Chen Zhang},
  \bibinfo{person}{Zhiqiang Xie}, \bibinfo{person}{Cong Guo},
  \bibinfo{person}{Yunxin Liu}, {and} \bibinfo{person}{Jingwen Leng}.}
  \bibinfo{year}{2021}\natexlab{}.
\newblock \showarticletitle{Dual-side sparse tensor core}. In
  \bibinfo{booktitle}{\emph{2021 ACM/IEEE 48th Annual International Symposium
  on Computer Architecture (ISCA)}}. IEEE, \bibinfo{pages}{1083--1095}.
\newblock


\bibitem[\protect\citeauthoryear{Wang, Lu, Tao, Zhou, and Tian}{Wang
  et~al\mbox{.}}{2019b}]%
        {wang2019learning}
\bibfield{author}{\bibinfo{person}{Ziwei Wang}, \bibinfo{person}{Jiwen Lu},
  \bibinfo{person}{Chenxin Tao}, \bibinfo{person}{Jie Zhou}, {and}
  \bibinfo{person}{Qi Tian}.} \bibinfo{year}{2019}\natexlab{b}.
\newblock \showarticletitle{Learning channel-wise interactions for binary
  convolutional neural networks}. In \bibinfo{booktitle}{\emph{Proceedings of
  the IEEE/CVF Conference on Computer Vision and Pattern Recognition}}.
  \bibinfo{pages}{568--577}.
\newblock


\bibitem[\protect\citeauthoryear{Wei, Zhang, Zhang, Gong, Zhang, Zhang, Yu, and
  Liu}{Wei et~al\mbox{.}}{2022}]%
        {wei2022outlier}
\bibfield{author}{\bibinfo{person}{Xiuying Wei}, \bibinfo{person}{Yunchen
  Zhang}, \bibinfo{person}{Xiangguo Zhang}, \bibinfo{person}{Ruihao Gong},
  \bibinfo{person}{Shanghang Zhang}, \bibinfo{person}{Qi Zhang},
  \bibinfo{person}{Fengwei Yu}, {and} \bibinfo{person}{Xianglong Liu}.}
  \bibinfo{year}{2022}\natexlab{}.
\newblock \showarticletitle{Outlier Suppression: Pushing the Limit of Low-bit
  Transformer Language Models}. In \bibinfo{booktitle}{\emph{Advances in Neural
  Information Processing Systems}}, \bibfield{editor}{\bibinfo{person}{Alice~H.
  Oh}, \bibinfo{person}{Alekh Agarwal}, \bibinfo{person}{Danielle Belgrave},
  {and} \bibinfo{person}{Kyunghyun Cho}} (Eds.).
\newblock
\urldef\tempurl%
\url{https://openreview.net/forum?id=yW5zeRSFdZ}
\showURL{%
\tempurl}


\bibitem[\protect\citeauthoryear{{Wikipedia contributors}}{{Wikipedia
  contributors}}{2022}]%
        {enwiki:1116981313}
\bibfield{author}{\bibinfo{person}{{Wikipedia contributors}}.}
  \bibinfo{year}{2022}\natexlab{}.
\newblock \bibinfo{title}{68–95–99.7 rule --- {Wikipedia}{,} The Free
  Encyclopedia}.
\newblock
\newblock
\newblock
\shownote{[Online].}


\bibitem[\protect\citeauthoryear{Yang, Breslow, Mars, and Tang}{Yang
  et~al\mbox{.}}{2013}]%
        {BubbleFlux}
\bibfield{author}{\bibinfo{person}{Hailong Yang}, \bibinfo{person}{Alex~D.
  Breslow}, \bibinfo{person}{Jason Mars}, {and} \bibinfo{person}{Lingjia
  Tang}.} \bibinfo{year}{2013}\natexlab{}.
\newblock \showarticletitle{Bubble-flux: precise online QoS management for
  increased utilization in warehouse scale computers}. In
  \bibinfo{booktitle}{\emph{The 40th Annual International Symposium on Computer
  Architecture ({ISCA})}}.
\newblock
\urldef\tempurl%
\url{https://doi.org/10.1145/2485922.2485974}
\showDOI{\tempurl}


\bibitem[\protect\citeauthoryear{Zadeh, Edo, Awad, and Moshovos}{Zadeh
  et~al\mbox{.}}{2020}]%
        {zadeh2020gobo}
\bibfield{author}{\bibinfo{person}{Ali~Hadi Zadeh}, \bibinfo{person}{Isak Edo},
  \bibinfo{person}{Omar~Mohamed Awad}, {and} \bibinfo{person}{Andreas
  Moshovos}.} \bibinfo{year}{2020}\natexlab{}.
\newblock \showarticletitle{Gobo: Quantizing attention-based nlp models for low
  latency and energy efficient inference}. In \bibinfo{booktitle}{\emph{2020
  53rd Annual IEEE/ACM International Symposium on Microarchitecture (MICRO)}}.
  IEEE, \bibinfo{pages}{811--824}.
\newblock


\bibitem[\protect\citeauthoryear{Zafrir, Boudoukh, Izsak, and
  Wasserblat}{Zafrir et~al\mbox{.}}{2019}]%
        {zafrir2019q8bert}
\bibfield{author}{\bibinfo{person}{Ofir Zafrir}, \bibinfo{person}{Guy
  Boudoukh}, \bibinfo{person}{Peter Izsak}, {and} \bibinfo{person}{Moshe
  Wasserblat}.} \bibinfo{year}{2019}\natexlab{}.
\newblock \showarticletitle{Q8bert: Quantized 8bit bert}. In
  \bibinfo{booktitle}{\emph{2019 Fifth Workshop on Energy Efficient Machine
  Learning and Cognitive Computing-NeurIPS Edition (EMC2-NIPS)}}. IEEE,
  \bibinfo{pages}{36--39}.
\newblock


\bibitem[\protect\citeauthoryear{Zhang, Li, Sun, Guan, Xiao, and Cong}{Zhang
  et~al\mbox{.}}{2015}]%
        {zhang2015optimizing}
\bibfield{author}{\bibinfo{person}{Chen Zhang}, \bibinfo{person}{Peng Li},
  \bibinfo{person}{Guangyu Sun}, \bibinfo{person}{Yijin Guan},
  \bibinfo{person}{Bingjun Xiao}, {and} \bibinfo{person}{Jason Cong}.}
  \bibinfo{year}{2015}\natexlab{}.
\newblock \showarticletitle{Optimizing fpga-based accelerator design for deep
  convolutional neural networks}. In \bibinfo{booktitle}{\emph{Proceedings of
  the 2015 ACM/SIGDA international symposium on field-programmable gate
  arrays}}. \bibinfo{pages}{161--170}.
\newblock


\bibitem[\protect\citeauthoryear{Zhang, Yang, Ye, and Hua}{Zhang
  et~al\mbox{.}}{2018}]%
        {zhang2018lq}
\bibfield{author}{\bibinfo{person}{Dongqing Zhang}, \bibinfo{person}{Jiaolong
  Yang}, \bibinfo{person}{Dongqiangzi Ye}, {and} \bibinfo{person}{Gang Hua}.}
  \bibinfo{year}{2018}\natexlab{}.
\newblock \showarticletitle{Lq-nets: Learned quantization for highly accurate
  and compact deep neural networks}. In \bibinfo{booktitle}{\emph{Proceedings
  of the European conference on computer vision (ECCV)}}.
  \bibinfo{pages}{365--382}.
\newblock


\bibitem[\protect\citeauthoryear{Zhang, Du, Zhang, Lan, Liu, Li, Guo, Chen, and
  Chen}{Zhang et~al\mbox{.}}{2016}]%
        {zhang2016cambricon}
\bibfield{author}{\bibinfo{person}{Shijin Zhang}, \bibinfo{person}{Zidong Du},
  \bibinfo{person}{Lei Zhang}, \bibinfo{person}{Huiying Lan},
  \bibinfo{person}{Shaoli Liu}, \bibinfo{person}{Ling Li}, \bibinfo{person}{Qi
  Guo}, \bibinfo{person}{Tianshi Chen}, {and} \bibinfo{person}{Yunji Chen}.}
  \bibinfo{year}{2016}\natexlab{}.
\newblock \showarticletitle{Cambricon-X: An accelerator for sparse neural
  networks}. In \bibinfo{booktitle}{\emph{2016 49th Annual IEEE/ACM
  International Symposium on Microarchitecture (MICRO)}}. IEEE,
  \bibinfo{pages}{1--12}.
\newblock


\bibitem[\protect\citeauthoryear{Zhang, Roller, Goyal, Artetxe, Chen, Chen,
  Dewan, Diab, Li, Lin, et~al\mbox{.}}{Zhang et~al\mbox{.}}{2022}]%
        {zhang2022opt}
\bibfield{author}{\bibinfo{person}{Susan Zhang}, \bibinfo{person}{Stephen
  Roller}, \bibinfo{person}{Naman Goyal}, \bibinfo{person}{Mikel Artetxe},
  \bibinfo{person}{Moya Chen}, \bibinfo{person}{Shuohui Chen},
  \bibinfo{person}{Christopher Dewan}, \bibinfo{person}{Mona Diab},
  \bibinfo{person}{Xian Li}, \bibinfo{person}{Xi~Victoria Lin},
  {et~al\mbox{.}}} \bibinfo{year}{2022}\natexlab{}.
\newblock \showarticletitle{Opt: Open pre-trained transformer language models}.
\newblock \bibinfo{journal}{\emph{arXiv preprint arXiv:2205.01068}}
  (\bibinfo{year}{2022}).
\newblock


\bibitem[\protect\citeauthoryear{Zheng, Jia, Sun, Wu, Yu, Haj{-}Ali, Wang,
  Yang, Zhuo, Sen, Gonzalez, and Stoica}{Zheng et~al\mbox{.}}{2020a}]%
        {Ansor}
\bibfield{author}{\bibinfo{person}{Lianmin Zheng}, \bibinfo{person}{Chengfan
  Jia}, \bibinfo{person}{Minmin Sun}, \bibinfo{person}{Zhao Wu},
  \bibinfo{person}{Cody~Hao Yu}, \bibinfo{person}{Ameer Haj{-}Ali},
  \bibinfo{person}{Yida Wang}, \bibinfo{person}{Jun Yang},
  \bibinfo{person}{Danyang Zhuo}, \bibinfo{person}{Koushik Sen},
  \bibinfo{person}{Joseph~E. Gonzalez}, {and} \bibinfo{person}{Ion Stoica}.}
  \bibinfo{year}{2020}\natexlab{a}.
\newblock \showarticletitle{Ansor: Generating High-Performance Tensor Programs
  for Deep Learning}. In \bibinfo{booktitle}{\emph{14th {USENIX} Symposium on
  Operating Systems Design and Implementation ({OSDI})}}.
\newblock
\urldef\tempurl%
\url{https://doi.org/10.5555/3488766.3488815}
\showDOI{\tempurl}


\bibitem[\protect\citeauthoryear{Zheng, Liang, Wang, Chen, and Sheng}{Zheng
  et~al\mbox{.}}{2020b}]%
        {FlexTensor}
\bibfield{author}{\bibinfo{person}{Size Zheng}, \bibinfo{person}{Yun Liang},
  \bibinfo{person}{Shuo Wang}, \bibinfo{person}{Renze Chen}, {and}
  \bibinfo{person}{Kaiwen Sheng}.} \bibinfo{year}{2020}\natexlab{b}.
\newblock \showarticletitle{FlexTensor: An Automatic Schedule Exploration and
  Optimization Framework for Tensor Computation on Heterogeneous System}. In
  \bibinfo{booktitle}{\emph{Architectural Support for Programming Languages and
  Operating Systems, Lausanne ({ASPLOS})}}.
\newblock
\urldef\tempurl%
\url{https://doi.org/10.1145/3373376.3378508}
\showDOI{\tempurl}


\bibitem[\protect\citeauthoryear{Zhou, Wu, Ni, Zhou, Wen, and Zou}{Zhou
  et~al\mbox{.}}{2016}]%
        {zhou2016dorefa}
\bibfield{author}{\bibinfo{person}{Shuchang Zhou}, \bibinfo{person}{Yuxin Wu},
  \bibinfo{person}{Zekun Ni}, \bibinfo{person}{Xinyu Zhou}, \bibinfo{person}{He
  Wen}, {and} \bibinfo{person}{Yuheng Zou}.} \bibinfo{year}{2016}\natexlab{}.
\newblock \showarticletitle{Dorefa-net: Training low bitwidth convolutional
  neural networks with low bitwidth gradients}.
\newblock \bibinfo{journal}{\emph{arXiv preprint arXiv:1606.06160}}
  (\bibinfo{year}{2016}).
\newblock


\bibitem[\protect\citeauthoryear{Zhou, Du, Guo, Liu, Liu, Wang, Zhou, Li, Chen,
  and Chen}{Zhou et~al\mbox{.}}{2018}]%
        {zhou2018cambricon}
\bibfield{author}{\bibinfo{person}{Xuda Zhou}, \bibinfo{person}{Zidong Du},
  \bibinfo{person}{Qi Guo}, \bibinfo{person}{Shaoli Liu},
  \bibinfo{person}{Chengsi Liu}, \bibinfo{person}{Chao Wang},
  \bibinfo{person}{Xuehai Zhou}, \bibinfo{person}{Ling Li},
  \bibinfo{person}{Tianshi Chen}, {and} \bibinfo{person}{Yunji Chen}.}
  \bibinfo{year}{2018}\natexlab{}.
\newblock \showarticletitle{Cambricon-S: Addressing irregularity in sparse
  neural networks through a cooperative software/hardware approach}. In
  \bibinfo{booktitle}{\emph{2018 51st Annual IEEE/ACM International Symposium
  on Microarchitecture (MICRO)}}. IEEE, \bibinfo{pages}{15--28}.
\newblock


\bibitem[\protect\citeauthoryear{Zhou, Leng, Song, Lu, Wang, Li, Guo, Shen, Li,
  Lin, et~al\mbox{.}}{Zhou et~al\mbox{.}}{2023}]%
        {zhou2023ugrapher}
\bibfield{author}{\bibinfo{person}{Yangjie Zhou}, \bibinfo{person}{Jingwen
  Leng}, \bibinfo{person}{Yaoxu Song}, \bibinfo{person}{Shuwen Lu},
  \bibinfo{person}{Mian Wang}, \bibinfo{person}{Chao Li},
  \bibinfo{person}{Minyi Guo}, \bibinfo{person}{Wenting Shen},
  \bibinfo{person}{Yong Li}, \bibinfo{person}{Wei Lin}, {et~al\mbox{.}}}
  \bibinfo{year}{2023}\natexlab{}.
\newblock \showarticletitle{uGrapher: High-Performance Graph Operator
  Computation via Unified Abstraction for Graph Neural Networks}. In
  \bibinfo{booktitle}{\emph{Proceedings of the 28th ACM International
  Conference on Architectural Support for Programming Languages and Operating
  Systems, Volume 2}}. \bibinfo{pages}{878--891}.
\newblock


\bibitem[\protect\citeauthoryear{Zhou, Yang, Guo, Leng, Liang, Chen, Guo, and
  Zhu}{Zhou et~al\mbox{.}}{2021}]%
        {zhou2021characterizing}
\bibfield{author}{\bibinfo{person}{Yangjie Zhou}, \bibinfo{person}{Mengtian
  Yang}, \bibinfo{person}{Cong Guo}, \bibinfo{person}{Jingwen Leng},
  \bibinfo{person}{Yun Liang}, \bibinfo{person}{Quan Chen},
  \bibinfo{person}{Minyi Guo}, {and} \bibinfo{person}{Yuhao Zhu}.}
  \bibinfo{year}{2021}\natexlab{}.
\newblock \showarticletitle{Characterizing and demystifying the implicit
  convolution algorithm on commercial matrix-multiplication accelerators}. In
  \bibinfo{booktitle}{\emph{2021 IEEE International Symposium on Workload
  Characterization (IISWC)}}. IEEE, \bibinfo{pages}{214--225}.
\newblock


\bibitem[\protect\citeauthoryear{Zhu, Wu, Diao, Ke, Li, Zhang, Xue, Ma, Xia,
  Cui, Yang, Yang, Zhou, Cidon, and Pekhimenko}{Zhu et~al\mbox{.}}{2022}]%
        {Roller}
\bibfield{author}{\bibinfo{person}{Hongyu Zhu}, \bibinfo{person}{Ruofan Wu},
  \bibinfo{person}{Yijia Diao}, \bibinfo{person}{Shanbin Ke},
  \bibinfo{person}{Haoyu Li}, \bibinfo{person}{Chen Zhang},
  \bibinfo{person}{Jilong Xue}, \bibinfo{person}{Lingxiao Ma},
  \bibinfo{person}{Yuqing Xia}, \bibinfo{person}{Wei Cui}, \bibinfo{person}{Fan
  Yang}, \bibinfo{person}{Mao Yang}, \bibinfo{person}{Lidong Zhou},
  \bibinfo{person}{Asaf Cidon}, {and} \bibinfo{person}{Gennady Pekhimenko}.}
  \bibinfo{year}{2022}\natexlab{}.
\newblock \showarticletitle{{ROLLER}: Fast and Efficient Tensor Compilation for
  Deep Learning}. In \bibinfo{booktitle}{\emph{16th USENIX Symposium on
  Operating Systems Design and Implementation (OSDI 22)}}.
  \bibinfo{pages}{233--248}.
\newblock


\bibitem[\protect\citeauthoryear{Zhu, Zhang, Gu, and Xie}{Zhu
  et~al\mbox{.}}{2019}]%
        {zhu2019sparse}
\bibfield{author}{\bibinfo{person}{Maohua Zhu}, \bibinfo{person}{Tao Zhang},
  \bibinfo{person}{Zhenyu Gu}, {and} \bibinfo{person}{Yuan Xie}.}
  \bibinfo{year}{2019}\natexlab{}.
\newblock \showarticletitle{Sparse tensor core: Algorithm and hardware
  co-design for vector-wise sparse neural networks on modern gpus}. In
  \bibinfo{booktitle}{\emph{Proceedings of the 52nd Annual IEEE/ACM
  International Symposium on Microarchitecture}}. \bibinfo{pages}{359--371}.
\newblock


\bibitem[\protect\citeauthoryear{Zhuang, Tan, Liu, Liu, Reid, and Shen}{Zhuang
  et~al\mbox{.}}{2021}]%
        {zhuang2021effective}
\bibfield{author}{\bibinfo{person}{Bohan Zhuang}, \bibinfo{person}{Mingkui
  Tan}, \bibinfo{person}{Jing Liu}, \bibinfo{person}{Lingqiao Liu},
  \bibinfo{person}{Ian Reid}, {and} \bibinfo{person}{Chunhua Shen}.}
  \bibinfo{year}{2021}\natexlab{}.
\newblock \showarticletitle{Effective training of convolutional neural networks
  with low-bitwidth weights and activations}.
\newblock \bibinfo{journal}{\emph{IEEE Transactions on Pattern Analysis and
  Machine Intelligence}} (\bibinfo{year}{2021}).
\newblock


\end{thebibliography}
